\documentclass{article}

\usepackage{PRIMEarxiv}
\usepackage{appendix}
\usepackage[utf8]{inputenc} % allow utf-8 input
\usepackage[T1]{fontenc}    % use 8-bit T1 fonts
\usepackage{hyperref}       % hyperlinks
\usepackage{url}            % simple URL typesetting
\usepackage{booktabs}       % professional-quality tables
\usepackage{amsfonts}       % blackboard math symbols
\usepackage{nicefrac}       % compact symbols for 1/2, etc.
\usepackage{microtype}      % microtypography
\usepackage{lipsum}
\usepackage{fancyhdr}       % header
\usepackage{graphicx}       % graphics
\graphicspath{{media/}}     % organize your images and other figures under media/ folder
\usepackage{cite}
\usepackage{array}
\usepackage{amsmath,amssymb,amsfonts}
\usepackage{amssymb}
\usepackage{algorithm}
\usepackage{algorithmic}
\usepackage{makecell}
\usepackage{colortbl}
\usepackage{multirow}
\usepackage{threeparttable}
\usepackage[normalem]{ulem}
\usepackage{graphicx}
\usepackage{bm}
\usepackage[caption=false,font=scriptsize,labelfont=rm,textfont=rm]{subfig}

\usepackage{graphicx}
\usepackage{pifont}
\usepackage{textcomp}
\usepackage{xcolor}
\def\BibTeX{{\rm B\kern-.05em{\sc i\kern-.025em b}\kern-.08em
    T\kern-.1667em\lower.7ex\hbox{E}\kern-.125emX}}
\usepackage{acronym}
\usepackage{xspace}

\newcommand{\ie}{i.e.,\xspace}
\newcommand{\eg}{e.g.,\xspace}

\newcommand{\etal}{\textit{et al.}\xspace}
\newcommand{\eat}[1]{}

\acrodef{iot}[IoT]{Internet of Things}

\acrodef{sgd}[FedSGD]{Federated Stochastic Gradient Descent}
\newcommand{\sgd}{\ac{sgd}\xspace}
\acrodef{avg}[FedAvg]{Federated Averaging}
\newcommand{\avg}{\ac{avg}\xspace}

\makeatletter

\newcommand{\Rmnum}[1]{\expandafter\@slowromancap\romannumeral #1@}
\title{An Experimental Study of Different Aggregation Schemes in Semi-Asynchronous Federated Learning
}

\author{
  Yunbo Li, Jiaping Gui*, Yue Wu* \\
  School of Cyber Science and Engineering \\
  Shanghai Jiao Tong University \\
  Shanghai, China\\
  \texttt{\{li-yun-bo, jgui, wuyue\}@sjtu.edu.cn} \\
}

\begin{document}
\maketitle

\begin{abstract}
Federated learning is highly valued due to its high-performance computing in distributed environments while safeguarding data privacy. To address resource heterogeneity, researchers have proposed a semi-asynchronous federated learning (SAFL) architecture. However, the performance gap between different aggregation targets in SAFL remain unexplored. 

In this paper, we systematically compare the performance between two algorithm modes, FedSGD and FedAvg that correspond to aggregating gradients and models, respectively. Our results across various task scenarios indicate these two modes exhibit a substantial performance gap. Specifically, FedSGD achieves higher accuracy and faster convergence but experiences more severe fluctuates in accuracy, whereas FedAvg excels in handling straggler issues but converges slower with reduced accuracy.
\end{abstract}

% keywords can be removed
\keywords{Federated Learning \and Semi-Asynchronous Distribution Learning \and Performance Gap \and Prediction Accuracy \and Convergence Stability}

\section{Introduction}
% Federated Learning backgroud
In recent years, federated learning (FL) has emerged as a popular distributed artificial intelligence paradigm, which enables a federation of participating devices (which we refer to as clients) to collaboratively train a machine learning model under the coordination of a central server~\cite{mcmahan2017communication, li2019convergence, li2021fedbn, zhu2021federated, guo2021lightfed,huang2023distributed,qi2023hwamei}. In FL, each client leverages its local data to train a model and periodically shares the local model updates with the central server. And then, the server aggregates model parameters from parts of clients to obtain a global model~\cite{mcmahan2017communication}. In this process, the client does not directly transmit the raw data to the server, thus preserving client (user) privacy. FL has been widely studied and applied to real-world scenarios, such as enterprise infrastructures~\cite{ludwig2020ibm}, online banks~\cite{long2020federated} and medical health~\cite{wu2020fedhome}. 

FL is operating in two foundational modes: \sgd algorithm mode and \avg algorithm mode~\cite{mcmahan2017communication}. \sgd differs from \avg in what/how information is aggregated from the local model. In particular, \sgd shares with the central server the local gradients while \avg shares the local model weights after one or more epochs of training. If clients start from the same initialization with some constraints on the model training~\cite{nasirigerdeh2023utility}, \sgd and \avg achieve comparable performance w.r.t. accuracy and convergence. Hence, a conventional procedure in real-world scenarios is to select either algorithm mode without discerning subtleties between them. 

A key problem in this procedure is that it depends on the perception that, aside from aggregation strategy, there are no additional differences between two algorithm modes. While this is true for synchronous federated learning (SFL)~\cite{nasirigerdeh2023utility} where clients communicate with the server synchronously under the same condition, this fails to account for the system heterogeneity among clients. Their differences come in several forms: computation data scales, for which the training data is not independent and identically distributed (Non-IID) on the local devices~\cite{li2019convergence}; computation resources which affect the data processing rate~\cite{zhao2018federated}; and communication environments which vary significantly in a distributed physical world~\cite{sattler2019robust}. To train a global model on the server side, it is necessary for faster clients to wait for those with poorer training performance and communication quality, resulting in the so-called straggler problem. Consequently, these faster clients waste their time and resources for waiting. To address these issues, researchers propose semi-asynchronous federated learning (SAFL) to improve the system's communication efficiency and reduce the idle time of high-performance participants~\cite{wu2020safa,ma2021fedsa,cheng2022aafl}.

However, our understanding of the aforementioned straggler problem, its characterization and the extent to which it can impact on the overall SAFL between \sgd and \avg, is still quite limited. In this paper, we address these gaps by formally defining a set of metrics for characterizing two algorithm modes. We ground our quantitative assessments using external measurements to qualitatively investigate distinctness (\eg accuracy and loss). To the best of our knowledge, this is the first piece of work that systematically investigates the performance differences between \sgd and \avg in SAFL. Notice that we do not evaluate fully asynchronous FL~\cite{xie2019asynchronous,zhu2022online,baccarelli2022afafed} due to its massive communication demands which may cause the central server to crash, and its severe performance degradation especially on Non-IID data.

We first present the results of our investigation into the performance of \sgd and \avg in the SAFL paradigm. To carry out this investigation, we performed an extensive empirical analysis of different scenarios that span across computer vision (CV) and natural language processing (NLP) tasks, with five datasets, four models, and six data distributions. The results of our investigation show that there is, in fact, a notable performance gap between \sgd and \avg for the training of a global model in the semi-asynchronous system. Particularly, our results show that in the presence of stragglers, \sgd demonstrates on average 8.27\% and 6.03\% higher accuracy in CV and NLP tasks, respectively. As for convergence, \sgd has a faster convergence speed but a less robust learning process with severe oscillations in model accuracy. In contrast, \avg excels in handling straggler issues with no severe oscillations, but at the cost of slower convergence of the global model and a noticeable decrease in model accuracy. In terms of resource consumption, on average \avg consumes 6.2\% more transmission channel load and 4.3\% more training duration than \sgd. %We believe that \sgd demonstrates a shorter training duration due to its reduced operations.
Overall, we believe that these findings are significant and will help to inform researchers so they can better weigh the tradeoffs of incorporating a specific aggregation strategy in their SAFL scenario.

Our main contributions can be summarized as follows.

\begin{itemize}
	%\item We analyze the similarities and differences between FedSGD and FedAvg under synchronous and semi-asynchronous FL paradigms.
	\item To our knowledge, we perform the first systematic investigation to quantify the performance gap between \sgd and \avg in SAFL.
	\item We conduct a wealth of experiments to measure the metrics that determine the impact of two foundational aggregation strategies on the global model training in SAFL.
	\item We provide a qualitative explanation to the performance gap, attributing to model staleness and data nature of aggregation in semi-asynchronous communication scenarios.
	% \item We further propose \tool to bridge the performance gap between \sgd and \avg. We show that \tool performs a more robust and consistent model training compared to SOTA algorithms.
\end{itemize}

\section{Problem Formulation}
In this section, we first introduce the basic FL mechanism. Then, we explore federated learning under different communication systems, \ie synchronous vs. semi-asynchronous, and provide a summarized overview of each of the corresponding algorithm modes.

\subsection{Basics of Federated Learning}

We consider an FL system consists of a set of clients $\mathcal{C} = \{C_1, C_2,...,C_N\}$ and a server $S$. The server has a global model $M_g$ whose weights are denoted as $w_g$ and a collection $\mathcal{S}$ that stores all the data uploaded by the clients. Each client $C_i$ has its own model $M_i$ and dataset $\mathcal{D}_i$ for training, where each data point $k \in \mathcal{D}_i$ is denoted by $(x_k, y_k)$.  In addition, some clients may use a different optimization method to train their local models, but the model architecture among clients is consistent. We use 
$w_i$ to denote the local model parameters and $L_i$ to represent the empirical risk function of the model in $C_i$. Therefore, $C_i$ can use a prediction loss function $l$ to train its local model as: 
\begin{equation}
    \min_{w_i} \ L_i(w_i) \triangleq \frac{1}{|\mathcal{D}_i|} \sum_{k \in \mathcal{D}_i} l(k;w_i) = \frac{1}{|\mathcal{D}_i|} \sum_{k \in \mathcal{D}_i} l(y_k, w_i(x_k))
\end{equation}

To optimize the local model, without loss of generality, we presume that each client employs the mini-batch SGD optimization method in the subsequent discussions in this paper since FL does not prescribe specific optimization algorithms for each client. The entire training process of client $C_i$ can be represented by the following equation:
\begin{equation}
\begin{split}
        w_{i,e+1} &= w_{i, e} - \eta_i\nabla L_{i,e} \\ &= w_{i, e} - \frac{\eta_i}{|\mathcal{D}_i|}  \sum_{B_i \in \mathcal{B}_i} \nabla l(w_{i,e}; B_i),
\end{split}
\label{equ:mini-batch}
\end{equation}
where $\eta_i$ is the learning rate utilized by this client.

% \begin{table}[!htp]
%     \normalsize
%     \caption{Summary of Key Notations. \jp{display issue of caption. move this table to appendix}}
%     \centering
%     \begin{tabular*}{\linewidth}{c l}
%     \hline
%        Symbol  & Description \\
%     \hline
%          $\mathcal{C}$ & the set of clients \\
%          $C_i$ & the \textit{i}-th client\\
%          $S$ & the server \\
%          $\mathcal{S}$ & the set of uploaded data stored on the server  \\
%          $w_g$ & parameters of global model \\
%          $\mathcal{D}_i$ & the dataset for training in $C_i$ \\
%          $D_i$ & the size of $\mathcal{D}_i$ \\
%          $w_i$ & parameters of the local model in $C_i$\\
%          $L_i$ & the loss function of $w_i$ \\
%          $\nabla L_i$ & the gradient of $L_i$ \\
%          \multirow{2}*{$K$} & the activation rate in synchronous FL \\
%          & the target amount in semi-asynchronous FL \\
%          $E$ & the maximum number of local epochs  \\
%          $T$ & the maximum number of global epochs \\
%          $\eta_i$ & the learning rate of $C_i$ \\
%          $\eta$ & the learning rate of $S$ \\
%          $B$ & a batch of data \\
%          $\mathcal{B}$ & the set of batch data \\
%          $\tau_s^t$ & the staleness of the data $s$ in $t$-th global epoch \\
%          $Acc_t$ & the target accuracy to evaluate convergence \\
%          $ost$ & the oscillation threshold \\
%          $T_s$ & the epochs that achieve the target accuracy firstly\\
%          $T_f$ & the epochs after which remains stable \\
%     \hline
%     \end{tabular*}
% \end{table}

\subsection{Communication Strategies}
\begin{figure*}[!htp]
\centering
\subfloat[Synchronous Federated Learning.]{\includegraphics[width=3.3in]{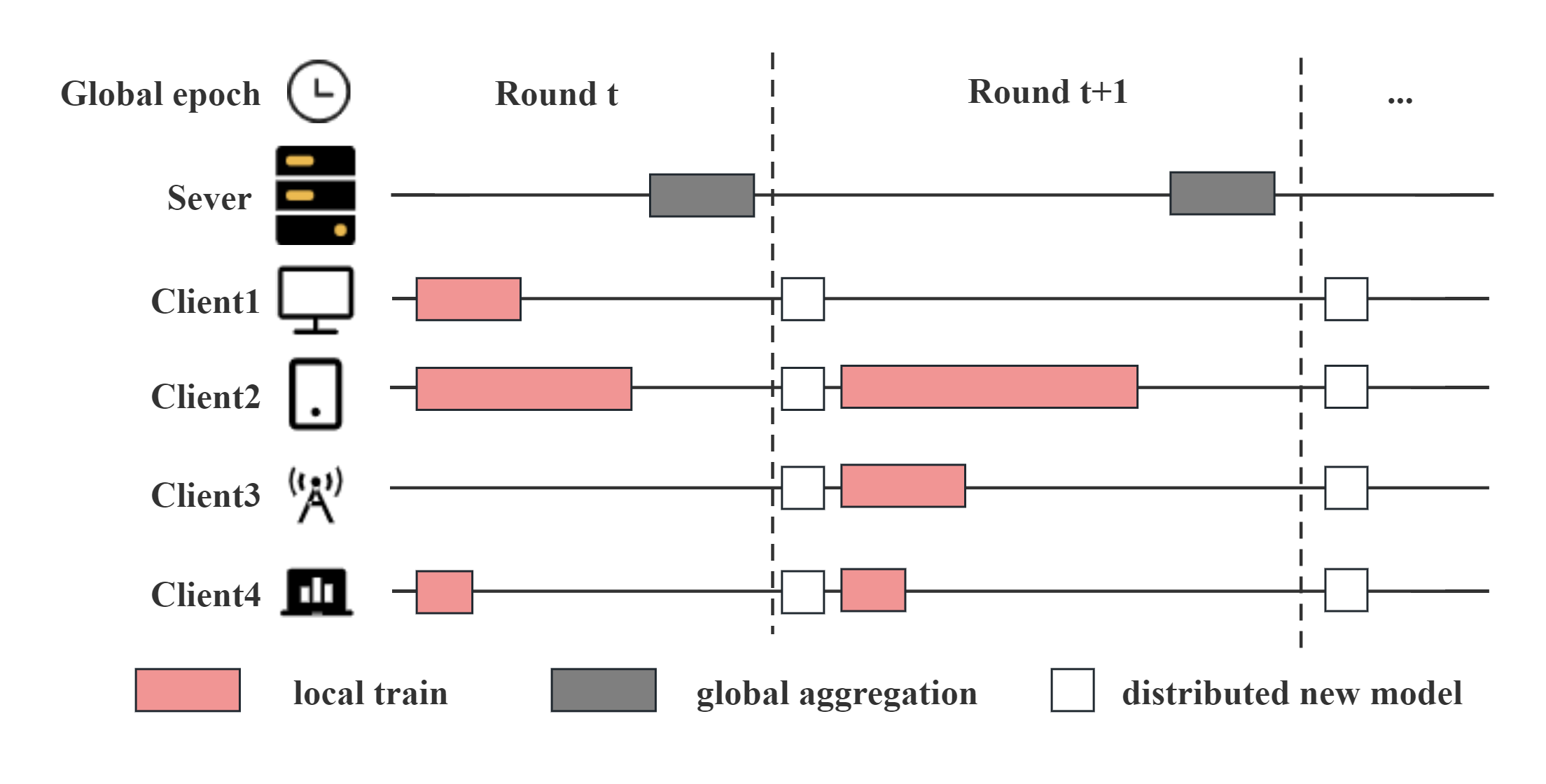}%
\label{fig_syn_fl}}
\hfil
\subfloat[Semi-Asynchronous Federated Learning.]{\includegraphics[width=3in]{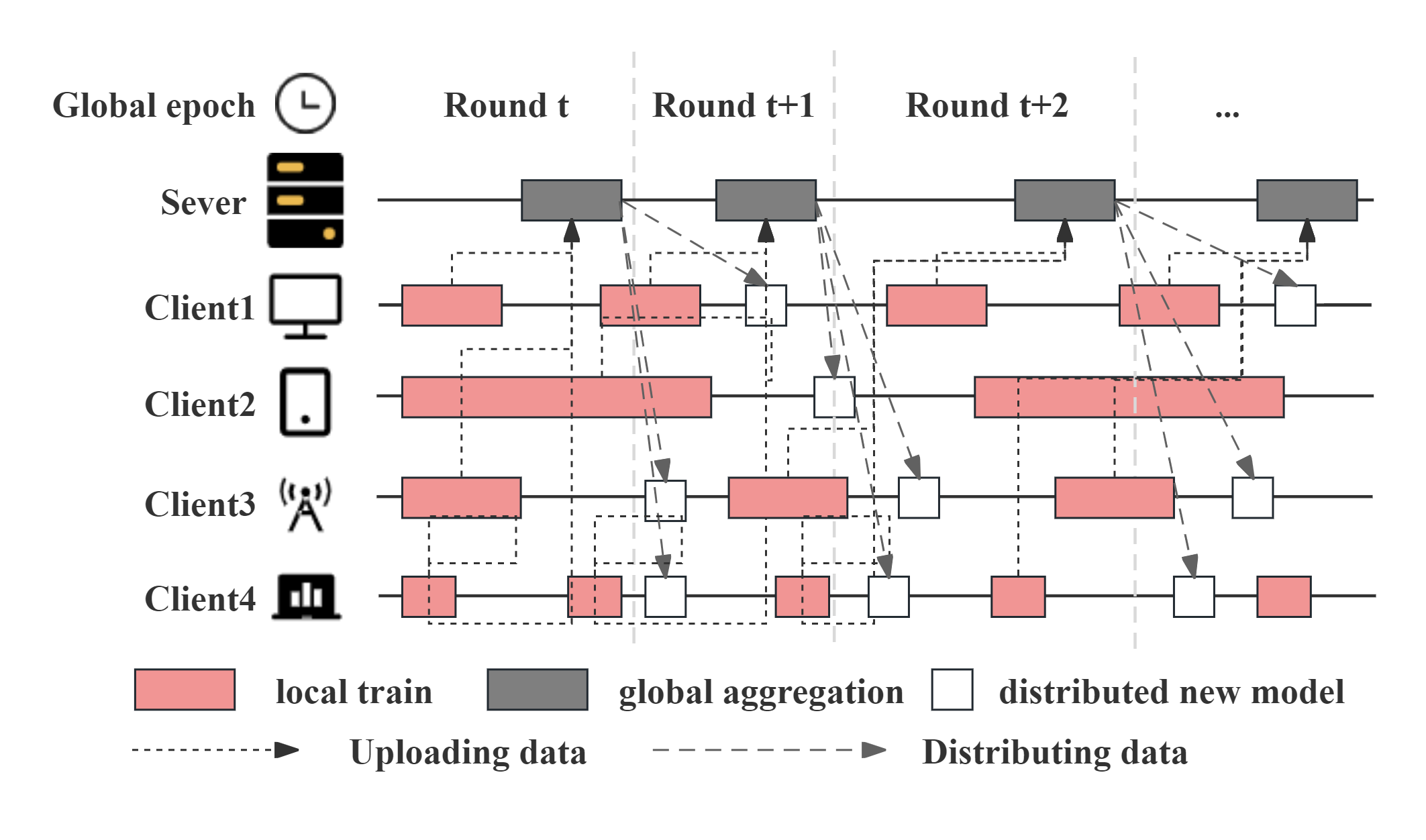}%
\label{fig_safl}}
\caption{Computational process in synchronous federated learning and semi-asynchronous federated learning with $K=3$.}
\label{fig_compare}
% \vspace*{-3ex}
\end{figure*}

Below we outline the differences of communication strategies between SFL and SAFL. Figure~\ref{fig_compare} illustrates the computational process of FL involving four clients with heterogeneous resources.

\subsubsection{Synchronous}
%In this subsection, we will provide the detailed process of synchronous FL. 
On the server side, in each round of SFL, the server randomly selects a set of active clients, denoted by $\mathcal{AC} \subseteq \mathcal{C}$, that occupy a specific proportion of all distributed clients. For example, in Figure~\ref{fig_syn_fl}, \texttt{Client1}, \texttt{Client2} and \texttt{Client4} are activated in round $t$. The server first leverages local model updates from these active clients to train the global model. We refer to the proportion of clients that are selected as the activation rate, denoted by $K$ that is equal to $|\mathcal{S}|$. The server will not update the global model until it receives shared data from all clients in $\mathcal{AC}$. Then, the server broadcasts the new model to all clients (\ie $\mathcal{C}$). On the client side, after receiving a new model, the client replaces the local model with the new one. If the new model is activated, the local training will be started. Otherwise, the client has to wait in idle state.
   
In SFL, the server has a strong control over individual clients due to the synchronized aggregation of local model updates. However, the training process could be slow since there exist runtime differences (\eg communication conditions and hardware variations) among active clients. Consequently, some devices may be forced to wait (in idle) for late-arriving clients (stragglers) such that the server can finish one round of the global model training. As shown in Figure~\ref{fig_syn_fl}, \texttt{Client4} is forced to wait for \texttt{Client2} to complete the local training in each round. This results in a waste of both computational resources and processing time on \texttt{Client4}.

\subsubsection{Semi-asynchronous}
%We will outline the detailed process of semi-asynchronous FL below. 
In contrast, SAFL provides a better solution to the straggler issue. As shown in Figure~\ref{fig_safl}, in a semi-asynchronous flavor, all clients train their local model at their own pace and the server passively accepts local updates that are uploaded by clients. 
For example, at round $t$, the server receives a sufficient amount of data from \texttt{Client1}, \texttt{Client3}, and \texttt{Client4} (\textit{K} = 3). Then, the server performs an aggregation and deploys the latest global model across distributed clients. On the client side, each individual device (\eg \texttt{Client2}) undergoes continuous training and uploads model updates (\eg at round $t+1$) after completing each local epoch without interruption. If no new global model is received, the client continues training (\eg \texttt{Client1} and \texttt{Client4} at round $t$); otherwise, it updates the local model to the latest global one.
 
\eat{  
\subsubsection{Comparison}
Synchronous FL implements that the server has a strong control over individual clients by performing a synchronized aggregation of model updates from these clients in synchronous FL. However, the training process can be slow since there exist runtime differences (\eg communication conditions and performance) among active clients. Consequently, some devices may be forced to wait (in idle time) for late-arriving clients (stragglers) such that the server finishes one round of the global model training. This results in a waste of both computational resources and processing time.

Instead, semi-asynchronous FL improves the communication efficiency of federated learning by addressing the straggler problem. However, models trained through semi-asynchronous FL typically do not achieve the same quality as those obtained through synchronous FL. Due to the complexity of data distribution in the real world, semi-asynchronous FL can lead to model bias. That said, the global model trained through semi-asynchronous FL fits the data distribution of clients that have faster update speeds better than those with slower speeds.
}

\section{Aggregation Strategies in Federated Learning}

In this section, we first introduce two different aggregation strategies, targeting gradients and model parameters, respectively. These strategies involve the fundamental FedSGD/FedAvg, as well as their variants that are proposed to mitigate the performance impact of straggler clients on model aggregation. Finally, we outline our proposed optimization algorithm.

\subsection{Aggregation Strategy with Gradients as Targets}
\subsubsection{FedSGD}
In FedSGD, the server aggregates gradients from clients to train the global model. Specifically, each client $C_i$ first computes the cumulative gradient $\nabla L_i$ while training its local model as:
\begin{equation}
    \nabla L_i  = \frac{1}{|\mathcal{D}_i|} \sum_{B_i \in \mathcal{B}_i} \nabla l(w_{i}; B_i)
\label{equ:local_accumu_gradient}
\end{equation}

Then, $C_i$ transfers $\nabla L_i$ to the server, upon completion of local epoch computations. The server stores $\nabla L_i$ in the collection $\mathcal{S}$. When $\mathcal{S}$ meets specific conditions (\eg $|\mathcal{S}| = K$ or all activated clients have uploaded gradients), the server aggregates all gradients and updates the global model at the $t$-th global epoch according to
\begin{equation}
    \nabla L = \frac{1}{|\mathcal{S}|} \sum_{\nabla L_i \in \mathcal{S}} \nabla L_i 
\label{equ:fedsgd_agg}
\end{equation}
\begin{equation}
    w_g^t = w_g^{t-1} - \eta \nabla L,
\label{equ:fedsgd_update}
\end{equation}
where $\eta$ is the learning rate for the global model training. 

\subsection{Aggregation Strategy with Models as Targets}
\subsubsection{FedAvg}
In FedAvg, the clients achieve communication efficiency by making multiple local updates before sending their local models to the server, which averages the local model weights to compute the next round of global model. Under this strategy, each client uploads the weight parameters $w_i$ of its local model to the server instead of accumulated gradients. The server stores all weight parameters in a collection $\mathcal{S}$. Once $\mathcal{S}$ meets certain conditions (\eg  all activated clients have uploaded model weights) in the $t$-th global epoch, the server aggregates these parameters to update the global model as performed by:
\begin{equation}
    w_g^{t} = \frac{1}{D} \sum_{w_i \in \mathcal{S}} |\mathcal{D}_i| w_i^t,
\label{equ:fedavg_update}
\end{equation}
where $D \triangleq \sum_{w_i \in \mathcal{S}} |\mathcal{D}_i|$.

FedAvg is favored by researchers in synchronous FL thanks to its faster convergence speed. In additional, FedAvg exhibits greater resilience against gradient leakage attacks than FedSGD, offering enhanced privacy protection for clients~\cite{melis2019exploiting,zhu2019deep,wainakh2021user,chai2020fedeval}.

\section{Experimental Set-up}
In this section, we will introduce the datasets and their data distributions for local training, models for comparative experimental analysis, and the metrics we design for performance measurement.

\subsection{Datasets}
In our evaluation, we include three types of image data and two types of language text data. We explain each of the datasets in more detail below.

\subsubsection{CIFAR-10}
CIFAR-10~\cite{krizhevsky2009learning} is a widely used benchmark dataset for image classification tasks in machine learning. This dataset consists of ten labeled classes of images. Each class corresponds to $6,000$ images, each with a size of $32\times32$ pixels and in RGB format. 
\subsubsection{CIFAR-100}
Compared to CIFAR-10, CIFAR-100~\cite{krizhevsky2009learning} is more complex, encompassing a total of 100 categories. Each category consists of 600 images, each with a size of $32\times32$ pixels and in RGB format.
\subsubsection{FEMNIST}
FEMNIST~\cite{caldas2018leaf} is an enhanced dataset derived from the EMNIST dataset~\cite{cohen2017emnist}, specifically tailored for the characteristics of FL scenarios. It is utilized for handwritten character recognition tasks. This dataset comprises 805,263 samples across 62 classes, with each image presented as a gray-scale $28\times28$ pixel image.
\subsubsection{Shakespeare}
This dataset is a text dataset utilized for natural language processing tasks, constructed from \textit{The Complete Works of William Shakespeare}. We embedded 80 characters, encompassing 26 uppercase English characters, 26 lowercase English characters, 10 numeric characters, and 18 special characters, into 80 labels,. The specific details of data partitioning will be elaborated in Section~\ref{sec:data_distri}.
\subsubsection{Sentiment140}
The Sentiment140~\cite{go2009twitter} dataset is a publicly available dataset extensively utilized for sentiment analysis research. This dataset comprises approximately 1,600,000 tweets collected from the Twitter platform, with a portion of the tweets annotated with sentiment labels.

\subsection{Data Distributions}
\label{sec:data_distri}
For performance analysis, we evaluate representative data distributions across datasets, includes Independent and Identical Distribution (IID) and non-IID. We include two IIDs, one for image data and the other for text data. Below we present details about non-IID data distributions, with the first three and last one corresponding to image~\cite{JMLR:v24:22-0440} and text data~\cite{caldas2018leaf}, respectively.
%In real-world scenarios, federated learning may encounter diverse data distribution situations. This section introduces the data distribution scenarios employed in this paper.

%Firstly, we will introduce the distribution of image data based on an FL benchmarking framework~\cite{JMLR:v24:22-0440}.

% \subsubsection{Balanced, Independent, and Identical Distribution}
% In an IID scenario, each participant possesses an equal amount of image data covering all labels. Simultaneously, the data distributions among clients are entirely identical. Under these conditions, FL is virtually indistinguishable from centralized learning.

\subsubsection{Shards Distribution}
Each client possesses an equal quantity of image data but only pertains to specific labels. In extreme instances, the distribution of data among participants is completely distinct. As an example, two participants $\mathcal{C} = \{C_1, C_2\}$ are planning to collaborate for FL. Nonetheless, $C_1$ possesses 500 data points labeled as $a$, whereas $C_2$ possesses 500 data points labeled as $b$. Despite having an equal quantity of data, their data distributions are entirely dissimilar.
% \begin{figure}[!htp]
% \centering
% \subfloat[Shards = 5]{\includegraphics[width=1.5in]{pic/distribution/Shards/5.png}%
% \label{fig_Distribution_Shards5}}
% \hfil
% \subfloat[Shards = 10]{\includegraphics[width=1.5in]{pic/distribution/Shards/10.png}%
% \label{fig_Distribution_Shards10}}
% \caption{Data Distribution under Shards Distribution.}
% \label{fig_Distribution_Shards}
% \end{figure}
\subsubsection{Unbalanced Dirichlet Distribution}
Each client possesses image data for all labels, and the data distribution is identical among participants. However, there are variations in the quantity of data each participant holds. The data distribution for each participant follows a Dirichlet distribution $D(\alpha)$ as:
\begin{equation}
    D(p|\alpha) = \frac{\Gamma(\sum_{i=1}^N \alpha_i)}{\Pi_{i=1}^N\Gamma(\alpha_i)}\Pi_{i=1}^Np_i^{\alpha_i-1},
\end{equation}
where $\alpha$ is a parameter controlling the distribution, $p_i$ represents the probability of having data from the $i$-th class and $\Gamma(x)$ is the Gamma-function.

The distribution of data quantity among clients follows a log-normal distribution $Log-N(0,\sigma^2)$.

% \begin{figure}[!htp]
% \centering
% \subfloat[$\sigma^2 = 1$]{\includegraphics[width=1.5in]{pic/distribution/UD/1Client_distribution.png}%
% \label{fig_Distribution_UD1_amount}}
% \hfil
% \subfloat[$\sigma^2 = 0.1$]{\includegraphics[width=1.5in]{pic/distribution/UD/0.1Client_distribution.png}%
% \label{fig_Distribution_UD_01_amount}}
% \hfil
% \subfloat[$\sigma^2 = 1$]{\includegraphics[width=1.5in]{pic/distribution/UD/1.png}%
% \label{fig_Distribution_UD1}}
% \hfil
% \subfloat[$\sigma^2 = 0.1$]{\includegraphics[width=1.5in]{pic/distribution/UD/0.1.png}%
% \label{fig_Distribution_UD_01}}
% \caption{Data Distribution under Unbalanced Dirichlet Distribution. (a)(b) Distribution of the amount of data; (c)(d) Distribution of the class of each client }
% \label{fig_Distribution_UD}
% \end{figure}

\subsubsection{Hetero Dirichlet Distribution}
This type of distribution is more frequently encountered in real-world scenarios. Each client holds image data pertaining to a subset of labels, exhibiting unequal data quantities and diverse data distributions among the participants. The data for each participant forms a Dirichlet distribution $D_k(\alpha)$ based on categories,
% \begin{equation}
%     D(p_k|\alpha) = \frac{\Gamma(\sum_{i=1}^N \alpha_i)}{\Pi_{i=1}^N\Gamma(\alpha_i)}\Pi_{i=1}^Np_{k,i}^{\alpha_i-1},
% \end{equation}
where $k$ represents the $k$-th client.
% \begin{figure}[!htp]
% \centering
% \subfloat[$\alpha = 1$]{\includegraphics[width=1.5in]{pic/distribution/HD/1Client_distribution.png}%
% \label{fig_Distribution_HD1_amount}}
% \hfil
% \subfloat[$\alpha = 0.3$]{\includegraphics[width=1.5in]{pic/distribution/HD/0.3Client_distribution.png}%
% \label{fig_Distribution_HD_03_amount}}
% \hfil
% \subfloat[$\alpha = 1$]{\includegraphics[width=1.5in]{pic/distribution/HD/1.png}%
% \label{fig_Distribution_HD1}}                
% \hfil
% \subfloat[$\alpha = 0.3$]{\includegraphics[width=1.5in]{pic/distribution/HD/0.3.png}%
% \label{fig_Distribution_HD_03}}
% \caption{Data Distribution under Unbalanced Dirichlet Distribution. (a)(b) Distribution of the amount of data; (c)(d) Distribution of the class of each client }
% \label{fig_Distribution_HD}
% \end{figure}

%Next, we will introduce the distribution of text data, where we followed a benchmark approach~\cite{caldas2018leaf} and partitioned the text data.

% \subsubsection{IID text data}
% We consider the text data to be IID when all participants uniformly sample without distinguishing character roles, indicating that each participant's data is independently and identically distributed compared to others.

\subsubsection{non-IID text data}
We consider the dialogues of distinct characters in different scripts in Shakespeare to be non-IID. As long as each participant is assigned data stemming from dialogue lines of different characters across diverse scripts, the data distribution among these participants is non-IID. 
For the Sentiment140 dataset, we distribute the data volume for each client according to a log-normal distribution $Log-N(0,\sigma^2)$.

\subsection{Models}
To accomplish the image classification task, we utilize three models: CNN, ResNet-18, and VGG-16. We employ the LSTM model to fulfill the text prediction task.

\subsubsection{CNN}
CNN (Convolutional Neural Network) is a classic deep learning neural network model. The CNN leveraged in this paper comprises three convolutional layers employing 3x3 kernels with a stride of 1, one max-pooling layer, and two fully connected layers, with the ReLU activation function being applied.

\subsubsection{Resnet-18}
ResNet~\cite{he2016deep} refers to a series of residual networks that introduce the concept of residual blocks. ResNet-18 is one of the smaller structures within the ResNet family, consisting of 18 layers with four residual blocks. Each residual block is composed of two convolutional layers utilizing 3x3 kernels with a stride of 1.

\subsubsection{VGG-16}
VGG~\cite{simonyan2014very} is a specific architecture of CNN designed for computer vision tasks. The model employed in this paper is VGG-16, which comprises 13 convolutional layers and 3 fully connected layers, and features a relatively large parameter scale.

\subsubsection{LSTM}
LSTM~\cite{hochreiter1997long} is a specific type of recurrent neural network (RNN) tailored for processing and learning sequential data. In our experiments, a straightforward LSTM model was utilized, comprising an embedding layer, an LSTM layer, and a fully connected layer.

\subsection{Metrics}
Our primary approach involves conducting experimental evaluations of the performance gap in FL across diverse scenarios and aggregation targets. To accomplish this, we have designed four categories of metrics, including accuracy and loss, resource utilization, convergence, and oscillation. We explain each of these metrics in more detail below.

\subsubsection{Accuracy and Loss}
We evaluate the accuracy and loss of the global model on the test dataset, documenting the training epochs and total training time after each global aggregation. We ensure that all data points in the test dataset are distinct from those in the training set to uphold the integrity of the evaluation process. 

\subsubsection{Resource utilization}
We evaluate the resource consumption of FL in diverse scenarios from three perspectives: memory usage, training duration, and channel transmission load. These three metrics can be directly measured during the training process.

% \subsubsection{Accumulated staleness}
% First, we will provide a definition for staleness in semi-asynchronous federated learning: for a client $C_i$, it starts its local train with the newest global model $w_g^{t_0}$ and finish its local train when getting the model $w_{i,E}^{t_0}$. If the gradient or the model it sends to the server is used to aggregate the global model under the $t$-th global epoch, then we say that this data uploaded from $C_i$ has a staleness as $\tau_s^t$ under the $t$-th global epoch, where:
% \begin{equation}
%     \tau_s^t = t - t_0 - 1
% \end{equation}

% It's obvious that $\tau_s^t \geq 0$ and when a participant can complete local training rapidly enough, meaning that it uploads data for aggregation in the $t$-th round immediately after training its local model with the $(t-1)$-th round's model, the staleness of data is considered to be $0$ since $\tau_s^t = t - (t-1) -1 = 0$.

% Based on the definition of staleness, we calculate the cumulative staleness of the clients participating in each global aggregation as:
% \begin{equation}
%     \tau^t = \sum_{s \in \mathcal{S}} \tau_s^t
% \end{equation}

% We believe that such cumulative staleness can reflect the overall impact of stragglers on the aggregation process in a given global epoch. A higher cumulative staleness indicates a greater degree of straggling among participants and a higher likelihood of negative effects on model aggregation.

\subsubsection{Convergence}
Although accuracy and loss function curves across training epochs offer a qualitative analysis of training convergence, we introduce additional quantitative indicators to determine convergence. These include the number of epochs (denoted as $T_f$) required to achieve the target accuracy, $Acc_t$, for the first time, and the epochs (denoted as $T_s$) after which the model remains stable above $Acc_t$. Evidently, the utilization of these two metrics can indicate the convergence speed and stability of federated models in diverse scenarios. Specifically, a smaller $T_f$ signifies a faster convergence speed, while a smaller $T_s - T_f$ implies a more stable convergence (\ie a faster stability speed).

\subsubsection{Oscillation}
We also tally the instances where the training process exhibits severe oscillations, denoted by $O_{ots}$ where $ots$ represents an oscillation threshold. We define severe oscillations as follows: if the accuracy in the current round falls below the accuracy in the previous round by a specific threshold, it is deemed to have undergone a severe oscillation.

In our study, we empirically chose the values of target accuracy $Acc_t$ and oscillation threshold $ots$ based on the experimental data obtained.

%In our definition, determining the target accuracy $Acc_t$ and oscillation threshold $ots$ is a crucial aspect. The selection of these values often varies across different experimental scenarios. In our study, we empirically chose these values based on the obtained experimental data.

\begin{table*}[!htp]
    \caption{Comparison of the best prediction performance for image classification tasks and text prediction tasks under different experimental settings. Each model is trained with the same hyper-parameters, except for the degree of non-IID.}
    \label{tab:acc}
    \centering
    \scriptsize
    \begin{threeparttable}
    \begin{tabular}{c|c l|c c c c|c c c c|c c c c}
    \hline
         \multicolumn{15}{c}{Prediction Accuracy for Computer Vision Tasks}\\ \hline
         \multirow{2}{*}{Dataset} & \multicolumn{2}{c|}{Distribution} & \multicolumn{4}{c|}{ResNet-18}& \multicolumn{4}{c|}{VGG-16}& \multicolumn{4}{c}{CNN}\\ 
         & Type & Parameter & SS & SA & AS & AA & SS & SA & AS & AA & SS & SA & AS & AA\\ \hline 
         
        \multirow{9}{*}{CIFAR-10} & \multirow{3}{*}{HD} & $\alpha = 0.1$ & 83.32 & 84.21 & 80.05 & \cellcolor[gray]{0.9}64.67 & 65.6 & 67.3 & 59.35 & \cellcolor[gray]{0.9}37 & 65.53 & 64.18 & 65.94 & \cellcolor[gray]{0.9}48.51\\ 
         &  & $\alpha = 0.5$ & 88.78 & 88.74 & 84.7 & \cellcolor[gray]{0.9}74.33 & 82.45 & 83.76 & 75.88 & \cellcolor[gray]{0.9}61.07 & 72.92 & 73.32 & 71.03 & \cellcolor[gray]{0.9}52.94\\ 
         &  & $\alpha = 1$ & 90.35 & 89.95 & 87.81 & \cellcolor[gray]{0.9}80.11 & 86.04 & 86.95 & 82.61 & \cellcolor[gray]{0.9}66.69  & 74.43 & 74.4 & 73.25 & \cellcolor[gray]{0.9}56.91\\ \cline{2-15} 
         & \multirow{3}{*}{SD} & $N = 2$ & 65.33 & 68.67 & 59.0 & \cellcolor[gray]{0.9}54.14 & 36.08 & 37.52 & 33.23 & \cellcolor[gray]{0.9}27.12 & 60.54 & 59.23 & 59.08 & \cellcolor[gray]{0.9}42.27\\ 
         &  & $N = 5$ & 87.75 & 88.48 & 83.94 & \cellcolor[gray]{0.9}73.12 & 80.35 & 76.89 & 69.47 & \cellcolor[gray]{0.9}53.32 & 70.83 & 69.26 & 63.39 & \cellcolor[gray]{0.9}52.5\\ 
         &  & $N = 10$ & 90.5 & 90.48 & 87.17 & \cellcolor[gray]{0.9}80.43 & 87.01 & 85.81 & 79.43 & \cellcolor[gray]{0.9}66.74 & 73.83 & 73.99 & 67.49 & \cellcolor[gray]{0.9}55.24\\ \cline{2-15}
         & \multirow{3}{*}{UD} & $\sigma = 0.1$ & 75.3 & 75.3 & 65.0 & \cellcolor[gray]{0.9}54.11 & 61.57 & 61.57 & 42.78 & \cellcolor[gray]{0.9}33.39 & 42.89 & 43.11 & 36.16 & \cellcolor[gray]{0.9}22.97\\
         &  & $\sigma = 0.5$ & 73.31 & 73.2 & 62.09 & \cellcolor[gray]{0.9}50.92 & 79.91 & 80.91 & 72.89 & \cellcolor[gray]{0.9}62.89 & 64.23 & 63.98 & 57.46 & \cellcolor[gray]{0.9}43.1\\
         &  & $\sigma = 1$ & 85.9 & 85.27 & 72.14 & \cellcolor[gray]{0.9}66.25 & 80.61 & 79.61 & 73.07 & \cellcolor[gray]{0.9}\cellcolor[gray]{0.9}56.83 & 51.73 & 53.41 & 46.08 & \cellcolor[gray]{0.9}34.33\\ \hline 
         \multirow{9}{*}{CIFAR-100} & \multirow{3}{*}{HD} & $\alpha = 0.1$ & 62.86 & 62.65 & 57.95 & \cellcolor[gray]{0.9}40.15 & 48.20 & 48.46 & 36.38 & \cellcolor[gray]{0.9}22.08 & 37.43 & 36.14 & 32.55 & \cellcolor[gray]{0.9}19.08\\ 
         &  & $\alpha = 0.5$ & 67.16 & 68.07 & 62.08 & 4\cellcolor[gray]{0.9}7.69 & 59.46 & 58.59 & 52.72 & \cellcolor[gray]{0.9}36.62 & 40.42 & 41.5 & 36.78 & \cellcolor[gray]{0.9}23.7\\ 
         &  & $\alpha = 1$ & 67.97 & 68.1 & 63.66 & \cellcolor[gray]{0.9}48.53 & 61.32 & 60.56 & 53.39 & \cellcolor[gray]{0.9}39.87 & 41.87 & 40.96 & 35.88 & \cellcolor[gray]{0.9}23.18\\ \cline{2-15} 
         & \multirow{3}{*}{SD} & $N = 2$ & 20.24 & 18.43 & 12.96 & \cellcolor[gray]{0.9}7.24 & 2.9 & 3.88 & 2.6 & \cellcolor[gray]{0.9}2.33 & 15.74 & 17.43 & 11.18 & \cellcolor[gray]{0.9}5.33\\ 
         &  & $N = 5$ & 49.66 & 48.19 & 38.79 & \cellcolor[gray]{0.9}20.9 & 11.64 & 13.85 & 11.46 & \cellcolor[gray]{0.9}7.68 & 29.85 & 31.0 & 22.36 & \cellcolor[gray]{0.9}11.81\\ 
         &  & $N = 10$ & 59.16 & 59.24 & 51.71 & \cellcolor[gray]{0.9}32.51 & 41.15 & 42.95 & 29.03 & \cellcolor[gray]{0.9}16.77 & 34.45 & 35.34 & 28.01 & \cellcolor[gray]{0.9}15.1\\ \cline{2-15}
         & \multirow{3}{*}{UD} & $\sigma = 0.1$ & 16.12 & 16.69 & 14.76 & \cellcolor[gray]{0.9}10.81 & 16.07 & 15.92 & 13.46 & \cellcolor[gray]{0.9}10.97 & 7.51 & 7.21 & 6.61 &\cellcolor[gray]{0.9} 3.03\\ 
         &  & $\sigma = 0.5$ & 17.88 & 17.95 & 13.4 & \cellcolor[gray]{0.9}11.12 & 16.2 & 16.82 & 11.85 & \cellcolor[gray]{0.9}8.21 & 9.75 & 9.43 & 7.57 & \cellcolor[gray]{0.9}3.43\\ 
         &  & $\sigma = 1$ & 24.05 & 24.05 & 19.55 & \cellcolor[gray]{0.9}16.05 & 23.34 & 23.34 & 19.71 & \cellcolor[gray]{0.9}15.57 & 13.7 & 12.62 & 11.89 & \cellcolor[gray]{0.9}5.89\\ \hline 
         \multirow{9}{*}{FEMNIST} & \multirow{3}{*}{HD} & $\alpha = 0.1$ & 82.91 & 83.02 & 82.66 & \cellcolor[gray]{0.9}82.43 & 84.40 & 84.43 & 83.64 & \cellcolor[gray]{0.9}82.04 & 83.36 & 83.25 & 82.8 & \cellcolor[gray]{0.9}81.55\\ 
         &  & $\alpha = 0.5$ & 84.1 & 84.24 & 83.52 & \cellcolor[gray]{0.9}83.27 & 86.16 & 86.04 & 85.64 &\cellcolor[gray]{0.9} 84.7 & 84.77 & 84.96 & 84.37 & \cellcolor[gray]{0.9}83.25\\ 
         &  & $\alpha = 1$ & 84.02 & 84.05 & 84 & \cellcolor[gray]{0.9}82.99 & 86.41 & 86.54 & 86.22 & \cellcolor[gray]{0.9}85.0 & 84.97 & 85.07 & 84.78 & \cellcolor[gray]{0.9}83.83\\ \cline{2-15} 
         & \multirow{3}{*}{SD} & $N = 2$ & 66.26 & 65.78 & 63.55 & \cellcolor[gray]{0.9}52.66 & 37.80 & 36.12 & 31.87 & \cellcolor[gray]{0.9}29.30 & 71.47 & 70.76 & 67.27 & \cellcolor[gray]{0.9}53.43\\ 
         &  & $N = 5$ & 77.80 & 78.12 & 74.42 & \cellcolor[gray]{0.9}70.38 & 71.07 & 71.50 & 64.11 & \cellcolor[gray]{0.9}59.66 & 78.93 & 79.49 & 77.59 & \cellcolor[gray]{0.9}72.57\\ 
         &  & $N = 10$ & 80.95 & 80.36 & 79.67 & \cellcolor[gray]{0.9}77.81 & 80.68 & 80.32 & 79.12 & \cellcolor[gray]{0.9}74.44 & 80.76 & 80.78 & 80.62 & \cellcolor[gray]{0.9}77.73\\ \cline{2-15}
         & \multirow{3}{*}{UD} & $\sigma = 0.1$ & 75.10 & 77.30 & 73.52 & \cellcolor[gray]{0.9}72.40 & 75.63 & 75.70 & 75.62 & \cellcolor[gray]{0.9}71.97 & 73.72 & 73.72 & 71.72 &\cellcolor[gray]{0.9}69.01\\ 
         &  & $\sigma = 0.5$ & 83.49 & 83.89 & 83.50 & \cellcolor[gray]{0.9}81.55 & 82.84 & 82.54 & 82.06 & \cellcolor[gray]{0.9}80.43 & 81.27 & 81.14 & 80.75 & \cellcolor[gray]{0.9}78.79\\ 
         &  & $\sigma = 1$ & 83.67 & 83.77 & 82.84 & \cellcolor[gray]{0.9}81.07 & 82.76 & 83.03 & 82.22 & \cellcolor[gray]{0.9}79.44 & 81.27 & 81.10 & 80.15 & \cellcolor[gray]{0.9}77.27\\ \hline 
         \multicolumn{15}{c}{Prediction Accuracy for Natural Language Processing tasks}\\ \hline
         \multirow{2}{*}{Dataset} & \multicolumn{2}{c|}{Distribution} & \multicolumn{12}{c}{LSTM} \\  
         & Type & Parameter & \multicolumn{3}{c}{SS} & \multicolumn{3}{c}{SA} & \multicolumn{3}{c}{AS} & \multicolumn{3}{c}{AA}\\ \hline 
         \multirow{2}{*}{Shakespeare} & \multirow{2}{*}{non-IID} & $N = 100$ & \multicolumn{3}{c}{42.44} & \multicolumn{3}{c}{42.53} & \multicolumn{3}{c}{41.32} & \multicolumn{3}{c}{\cellcolor[gray]{0.9}30.52}\\
         &  & $N = 300$ & \multicolumn{3}{c}{43.02} & \multicolumn{3}{c}{42.87} & \multicolumn{3}{c}{41.43} & \multicolumn{3}{c}{\cellcolor[gray]{0.9}29.18}\\\hline
         \multirow{3}{*}{Sentiment140} & \multirow{3}{*}{non-IID} & $\sigma = 0.1$ & \multicolumn{3}{c}{75.25} & \multicolumn{3}{c}{74.55} & \multicolumn{3}{c}{75.00} & \multicolumn{3}{c}{\cellcolor[gray]{0.9}72.49}\\
         &  & $\sigma = 0.5$ & \multicolumn{3}{c}{75.33} & \multicolumn{3}{c}{75.00} & \multicolumn{3}{c}{75.34} & \multicolumn{3}{c}{\cellcolor[gray]{0.9}72.82}\\
         &  & $\sigma = 1$ & \multicolumn{3}{c}{75.04} & \multicolumn{3}{c}{75.33} & \multicolumn{3}{c}{75.44} & \multicolumn{3}{c}{\cellcolor[gray]{0.9}73.33}\\
         \hline
         
    \end{tabular}
     \begin{tablenotes}
        \footnotesize
        \item[*] In accuracy, SS means \underline{S}FL-\underline{S}GD, SA means \underline{S}FL-\underline{A}vg, AS means S\underline{A}FL-\underline{S}GD, AA means S\underline{A}FL-\underline{A}vg.
        \item[*] In distributions, HD means \underline{H}etero \underline{D}irichlet, SD means \underline{S}har\underline{D}s, UD means \underline {U}nbalanced \underline{D}irichlet.
       \item[*] In HD, a larger $\alpha$ means a more even distribution. In SD, a larger $N$ means a more even distribution. In UD, a larger $\sigma$ means a more even distribution. In text prediction task, the parameter $N$ means the numbers of participants and a larger $N$ means more training data.
      \end{tablenotes}
\end{threeparttable}
\vspace*{-4ex}
\end{table*}
\section{Experimental Results and Discussion}
In this section,  we first present our experimental results of \sgd and \avg on various datasets and provide analyses of the outcomes, which show a notable performance gap between these two algorithm modes in SAFL. 
% Then, we qualitatively analyze this performance gap and compare the performance of \tool with that of SOTA approaches, demonstrating \tool's effectiveness in bridging the performance gap.
%Due to space constraints, we only present part of the typical experimental results to illustrate our conclusion.

\subsection{\sgd vs. \avg in SAFL}

\subsubsection{Accuracy and Loss}
\begin{figure}[!htp]
\vspace{-1ex}
\centering
\subfloat[$\alpha = 0.1$]{\includegraphics[width=1.5in]{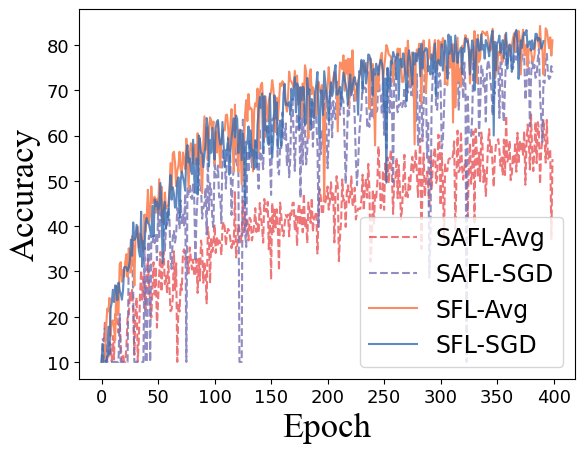}}
\hspace{1mm}
\subfloat[$\alpha = 0.1$]{\includegraphics[width=1.5in]{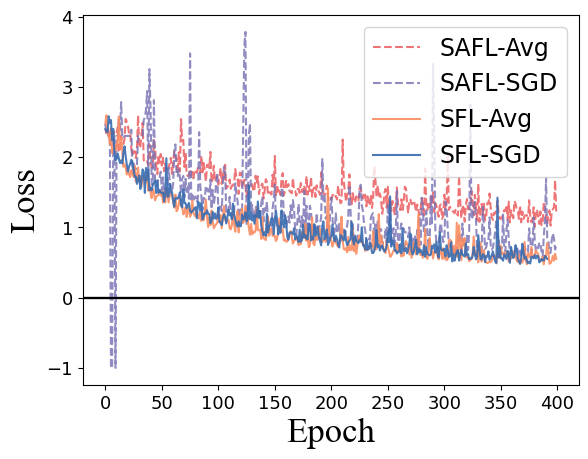}}
\hspace{1mm}
\subfloat[$\alpha = 0.5$]{\includegraphics[width=1.5in]{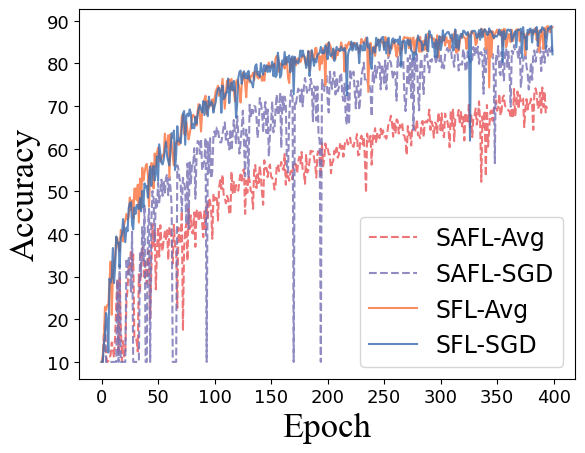}}
\hspace{1mm}
\subfloat[$\alpha = 0.5$]{\includegraphics[width=1.5in]{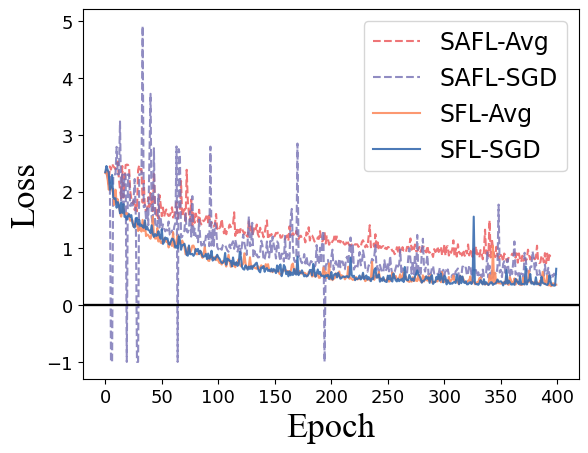}\label{fig_nan_concrete}}
\hspace{1mm}
\subfloat[$\alpha = 1$]{\includegraphics[width=1.5in]{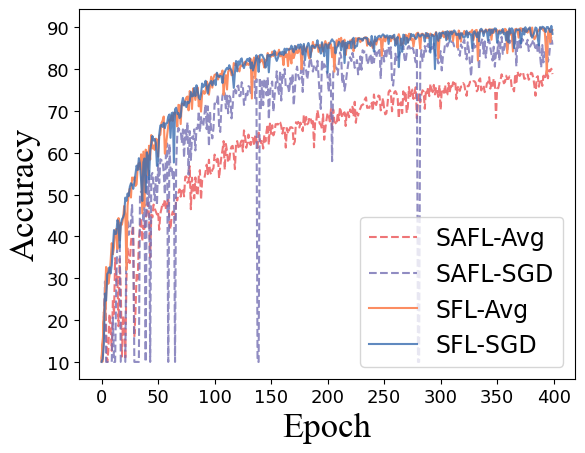}}
\hspace{1mm}
\subfloat[$\alpha = 1$]{\includegraphics[width=1.5in]{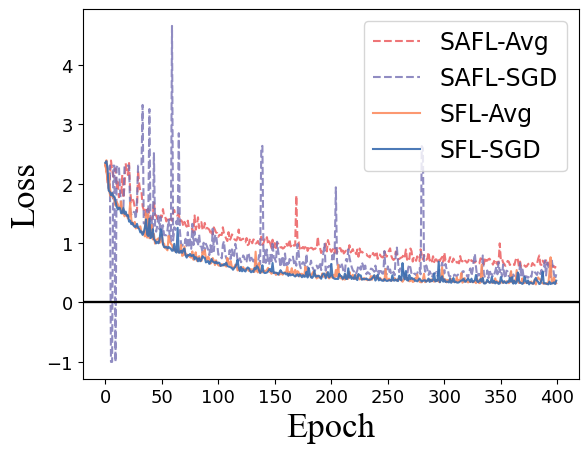}}
\hspace{1mm}
% \subfloat[$N = 2$]{\includegraphics[width=1.5in]{pic/ret/resnet18_cifar10/shards/acc/2.png}}
% \hspace{1mm}
% \subfloat[$N = 2$]{\includegraphics[width=1.5in]{pic/ret/resnet18_cifar10/shards/loss/2.png}}
% \hspace{1mm}
\subfloat[$N = 5$]{\includegraphics[width=1.5in]{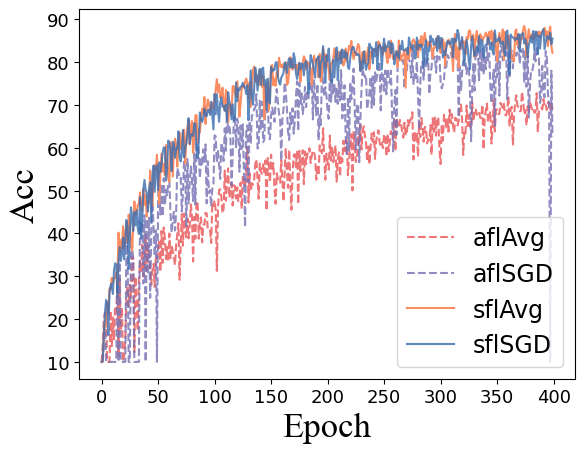}}
\hspace{1mm}
\subfloat[$N = 5$]{\includegraphics[width=1.5in]{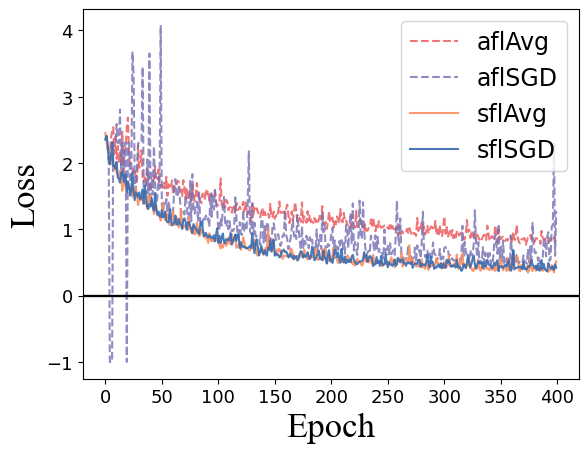}}
\hspace{1mm}
\subfloat[$N = 10$]{\includegraphics[width=1.5in]{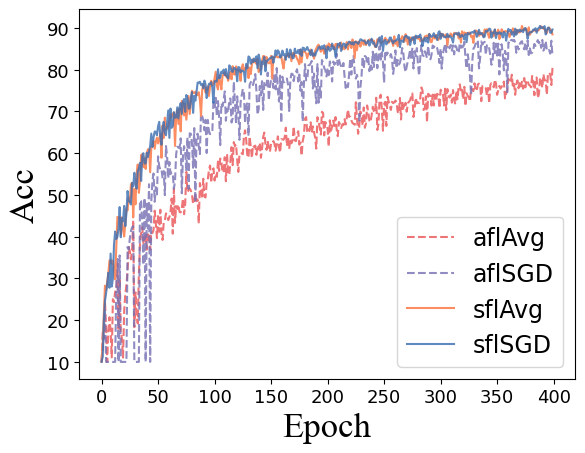}}
\hspace{1mm}
\subfloat[$N = 10$]{\includegraphics[width=1.5in]{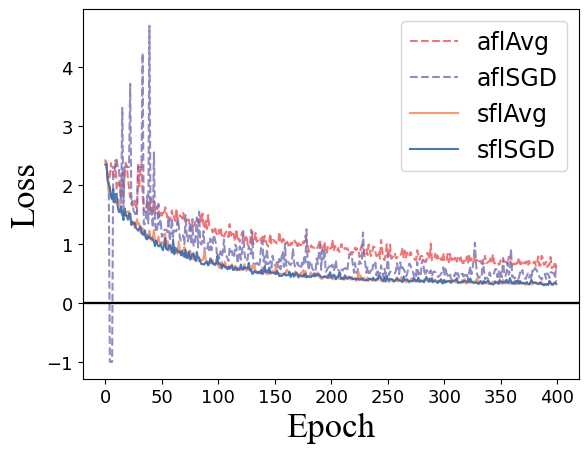}}
\hspace{1mm}
\subfloat[$\sigma = 0.1$]{\includegraphics[width=1.5in]{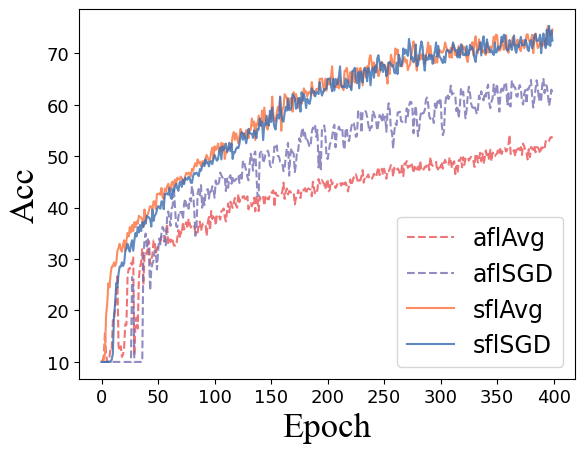}}
\hspace{1mm}
\subfloat[$\sigma = 0.1$]{\includegraphics[width=1.5in]{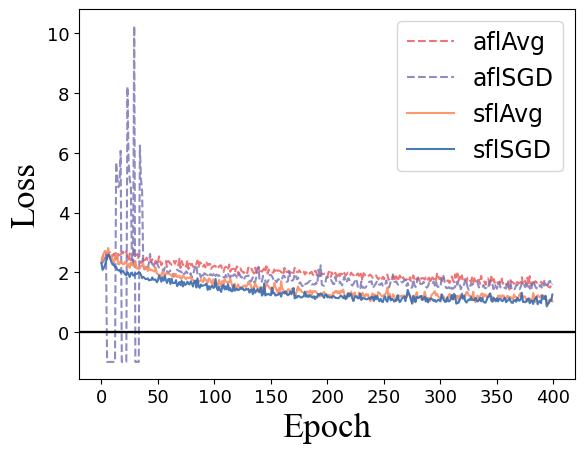}}
\hspace{1mm}
\subfloat[$\sigma = 0.5$]{\includegraphics[width=1.5in]{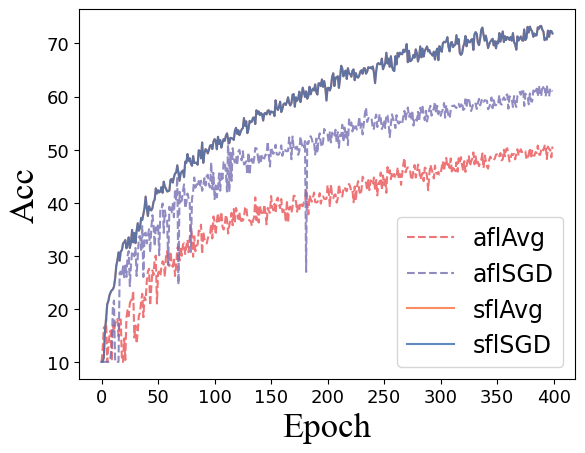}}
\hspace{1mm}
\subfloat[$\sigma = 0.5$]{\includegraphics[width=1.5in]{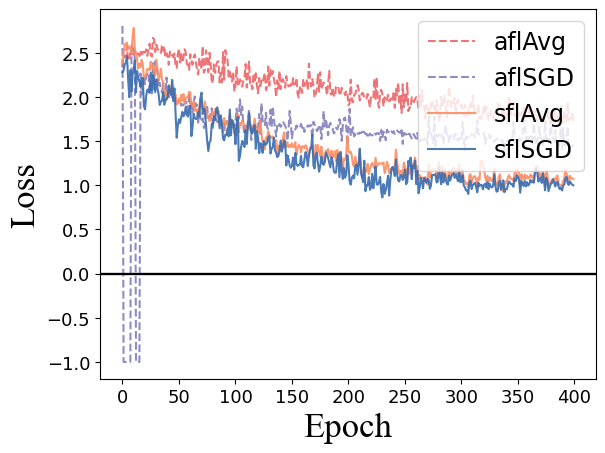}}
\hspace{1mm}
\subfloat[$\sigma = 1$]{\includegraphics[width=1.5in]{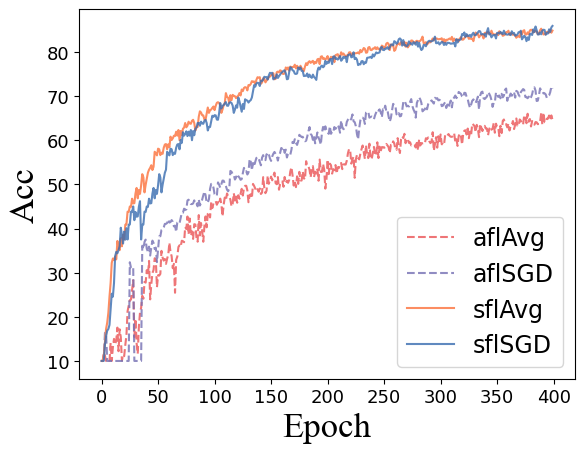}}
\hspace{1mm}
\subfloat[$\sigma = 1$]{\includegraphics[width=1.5in]{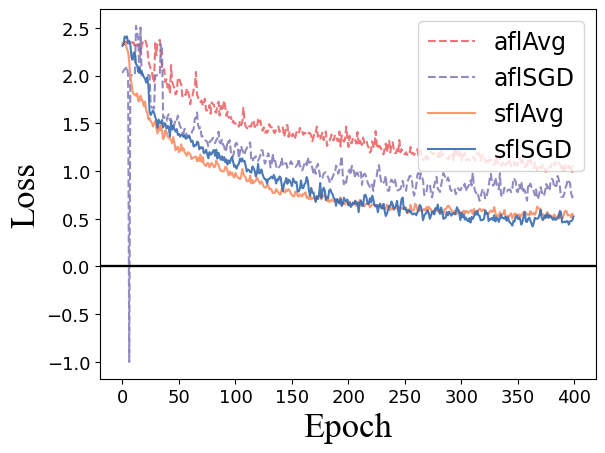}}
\caption{Global accuracy and loss of different models under CIFAR-10 dataset using ResNet-18 in SAFL. Note that -1 denotes the NAN value for loss. }
\label{fig_model_cifar10_HD}
\vspace{-1ex}
\end{figure}

Table~\ref{tab:acc} shows the results of the best prediction accuracy across all experimental settings after the training of the same number of epoches. These results show that when data distribution is uneven, in synchronous federated learning (SFL), \sgd (columns ``SS'') and \avg (columns ``SA'') have comparable performance in prediction accuracy. However, in semi-asynchronous federated learning (SAFL), \sgd (columns ``AS'') significantly outperforms \avg (shaded columns ``AA''). Additionally, in Figure~\ref{fig_model_cifar10_HD}\footnote{Note that all experimental scenarios share similar result patterns. Due to the page limit, we only include representative result figures or tables.} we present accuracy-epoch and loss-epoch graphs for representative scenarios to provide a more intuitive illustration of the performance gap between \sgd and \avg in SAFL.
% In Fig.~\ref{fig_resnet_cifar10_HD}, we demonstrate the performance gap under different degrees of heterogeneity for the same data distribution. We also respectively illustrate the performance gap under different datasets or different models in Fig.~\ref{fig_model_cifar10_HD} and Fig.~\ref{fig_resnet_dataset_HD} while show the comparison of different distributions in Fig.~\ref{fig_resnet_cifar10}.

Based on above results, it is evident that \sgd and \avg obtain comparable performance in SFL. However, \sgd consistently attains lower loss and higher accuracy compared to \avg in SAFL. Taking the CIFAR-10 dataset as an example, \avg experiences an accuracy drop of 4.86\%$\sim$22.32\% compared to \sgd. Regarding the Shakespeare dataset, this performance drop changes to 2\%$\sim$12.25\% with a median of 2.52\%. In addition, the performance of the same aggregation strategy is different between SFL and SAFL. For example, the accuracy of \sgd and \avg in SAFL is reduced by 5.41\%$\sim$18.79\% and 10.06\%$\sim$30.29\%, respectively compared to their performance in SFL for the CIFAR-10 dataset.
%\lyb{For the task of processing CIFAR-10, we observed that AA experiences a performance drop of 4.86\%$\sim$22.32\% compared to AS, AS exhibits a performance drop of 5.41\%$\sim$18.79\% compared to SS/SA, and AA demonstrates a 10.06\%$\sim$30.29\% performance drop compared to SS/SA. On the other hand, for text prediction using LSTM on the Shakespeare dataset, AA demonstrates an average performance degradation of around 2\%$\sim$12.25\% compared to AS with a median of 2.52\%, AS exhibits only an average performance drop of 1\% compared to SS/SA, and AA shows an average performance drop of approximately 6.62\% compared to SS/SA.} 

Nevertheless, employing \sgd as the aggregation strategy in SAFL renders the convergence process more volatile, suggesting a greater susceptibility to the influence of stragglers. This can result in notable performance gap across various stopping epochs. Furthermore, when \sgd is utilized for global model aggregation, the loss may occasionally exhibit an anomalous NAN phenomenon (represented as -1 in Figure~\ref{fig_model_cifar10_HD}) regardless of whether the data is IID, which is not observed in \avg. For instance, the loss function encountered NAN occurrences in multiple epochs (\eg the 5th and 6th epoch) during the processing of CIFAR-10 dataset, as shown in Figure~\ref{fig_nan_concrete}. In such outlier instances, the global model completely forfeits its predictive capabilities.

\subsubsection{Resource Utilization}

\begin{table}[!htp]
\vspace{-2ex}
\setlength{\tabcolsep}{3pt}
    \caption{Resource utilization on CIFAR-10 datasets and Shakespeare.}
    \label{Tab:resource_perform}
    \centering
    \scriptsize
    \begin{tabular}{c|c|c|c|c|c|c}
    \hline
         Dataset & Model &  Type & Strategy &  \makecell{Memory \\ usage /GB} &   \makecell{Training \\ duration /s}&   \makecell{Transmission \\ Load /GB}\\
    \hline
          \multirow{12}{*}{\makecell{CIFAR-10}}& \multirow{6}{*}{ResNet} &   \multirow{2}{*}{HD} & \avg & 34 & 61,832 & 193.64 \\
          & &  & \sgd & 34 & 61,529 & 167.77 \\\cline{3-7} 
          & &  \multirow{2}{*}{SD} & \avg & 33 & 35,371 & 193.64\\
          & &  & \sgd & 34 & 33,326 &167.77 \\\cline{3-7} 
          & &  \multirow{2}{*}{UD} & \avg & 30 & 45,323 &  193.64\\
          & &  & \sgd & 30 & 37,679 &167.77\\\cline{2-7} 

          & \multirow{6}{*}{VGG}  & \multirow{2}{*}{HD} & \avg & 45 & 87,265 & 503.00\\
          & &  & \sgd & 45 & 86,979 & 500.87 \\\cline{3-7}
          & &  \multirow{2}{*}{SD} & \avg & 45 & 84,541 & 503.00 \\
          & &  & \sgd & 44 & 84,516 & 500.87\\\cline{3-7}
          & &  \multirow{2}{*}{UD} & \avg & 40 & 68,426 & 503.00\\
          & &  & \sgd & 40 & 62,767 & 500.87\\\hline
          
          \multirow{4}{*}{\makecell{Shakespeare}}& \multirow{4}{*}{LSTM}   & \multirow{2}{*}{100} & \avg & 20 & 43,244 & 3.39\\
          & &  & \sgd & 20 & 43,116 & 3.35\\\cline{3-7}
          & &  \multirow{2}{*}{300} & \avg & 61& 87,655& 3.39\\
          & &  & \sgd & 61 & 87,583 & 3.35\\        
    \hline
    \end{tabular}
% \vspace{-4ex}
\end{table}

To determine the resource utilization of two aggregation strategies in SAFL, we measured memory usage, training time, and overall channel transmission payload, across various experimental scenarios. 

%\lyb{Specifically, both scheme exhibit similar performance in terms of memory consumption while the training duration is significantly influenced by both the model and the dataset. For example, in the CIFAR-10 image classification task with Unbalance Dirichlet distribution, when training with the ResNet-18 model, \sgd saves approximately 7,644 seconds compared to \avg, translating to roughly a 16\% performance improvement. Similarly, when training with the VGG-16 model in the same task, \sgd saves around 5,659 seconds, representing an approximately 8.2\% performance enhancement. Conversely, when employing the ResNet-18 model for training in the Shards distribution scenario of CIFAR-10 image classification, \sgd saves approximately 2,045 seconds, indicating about a 5\% performance increase. Additionally, the channel transmission load is only correlated with the model used. When utilizing ResNet-18, \sgd consumes approximately 15\% less channel load than \avg, whereas with LSTM, \sgd only reduces the load by 1\%.}

Table~\ref{Tab:resource_perform} shows the results of two representative datasets. Overall, \sgd exhibits superior performance in terms of resource utilization. For memory usage, there is minimal difference between the two algorithm modes. However, in terms of training time, \sgd outperforms \avg. For example, for the CIFAR-10 dataset with unbalance Dirichlet (UD) distribution, when training with the ResNet-18 model, \sgd saves 7,644 seconds compared to \avg, \ie 16\% less time for training. We attribute this improvement to the number of operations during aggregation. That said, in \sgd, the aggregation process simply involves retrieving the global learning rate from the configuration file. Conversely, \avg requires querying the total data volume possessed by the clients whose data is utilized for aggregation in that round, followed by the individual calculation of weighting coefficients for each participant. Consequently, this results in additional computational steps and longer calculation time. Regarding the channel transmission payload correlated with the model, \sgd places less pressure on the channel compared to \avg. This is because \sgd only necessitates the transmission of gradient data, whereas \avg requires the transmission of the entire model, including parameters of layers.

%We compared the resource consumption of the two aggregation strategies in several experimental scenarios in the Table~\ref{Tab:resource_perform}, including memory usage, total training duration, and overall channel transmission payload.  Overall, \sgd performs better in terms of resource consumption. Specifically, from the perspective of memory usage, there isn't much difference between the two aggregation algorithms. However, in terms of training duration, \sgd is significantly faster than \avg, with a speed improvement of several hundred seconds at least, and in some scenarios, even up to a thousand seconds improvement. We believe this pertains to the implementation of the aggregation algorithm. In \sgd, the aggregation process simply requires invoking the global learning rate from the config file. However, in \avg, it entails querying the total data volume possessed by the clients whose data is utilized for aggregation in this round, followed by the calculation of weighting coefficients for each participant individually. Consequently, this leads to additional computational steps and longer calculation time. Regarding the channel transmission payload, the channel pressure of \sgd is much lower than that of \avg, which is because \sgd solely necessitates transmitting gradient data, whereas \avg requires transmitting the entire model, encompassing parameters of layers that do not necessitate gradient calculation.

\subsubsection{Convergence}
\label{sec:conver}

\begin{table}[!htp]
\vspace{-2ex}
\setlength{\tabcolsep}{5pt}
    \caption{Convergence results w.r.t. ResNET-18 model and CIFAR-10 dataset.}
    \label{Tab:convergence_perform_resnet_cifar}
    \centering
    \scriptsize
    \begin{threeparttable}
    \begin{tabular}{c|c|c|c|c|c|c}
    \hline
         Type & $\alpha$ & Strategy &Threshold & $T_f \downarrow$ & $T_s$ & $T_s - T_f \downarrow$\\
    \hline
          \multirow{6}{*}{\makecell{Hetero \\ Dirichlet\\(HD)}} &   \multirow{2}{*}{0.1} & \avg & 50\% & 208 & 399 & \textbf{101}\\
          &  & \sgd & 50\% & \textbf{86} & 324 & 238\\\cline{2-7} 
          &  \multirow{2}{*}{0.5} & \avg & 60\% & 195 & 341& \textbf{146}\\
          &  & \sgd & 60\% & \textbf{82} &349 & 267\\\cline{2-7} 
          &  \multirow{2}{*}{1} & \avg & 60\% & 107 &  119& \textbf{12}\\
          &  & \sgd & 60\% & \textbf{57} &282& 225 \\\hline

          \multirow{6}{*}{\makecell{Unbalance \\ Dirichlet\\(UD)}}  & \multirow{2}{*}{0.1} & \avg & 50\% & 331 & 380& \textbf{49}\\
          &  & \sgd & 50\% & \textbf{132} & 201 & 69\\\cline{2-7} 
          &  \multirow{2}{*}{0.5} & \avg & 50\% & 358 & 399 & \textbf{41}\\
          &  & \sgd & 60\% & \textbf{254} & 397& 143\\\cline{2-7} 
          &  \multirow{2}{*}{1} & \avg & 60\% & 262 & 320&  \textbf{58}\\
          &  & \sgd & 60\% & \textbf{171} & 242& 71\\\hline
          
          \multirow{4}{*}{\makecell{Shards\\(SD)}}   & \multirow{2}{*}{5} & \avg & 60\% & 129 & 339& \textbf{210}\\
          &  & \sgd & 60\% & \textbf{84} & 398&  314\\\cline{2-7} 
          &  \multirow{2}{*}{10} & \avg & 70\%& 218& 289& \textbf{71}\\
          &  & \sgd & 70\% & \textbf{99} & 229& 130\\
            
    \hline
    \end{tabular}
    \begin{tablenotes}
        \footnotesize
        \item[*] In the table, $\downarrow$ indicates that the smaller the value, the better the performance.
      \end{tablenotes}
    \end{threeparttable}
% \vspace{-6ex}
\end{table}

To demonstrate the convergence performance of two aggregation strategies, we present the $T_f$ and $T_s$ metrics in Table~\ref{Tab:convergence_perform_resnet_cifar}, derived from training the CIFAR-10 dataset using the ResNet-18 model across distinct data distributions. In addition, we calculate the discrepancy (\ie $T_s - T_f$) between two metrics to highlight the convergence stability. The table clearly shows that the $T_f$ value of \sgd is consistently lower than that of \avg, indicating that \sgd converges faster and achieves higher prediction accuracy in the initial stages of training. Although \avg takes longer to reach the desired accuracy, its $T_s - T_f$ is smaller, suggesting that it requires fewer training epochs from initially reaching the target accuracy to stabilization. Therefore, \avg exhibits a more stable convergence process in SAFL. We reach the same conclusion in other experimental scenarios.

%To showcase the convergence performance of various aggregation strategies, we present the $T_f$ and $T_s$ metrics in the Table.~\ref{Tab:convergence_perform_resnet_cifar}, derived from training the CIFAR-10 dataset using the ResNet-18 model across distinct data distributions. Additionally, we calculate the discrepancy between them to demonstrate the convergence stability. From the table, it is evident that the $t_f$ value of \sgd is consistently lower than that of \avg, implying that \sgd converges more rapidly and attains higher prediction accuracy in the initial stages of training. Despite \avg requiring more time to achieve the desired accuracy, its $T_s - T_f$ is smaller, indicating that it necessitates fewer training epoches from the initial attainment of target accuracy to stabilization. Consequently, \avg's convergence process is more stable.

\subsubsection{Oscillation}

\begin{figure}[!htp]
\centering
\subfloat[$\alpha = 0.1$]{\includegraphics[width=1.5in]{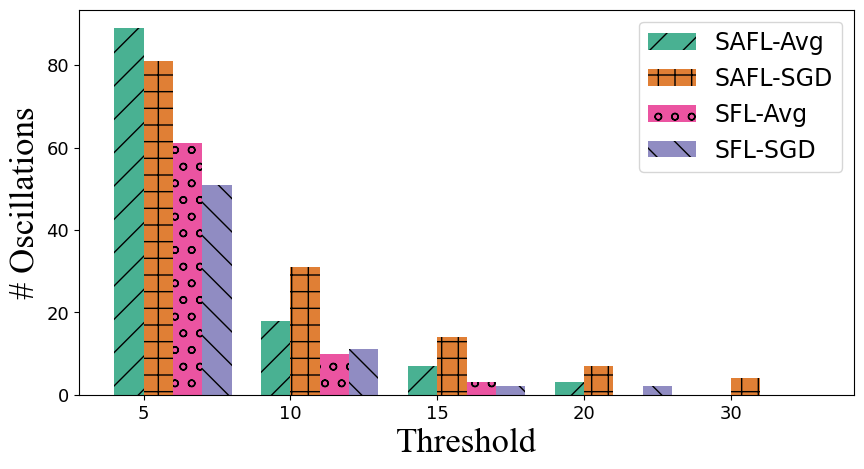}%
\label{fig_resnet18_cifar10_HD_ots_01}}
\hfil
\subfloat[$\alpha = 0.5$]{\includegraphics[width=1.5in]{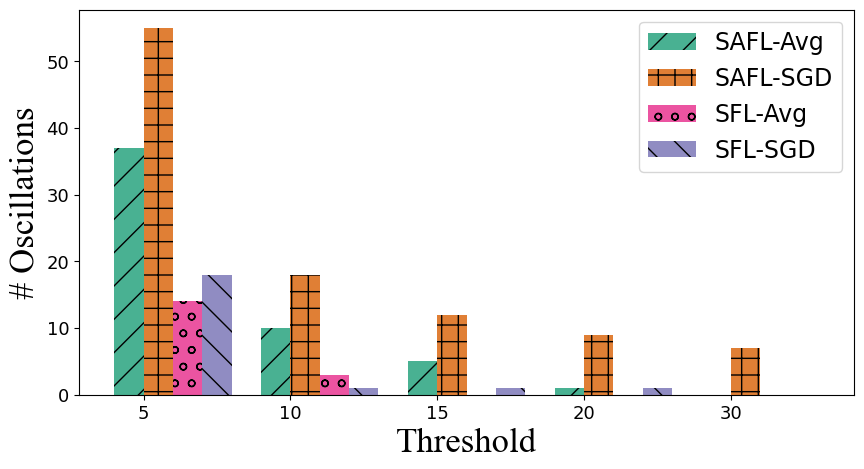}%
\label{fig_resnet18_cifar10_HD_ots_05}}
\hfil
\subfloat[$\alpha = 1$]{\includegraphics[width=1.5in]{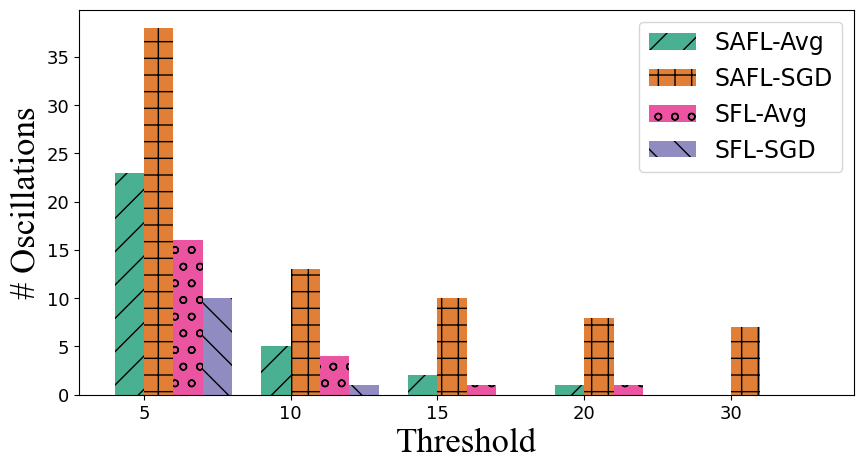}%
\label{fig_resnet18_cifar10_HD_ots_1}}                
\caption{Statistics of severe oscillations of the ResNet-18 model under the CIFAR-10 dataset with Hetero Dirichlet distribution.}
\label{fig_resnet18_cifar10_HD_ots}
% \vspace*{-4ex}
\end{figure}

In Section~\ref{sec:conver}, our primary focus was on the convergence speed and stability of various aggregation strategies. In this section, we utilize the $O_{ots}$ metric to offer a more intuitive assessment of the instability level during the convergence process of the two strategies. Figure~\ref{fig_resnet18_cifar10_HD_ots} shows the number of total oscillations under different thresholds when using ResNet-18 to train CIFAR-10.

The experimental results show that in CV tasks, SAFL tends to induce 70.12\%$\sim$86.77\% more pronounced oscillations (\ie larger oscillation amplitudes) compared to SFL across three distributions. Meanwhile, \sgd exhibits on average 58.3\% more significant oscillations than \avg in SAFL. Particularly, when the threshold is relatively small, both aggregation strategies display a notable degree of oscillations, attributed to the varying impacts of asynchrony in the update strategies. However, with a relatively large threshold, such as 15, \sgd generates on average 82.31\% more oscillations than \avg. These oscillations significantly affect the convergence process, leading to dramatic fluctuations in the model's performance across different rounds. As for NLP tasks, only \sgd in SAFL induces severe oscillations.
%Consequently, selecting an appropriate termination time becomes challenging for the server, and clients exiting at different intervals may receive models with markedly distinct performances.

% \subsection{\tool Performance}

% In this section, we first analyze the experimental results and provide a qualitative analysis that helps to elucidate the effectiveness of our algorithm. That is, our algorithm incorporates mechanisms to handle outliers and mitigates the impact of straggler clients.
% %thereby enhancing its capacity to improve convergence speed and stability. 
% Based on the qualitative analysis, we further demonstrate that \tool is able to achieve more stable convergence and improved performance compared to existing approaches in SAFL.

\begin{figure}[!htp]
\vspace{-1ex}
\centering
\subfloat[SFL]{\includegraphics[width=1.55in]{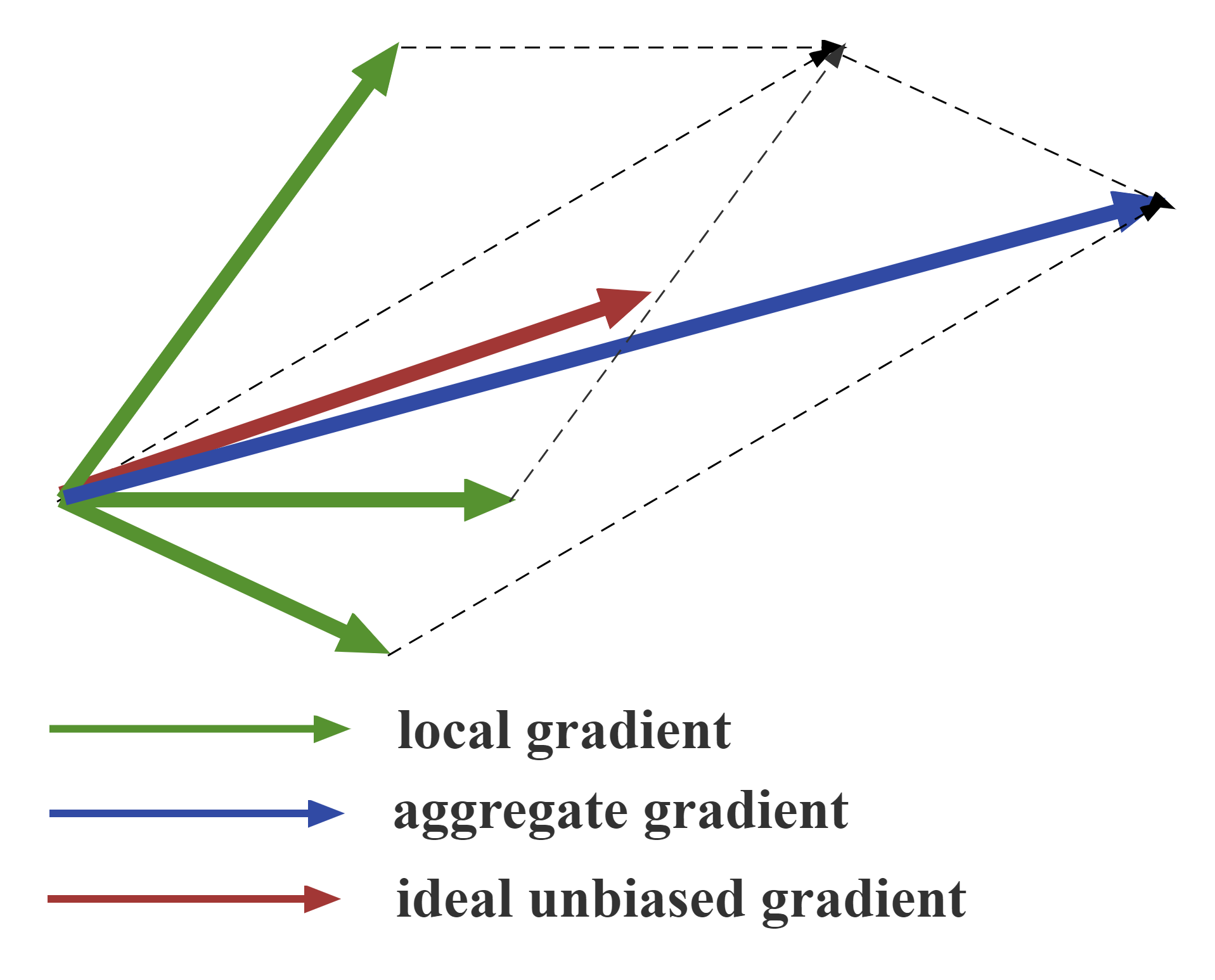}%
\label{fig_SFL_gradient}}
\hspace{0.5mm}
\subfloat[SAFL, a normal case]{\includegraphics[width=1.45in]{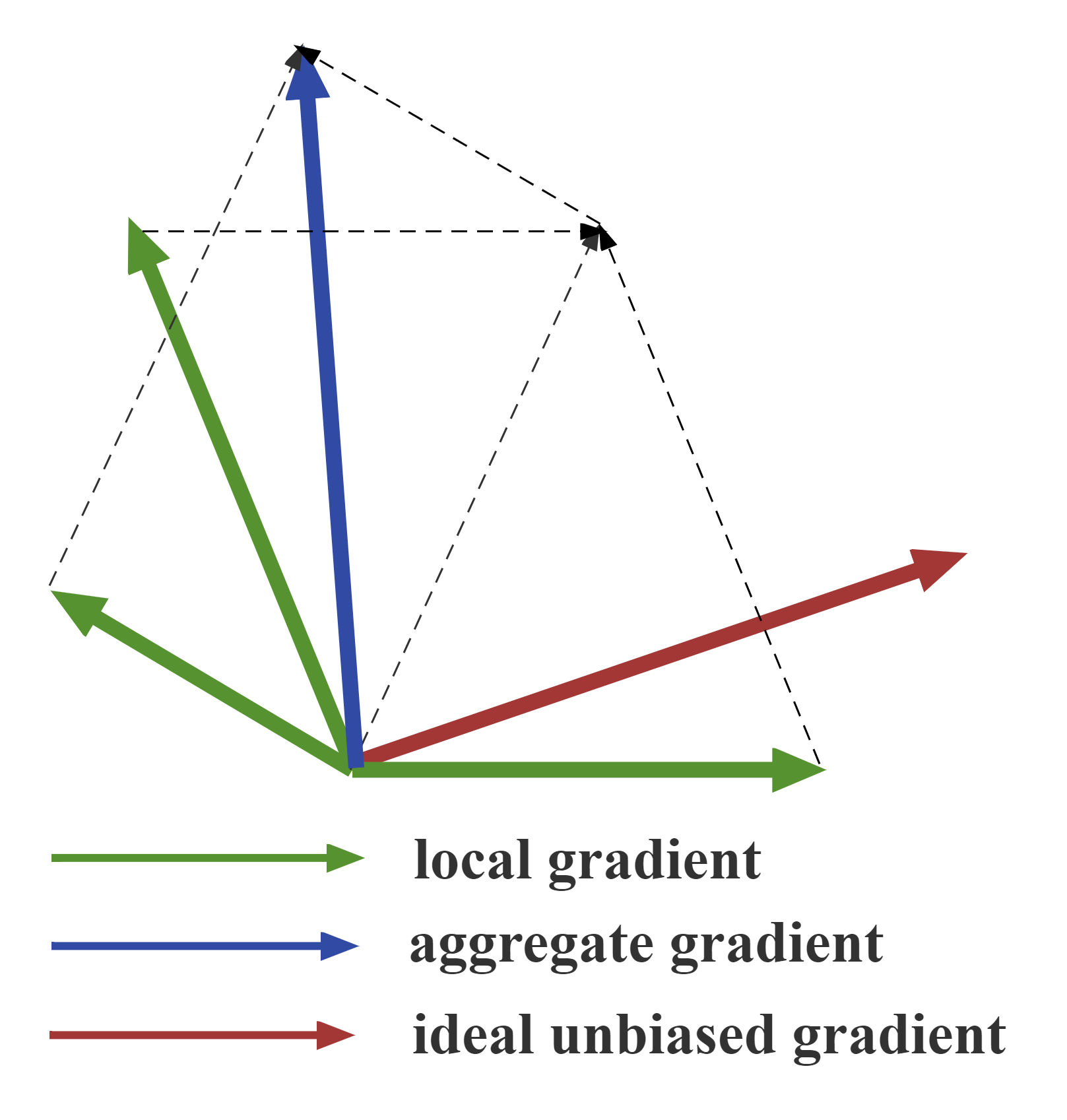}%
\label{fig_SAFL_gradient}}
\hspace{0.5mm}
\subfloat[SAFL, a special case]{\includegraphics[width=1.45in]{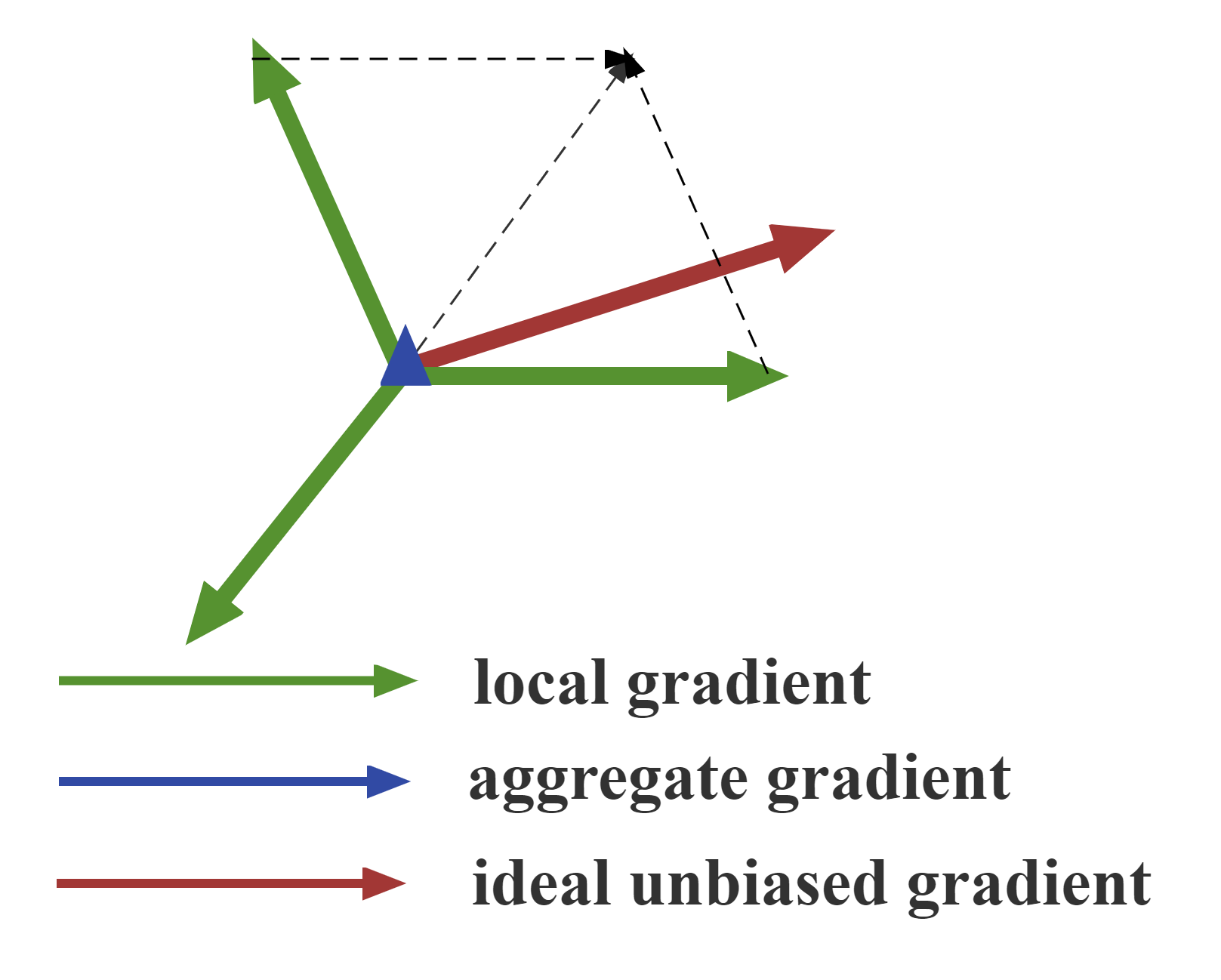}%
\label{fig_SAFL_gradient_NAN}}                
\caption{Directional Gradient Aggregation.}
\label{fig_gradient}
\vspace{-1ex}
\end{figure}

\subsubsection{Qualitative Analysis}

% \subsubsection{Directional gradients}
In \sgd, gradients are vectors with direction, while in \avg, model weights are scalars without direction. This implies that directional gradients provide a reasonable explanation for the difference in gredient aggregation between SFL and SAFL. We utilize Figure~\ref{fig_gradient} to explain more details. It has been suggested that as the number of local training rounds increases, the local model deviates further from the ideal state~\cite{wang2019adaptive}. In SFL, since the gradients in the set of active clients have no staleness, each gradient (green arrow) exhibits less deviation from the ideal gradient (red arrow), resulting in reduced deviation between the aggregated gradient and the ideal gradient, as shown in Figure~\ref{fig_SFL_gradient}. However, in SAFL, the server faces challenges in controlling the aggregation sequence, since the aggregated data contains gradients with varying degrees of staleness. There is a high probability of the existence of gradients with high staleness, leading to significant deviation from the unbiased direction (as shown in Figure~\ref{fig_SAFL_gradient}). This significant deviation is one of the reasons for the model's large oscillations, rendering the convergence process unstable (Problem\ding{172}). Furthermore, in the gradient aggregation of SAFL, there is an extreme case. As illustrated in Figure~\ref{fig_SAFL_gradient_NAN}, the aggregation of the three gradients used in a particular round results in zero, effectively causing the gradient to disappear completely. This renders the model incapable and leads to NaN values in loss computation during model testing due to division by zero.

% \subsubsection{Aggregation effectiveness}
The difference in aggregation effectiveness is also a factor contributing to the performance gap between gradient and model aggregation. Specifically, model aggregation is more likely to fall into local optima (\ie achieving lower accuracy as reflected in Figure~\ref{fig_model_cifar10_HD}a) than gradient aggregation (Problem\ding{173}). On one hand, as illustrated in Figure~\ref{fig_gradient}, using gradients to aggregate the global model disrupts its effectiveness frequently, implying that gradient aggregation is more likely to break local optima, and thus achieving global optima. This is similar to the chaos phenomenon~\cite{chengtian2021adaptive}, where chaotic variables constantly search and traverse the entire space for optimization. On the other hand, model aggregation is more stable in maintaining accuracy because it keeps effectiveness after aggregation, which results in less randomness and thus increases the likelihood of aggregated models converging to local optima.

\section{Related Works}
%Due to its privacy-preserving and the distributed learning characteristics, federated learning\cite{mcmahan2017communication} has ignited significant research interest among scholars. Nonetheless, the heterogeneity of data and resources within federated learning present more nuanced challenges compared to other deep learning frameworks \jp{change to a more objective description}, rendering it a more intricate area for research\cite{pfeiffer2023federated, chai2019towards, reisizadeh2022straggler, zhao2018federated}.

In this section we present related work in terms of performance analysis and improvement in federated learning (FL).

\textbf{Performance Analysis.} To better understand the performance of FL models, researchers have conducted extensive analyses. Xiao \etal~\cite{xiao2021experimental} provide a quantitative analysis of class label imbalance in FL for the first time, and experimentally validate that global class imbalance reduces the performance of the global model. Li \etal~\cite{li2022federated} statistically analyse the non-IID problem in FL and demonstrate that when the data distribution is extremely uneven, commonly used FL algorithms struggle to learn good predictive capabilities. Abdelmoniem \etal~\cite{abdelmoniem2022empirical} conduct extensive empirical research on five popular FL benchmarks, and determine the impact of device and behavioral heterogeneity on the performance and fairness of trained models. Rodriguez \etal~\cite{rodriguez2023survey} argue that the inaccessibility of data in FL exacerbates the risks to model integrity and data privacy. However, all these analyses of FL concentrate on SFL systems, with limited analysis related to SAFL. In contrast to the above studies, our work systematically investigates the performance gap between \sgd and \avg in SAFL.

\textbf{Performance Improvement.} To address the issue of resource heterogeneity, researchers have proposed asynchronous/semi-asynchronous FL frameworks to improve communication efficiency. Wang \etal~\cite{wang2020optimize} propose fully asynchronous FL, named FedAsync, and demonstrate that FedAsync can achieve linear convergence to global optima under certain constraints. Chai \etal~\cite{chai2020fedat} further propose FedAT to address the issue of weak robustness to stragglers in FedAysnc when employing model aggregation. Wu \etal~\cite{wu2020safa} propose SAFA, a SAFL protocal, to achieve fast federated optimization. Recently, some researchers have investigated the issue of stragglers in SAFL and proposed solutions for improvement~\cite{chen2020asynchronous, baccarelli2022afafed,hu2023scheduling,zhang2023fedmds}. KAFL~\cite{wu2022kafl} significantly improves the accuracy when using \avg in the SAFL scenario. Ma \etal~\cite{ma2021fedsa} design the FedSA algorithm to address the low turnover efficiency and slow convergence speed of \sgd in SAFL. Additionally, the work by Zhou \etal~\cite{zhou2022towards} makes \sgd more stable and achieve higher accuracy. However, none of these studies have focused on the performance gap between different aggregation strategies. 
While the above solutions can improve model performance for a specific aggregation strategy, they fail to work effectively for another aggregation strategy.
\section{Conclusion}

In this paper, we systematically study the performance of two aggregation strategies in SAFL systems across diverse scenarios. We find that the performance gap between the two strategies in terms of accuracy, loss, resource consumption, convergence speed, and oscillation amplitude is significant. 
%The experimental results demonstrate that in SAFL, gradient aggregation-based \sgd can facilitate faster convergence, higher accuracy, and less channel payload and training time than model aggregation-based \avg. However, the convergence process of \sgd exhibits instability, characterized by notable oscillations and occurrences of loss disappearance that leads to NAN values.
The experimental results demonstrate that in SAFL, gradient aggregation-based \sgd achieves higher accuracy, faster convergence, less channel payload, and less training time than model aggregation-based \avg. Specifically, \sgd improves the accuracy by 0.23\%$\sim$22.35\% (on average 8.27\%) and 2.10\%$\sim$12.25\% (on average 6.03\%) in CV and NLP tasks, respectively. In addition, \sgd consumes 1\%$\sim$15\% less channel payload and 2\%$\sim$16\% less training time than \avg. However, the convergence process of \sgd exhibits instability, on average characterized by 58\% more oscillations (threshold at 5) than \avg. In terms of severe oscillations (threshold at 15), this number even increases to 82\%. Furthermore, only \sgd experiences occurrences of loss disappearance that leads to NAN values during training.
We attribute these phenomena to the directional nature of gradients and the effectiveness of aggregation. 
% This motivates us to propose \tool that bridges the performance gap between \sgd and \avg in SAFL. Through server-side supervision of aggregation using small-scale datasets, our algorithm adeptly accommodates both types of aggregation strategies and outperforms SOTA approaches in diverse scenarios. In particular, our approach effectively mitigates the occurrences of NAN in loss function and severe oscillations, alleviating the impact of stragglers on the global model training in gradient aggregation. In addition, our approach reduces the 30\% accuracy gap on average across all scenarios between two aggregation strategies, with an accuracy improvement of 84.18\%$\sim$86.23\% and 73.04\%$\sim$78.54\% on gradient aggregation and model aggregation, respectively. To the best of our knowledge, this is the first study exploring the impact of aggregation objects on performance in SAFL. 
We believe that our study provides strong evidence of the performance gap between two aggregation strategies in SAFL. The community needs to meticulously select aggregation strategies to meet scenario requirements, especially when accounting for the system heterogeneity among clients.

\bibliographystyle{unsrt}  
% \bibliography{references}  
% \bibliographystyle{IEEEtran}
\bibliography{IEEEabrv,contents/ref}

\begin{thebibliography}{10}

\bibitem{mcmahan2017communication}
Brendan McMahan, Eider Moore, Daniel Ramage, Seth Hampson, and Blaise~Aguera
  y~Arcas.
\newblock Communication-efficient learning of deep networks from decentralized
  data.
\newblock In {\em Artificial intelligence and statistics}, pages 1273--1282.
  PMLR, 2017.

\bibitem{li2019convergence}
Xiang Li, Kaixuan Huang, Wenhao Yang, Shusen Wang, and Zhihua Zhang.
\newblock On the convergence of fedavg on non-iid data.
\newblock {\em arXiv preprint arXiv:1907.02189}, 2019.

\bibitem{li2021fedbn}
Xiaoxiao Li, Meirui Jiang, Xiaofei Zhang, Michael Kamp, and Qi~Dou.
\newblock Fedbn: Federated learning on non-iid features via local batch
  normalization.
\newblock {\em arXiv preprint arXiv:2102.07623}, 2021.

\bibitem{zhu2021federated}
Zhaowei Zhu, Jingxuan Zhu, Ji~Liu, and Yang Liu.
\newblock Federated bandit: A gossiping approach.
\newblock {\em Proceedings of the ACM on Measurement and Analysis of Computing
  Systems}, 5(1):1--29, 2021.

\bibitem{guo2021lightfed}
Jialin Guo, Jie Wu, Anfeng Liu, and Neal~N Xiong.
\newblock Lightfed: An efficient and secure federated edge learning system on
  model splitting.
\newblock {\em IEEE Transactions on Parallel and Distributed Systems},
  33(11):2701--2713, 2021.

\bibitem{huang2023distributed}
Hong Huang, Lan Zhang, Chaoyue Sun, Ruogu Fang, Xiaoyong Yuan, and Dapeng Wu.
\newblock Distributed pruning towards tiny neural networks in federated
  learning.
\newblock In {\em 2023 IEEE 43rd International Conference on Distributed
  Computing Systems (ICDCS)}, pages 190--201. IEEE, 2023.

\bibitem{qi2023hwamei}
Tianyu Qi, Yufeng Zhan, Peng Li, Jingcai Guo, and Yuanqing Xia.
\newblock Hwamei: A learning-based synchronization scheme for hierarchical
  federated learning.
\newblock In {\em 2023 IEEE 43rd International Conference on Distributed
  Computing Systems (ICDCS)}, pages 534--544. IEEE, 2023.

\bibitem{ludwig2020ibm}
Heiko Ludwig, Nathalie Baracaldo, Gegi Thomas, Yi~Zhou, Ali Anwar, Shashank
  Rajamoni, Yuya Ong, Jayaram Radhakrishnan, Ashish Verma, Mathieu Sinn, et~al.
\newblock Ibm federated learning: an enterprise framework white paper v0. 1.
\newblock {\em arXiv preprint arXiv:2007.10987}, 2020.

\bibitem{long2020federated}
Guodong Long, Yue Tan, Jing Jiang, and Chengqi Zhang.
\newblock Federated learning for open banking.
\newblock In {\em Federated Learning: Privacy and Incentive}, pages 240--254.
  Springer, 2020.

\bibitem{wu2020fedhome}
Qiong Wu, Xu~Chen, Zhi Zhou, and Junshan Zhang.
\newblock Fedhome: Cloud-edge based personalized federated learning for in-home
  health monitoring.
\newblock {\em IEEE Transactions on Mobile Computing}, 21(8):2818--2832, 2020.

\bibitem{nasirigerdeh2023utility}
Reza Nasirigerdeh, Daniel Rueckert, and Georgios Kaissis.
\newblock Utility-preserving federated learning.
\newblock In {\em Proceedings of the 16th ACM Workshop on Artificial
  Intelligence and Security}, pages 55--65, 2023.

\bibitem{zhao2018federated}
Yue Zhao, Meng Li, Liangzhen Lai, Naveen Suda, Damon Civin, and Vikas Chandra.
\newblock Federated learning with non-iid data.
\newblock {\em arXiv preprint arXiv:1806.00582}, 2018.

\bibitem{sattler2019robust}
Felix Sattler, Simon Wiedemann, Klaus-Robert M{\"u}ller, and Wojciech Samek.
\newblock Robust and communication-efficient federated learning from non-iid
  data.
\newblock {\em IEEE transactions on neural networks and learning systems},
  31(9):3400--3413, 2019.

\bibitem{wu2020safa}
Wentai Wu, Ligang He, Weiwei Lin, Rui Mao, Carsten Maple, and Stephen Jarvis.
\newblock Safa: A semi-asynchronous protocol for fast federated learning with
  low overhead.
\newblock {\em IEEE Transactions on Computers}, 70(5):655--668, 2020.

\bibitem{ma2021fedsa}
Qianpiao Ma, Yang Xu, Hongli Xu, Zhida Jiang, Liusheng Huang, and He~Huang.
\newblock Fedsa: A semi-asynchronous federated learning mechanism in
  heterogeneous edge computing.
\newblock {\em IEEE Journal on Selected Areas in Communications},
  39(12):3654--3672, 2021.

\bibitem{cheng2022aafl}
Jieren Cheng, Ping Luo, N~Xiong, and Jie Wu.
\newblock Aafl: Asynchronous-adaptive federated learning in edge-based wireless
  communication systems for countering communicable infectious diseasess.
\newblock {\em IEEE Journal on Selected Areas in Communications},
  40(11):3172--3190, 2022.

\bibitem{xie2019asynchronous}
Cong Xie, Sanmi Koyejo, and Indranil Gupta.
\newblock Asynchronous federated optimization.
\newblock {\em arXiv preprint arXiv:1903.03934}, 2019.

\bibitem{zhu2022online}
Hongbin Zhu, Yong Zhou, Hua Qian, Yuanming Shi, Xu~Chen, and Yang Yang.
\newblock Online client selection for asynchronous federated learning with
  fairness consideration.
\newblock {\em IEEE Transactions on Wireless Communications}, 22(4):2493--2506,
  2022.

\bibitem{baccarelli2022afafed}
Enzo Baccarelli, Michele Scarpiniti, Alireza Momenzadeh, and Sima~Sarv Ahrabi.
\newblock Afafed—asynchronous fair adaptive federated learning for iot stream
  applications.
\newblock {\em Computer Communications}, 195:376--402, 2022.

\bibitem{melis2019exploiting}
Luca Melis, Congzheng Song, Emiliano De~Cristofaro, and Vitaly Shmatikov.
\newblock Exploiting unintended feature leakage in collaborative learning.
\newblock In {\em 2019 IEEE symposium on security and privacy (SP)}, pages
  691--706. IEEE, 2019.

\bibitem{zhu2019deep}
Ligeng Zhu, Zhijian Liu, and Song Han.
\newblock Deep leakage from gradients.
\newblock {\em Advances in neural information processing systems}, 32, 2019.

\bibitem{wainakh2021user}
Aidmar Wainakh, Fabrizio Ventola, Till M{\"u}{\ss}ig, Jens Keim, Carlos~Garcia
  Cordero, Ephraim Zimmer, Tim Grube, Kristian Kersting, and Max
  M{\"u}hlh{\"a}user.
\newblock User-level label leakage from gradients in federated learning.
\newblock {\em arXiv preprint arXiv:2105.09369}, 2021.

\bibitem{chai2020fedeval}
Di~Chai, Leye Wang, Liu Yang, Junxue Zhang, Kai Chen, and Qiang Yang.
\newblock Fedeval: A holistic evaluation framework for federated learning.
\newblock {\em arXiv preprint arXiv:2011.09655}, 2020.

\bibitem{krizhevsky2009learning}
Alex Krizhevsky, Geoffrey Hinton, et~al.
\newblock Learning multiple layers of features from tiny images.
\newblock 2009.

\bibitem{caldas2018leaf}
Sebastian Caldas, Sai Meher~Karthik Duddu, Peter Wu, Tian Li, Jakub
  Kone{\v{c}}n{\`y}, H~Brendan McMahan, Virginia Smith, and Ameet Talwalkar.
\newblock Leaf: A benchmark for federated settings.
\newblock {\em arXiv preprint arXiv:1812.01097}, 2018.

\bibitem{cohen2017emnist}
Gregory Cohen, Saeed Afshar, Jonathan Tapson, and Andre Van~Schaik.
\newblock Emnist: Extending mnist to handwritten letters.
\newblock In {\em 2017 international joint conference on neural networks
  (IJCNN)}, pages 2921--2926. IEEE, 2017.

\bibitem{go2009twitter}
Alec Go, Richa Bhayani, and Lei Huang.
\newblock Twitter sentiment classification using distant supervision.
\newblock {\em CS224N project report, Stanford}, 1(12):2009, 2009.

\bibitem{JMLR:v24:22-0440}
Dun Zeng, Siqi Liang, Xiangjing Hu, Hui Wang, and Zenglin Xu.
\newblock Fedlab: A flexible federated learning framework.
\newblock {\em Journal of Machine Learning Research}, 24(100):1--7, 2023.

\bibitem{he2016deep}
Kaiming He, Xiangyu Zhang, Shaoqing Ren, and Jian Sun.
\newblock Deep residual learning for image recognition.
\newblock In {\em Proceedings of the IEEE conference on computer vision and
  pattern recognition}, pages 770--778, 2016.

\bibitem{simonyan2014very}
Karen Simonyan and Andrew Zisserman.
\newblock Very deep convolutional networks for large-scale image recognition.
\newblock {\em arXiv preprint arXiv:1409.1556}, 2014.

\bibitem{hochreiter1997long}
Sepp Hochreiter and J{\"u}rgen Schmidhuber.
\newblock Long short-term memory.
\newblock {\em Neural computation}, 9(8):1735--1780, 1997.

\bibitem{wang2019adaptive}
Shiqiang Wang, Tiffany Tuor, Theodoros Salonidis, Kin~K Leung, Christian
  Makaya, Ting He, and Kevin Chan.
\newblock Adaptive federated learning in resource constrained edge computing
  systems.
\newblock {\em IEEE journal on selected areas in communications},
  37(6):1205--1221, 2019.

\bibitem{chengtian2021adaptive}
Ouyang Chengtian, Liu Yujia, and Zhu Donglin.
\newblock An adaptive chaotic sparrow search optimization algorithm.
\newblock In {\em 2021 IEEE 2nd international conference on Big data,
  artificial intelligence and internet of things engineering (ICBAIE)}, pages
  76--82. IEEE, 2021.

\bibitem{xiao2021experimental}
Chenguang Xiao and Shuo Wang.
\newblock An experimental study of class imbalance in federated learning.
\newblock In {\em 2021 IEEE Symposium Series on Computational Intelligence
  (SSCI)}, pages 1--7. IEEE, 2021.

\bibitem{li2022federated}
Qinbin Li, Yiqun Diao, Quan Chen, and Bingsheng He.
\newblock Federated learning on non-iid data silos: An experimental study.
\newblock In {\em 2022 IEEE 38th international conference on data engineering
  (ICDE)}, pages 965--978. IEEE, 2022.

\bibitem{abdelmoniem2022empirical}
Ahmed~M Abdelmoniem, Chen-Yu Ho, Pantelis Papageorgiou, and Marco Canini.
\newblock Empirical analysis of federated learning in heterogeneous
  environments.
\newblock In {\em Proceedings of the 2nd European Workshop on Machine Learning
  and Systems}, pages 1--9, 2022.

\bibitem{rodriguez2023survey}
Nuria Rodr{\'\i}guez-Barroso, Daniel Jim{\'e}nez-L{\'o}pez, M~Victoria
  Luz{\'o}n, Francisco Herrera, and Eugenio Mart{\'\i}nez-C{\'a}mara.
\newblock Survey on federated learning threats: Concepts, taxonomy on attacks
  and defences, experimental study and challenges.
\newblock {\em Information Fusion}, 90:148--173, 2023.

\bibitem{wang2020optimize}
Cong Wang, Xin Wei, and Pengzhan Zhou.
\newblock Optimize scheduling of federated learning on battery-powered mobile
  devices.
\newblock In {\em 2020 IEEE International Parallel and Distributed Processing
  Symposium (IPDPS)}, pages 212--221. IEEE, 2020.

\bibitem{chai2020fedat}
Zheng Chai, Yujing Chen, Liang Zhao, Yue Cheng, and Huzefa Rangwala.
\newblock Fedat: A communication-efficient federated learning method with
  asynchronous tiers under non-iid data.
\newblock {\em ArXivorg}, 2020.

\bibitem{chen2020asynchronous}
Yujing Chen, Yue Ning, Martin Slawski, and Huzefa Rangwala.
\newblock Asynchronous online federated learning for edge devices with non-iid
  data.
\newblock In {\em 2020 IEEE International Conference on Big Data (Big Data)},
  pages 15--24. IEEE, 2020.

\bibitem{hu2023scheduling}
Chung-Hsuan Hu, Zheng Chen, and Erik~G Larsson.
\newblock Scheduling and aggregation design for asynchronous federated learning
  over wireless networks.
\newblock {\em IEEE Journal on Selected Areas in Communications},
  41(4):874--886, 2023.

\bibitem{zhang2023fedmds}
Yu~Zhang, Duo Liu, Moming Duan, Li~Li, Xianzhang Chen, Ao~Ren, Yujuan Tan, and
  Chengliang Wang.
\newblock Fedmds: An efficient model discrepancy-aware semi-asynchronous
  clustered federated learning framework.
\newblock {\em IEEE Transactions on Parallel and Distributed Systems},
  34(3):1007--1019, 2023.

\bibitem{wu2022kafl}
Xueyu Wu and Cho-Li Wang.
\newblock Kafl: Achieving high training efficiency for fast-k asynchronous
  federated learning.
\newblock In {\em 2022 IEEE 42nd International Conference on Distributed
  Computing Systems (ICDCS)}, pages 873--883. IEEE, 2022.

\bibitem{zhou2022towards}
Zihao Zhou, Yanan Li, Xuebin Ren, and Shusen Yang.
\newblock Towards efficient and stable k-asynchronous federated learning with
  unbounded stale gradients on non-iid data.
\newblock {\em IEEE Transactions on Parallel and Distributed Systems},
  33(12):3291--3305, 2022.

\end{thebibliography}
\appendices
\section{More experiment results} 

\subsection{CIFAR-100 @ ResNet-18}
% The accuracy and loss under CIFAR-100 using ResNet-18 in SAFL:

\begin{figure}[!h]
\vspace{-1ex}
\centering
\subfloat[$\alpha = 0.1$]{\includegraphics[width=1.5in]{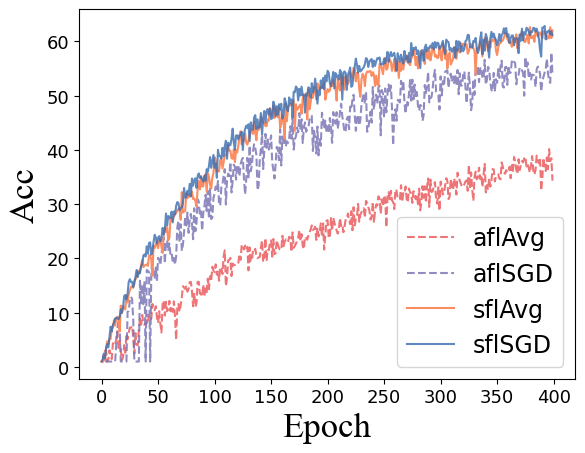}}
\hspace{1mm}
\subfloat[$\alpha = 0.1$]{\includegraphics[width=1.5in]{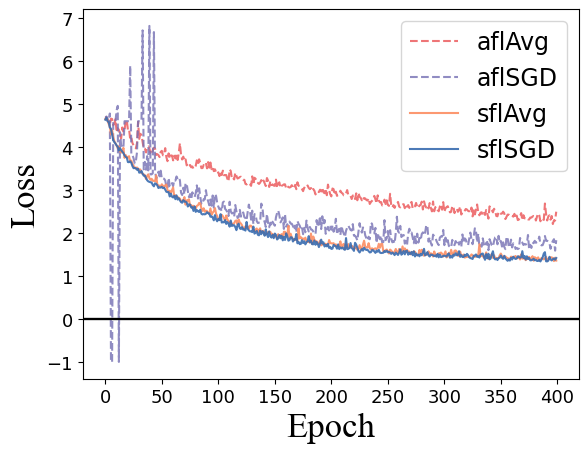}}
\hspace{1mm}
\subfloat[$\alpha = 0.5$]{\includegraphics[width=1.5in]{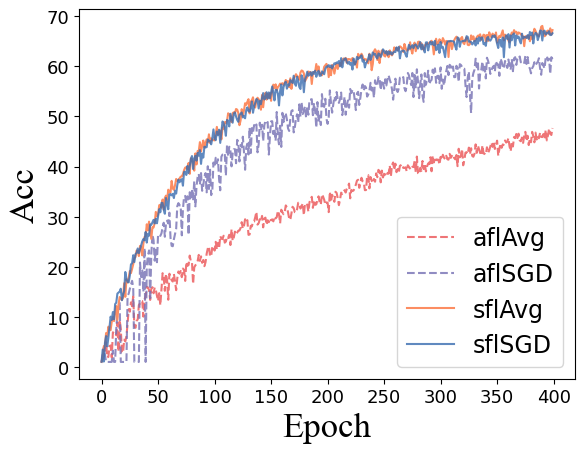}}
\hspace{1mm}
\subfloat[$\alpha = 0.5$]{\includegraphics[width=1.5in]{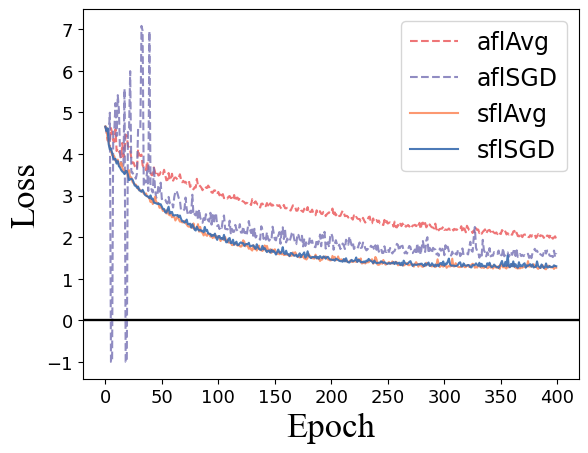}\label{fig_nan_concrete}}
\hspace{1mm}
\subfloat[$\alpha = 1$]{\includegraphics[width=1.5in]{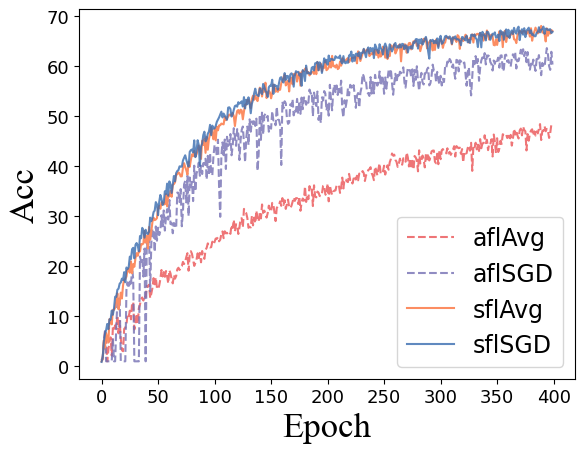}}
\hspace{1mm}
\subfloat[$\alpha = 1$]{\includegraphics[width=1.5in]{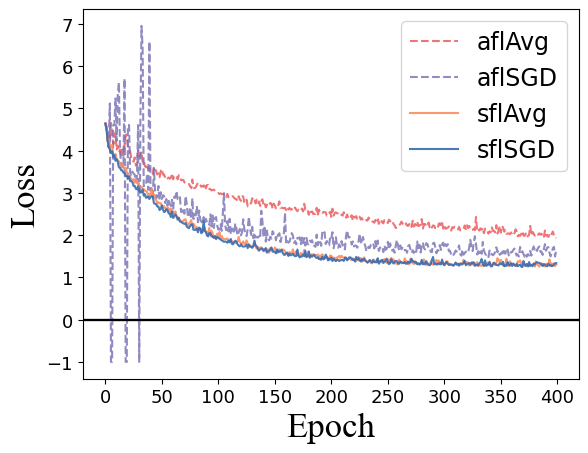}}
\hspace{1mm}
% \subfloat[$N = 2$]{\includegraphics[width=1.5in]{pic/ret/resnet18_cifar100/shards/acc/2.png}}
% \hspace{1mm}
% \subfloat[$N = 2$]{\includegraphics[width=1.5in]{pic/ret/resnet18_cifar100/shards/loss/2.png}}
% \hspace{1mm}
\subfloat[$N = 5$]{\includegraphics[width=1.5in]{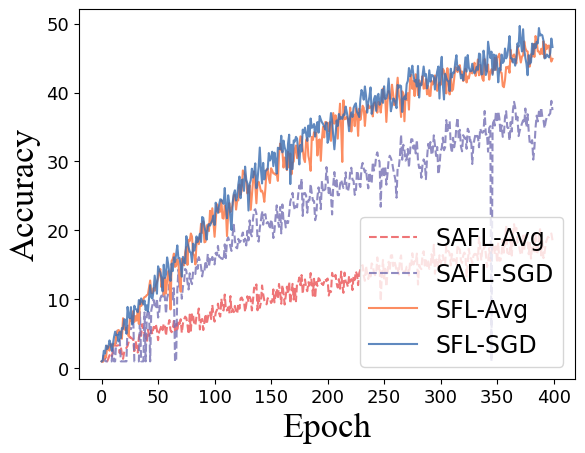}}
\hspace{1mm}
\subfloat[$N = 5$]{\includegraphics[width=1.5in]{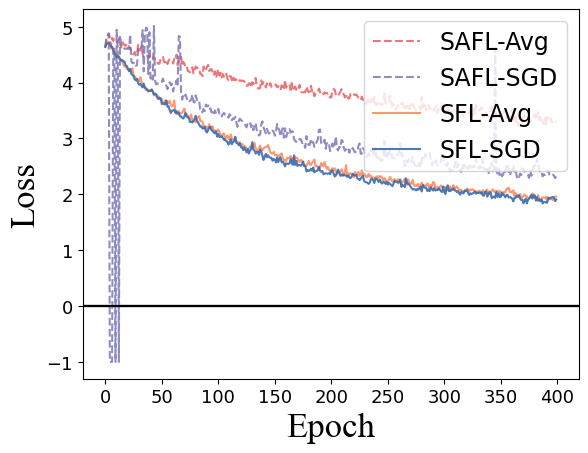}}
\hspace{1mm}
\subfloat[$N = 10$]{\includegraphics[width=1.5in]{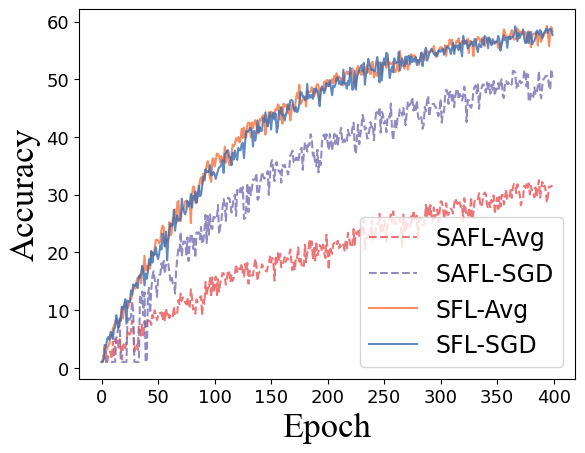}}
\hspace{1mm}
\subfloat[$N = 10$]{\includegraphics[width=1.5in]{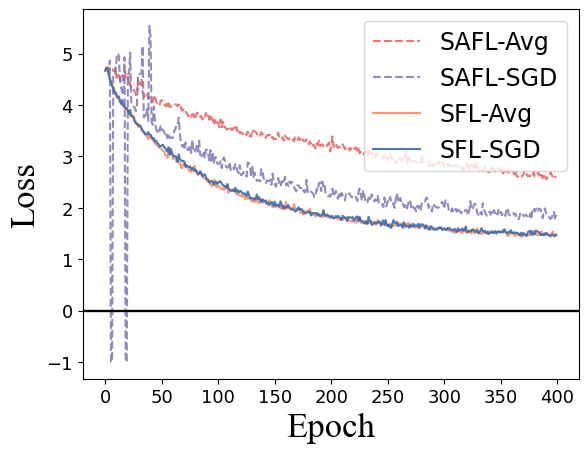}}
\hspace{1mm}
\subfloat[$\sigma = 0.1$]{\includegraphics[width=1.5in]{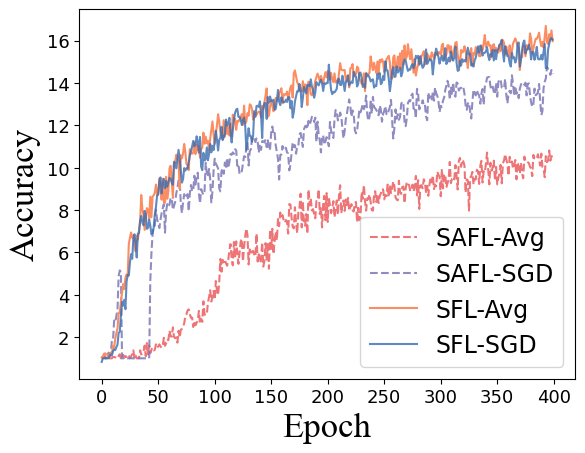}}
\hspace{1mm}
\subfloat[$\sigma = 0.1$]{\includegraphics[width=1.5in]{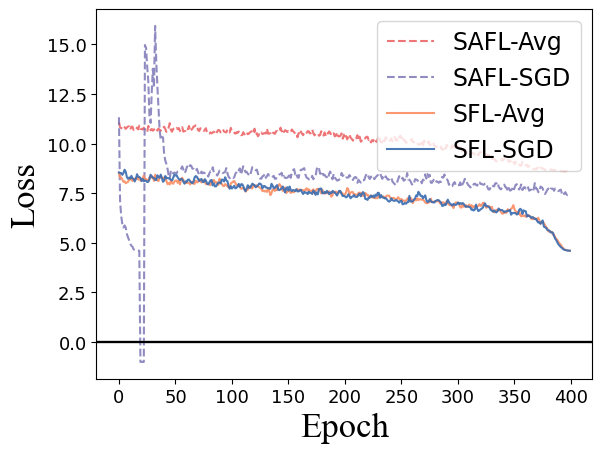}}
\hspace{1mm}
\subfloat[$\sigma = 0.5$]{\includegraphics[width=1.5in]{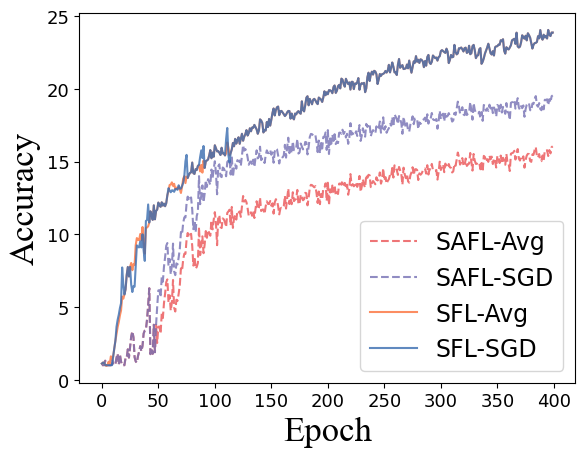}}
\hspace{1mm}
\subfloat[$\sigma = 0.5$]{\includegraphics[width=1.5in]{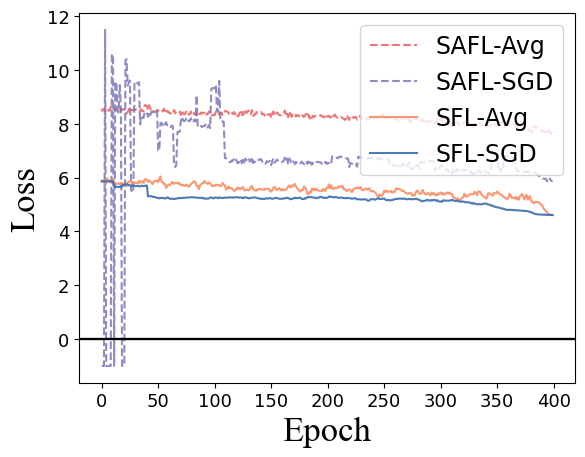}}
\hspace{1mm}
\subfloat[$\sigma = 1$]{\includegraphics[width=1.5in]{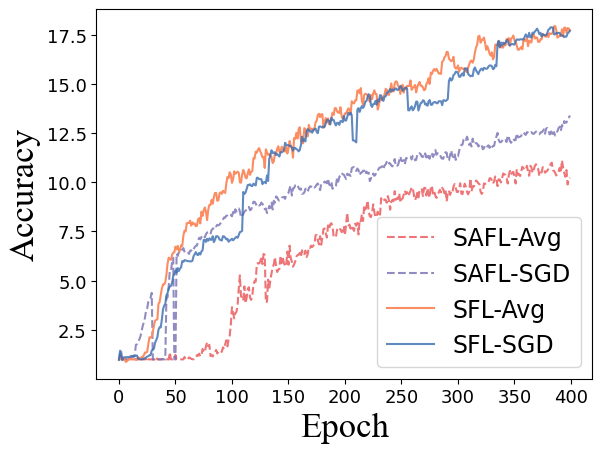}}
\hspace{1mm}
\subfloat[$\sigma = 1$]{\includegraphics[width=1.5in]{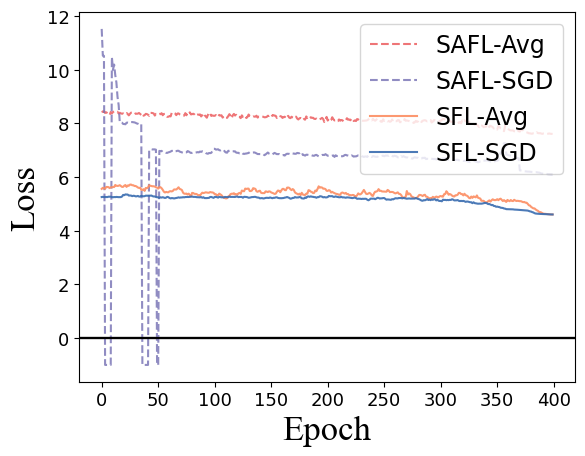}}
\caption{Global accuracy and loss of different models under CIFAR-100 dataset using ResNet-18 in SAFL. Note that -1 denotes the NAN value for loss. }
\vspace{-1ex}
\end{figure}

\begin{figure}[!h]
\vspace{-1ex}
\centering
\subfloat[$\alpha = 0.1$]{\includegraphics[width=1.5in]{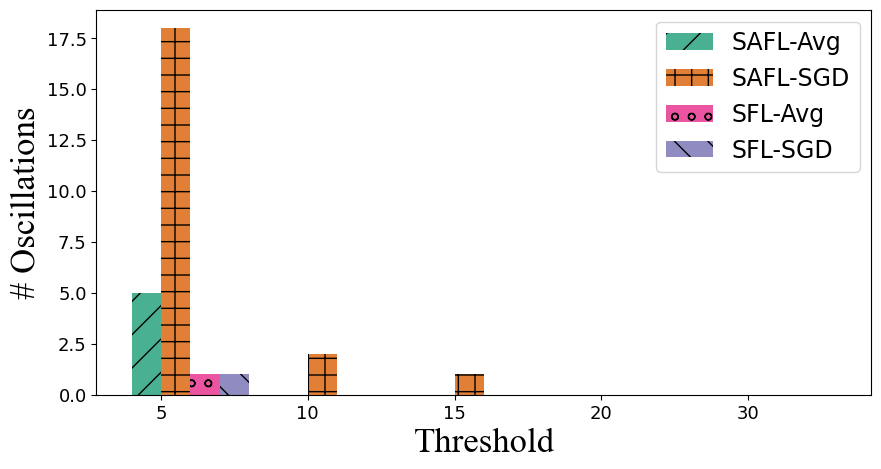}%
}
\hfil
\subfloat[$\alpha = 0.5$]{\includegraphics[width=1.5in]{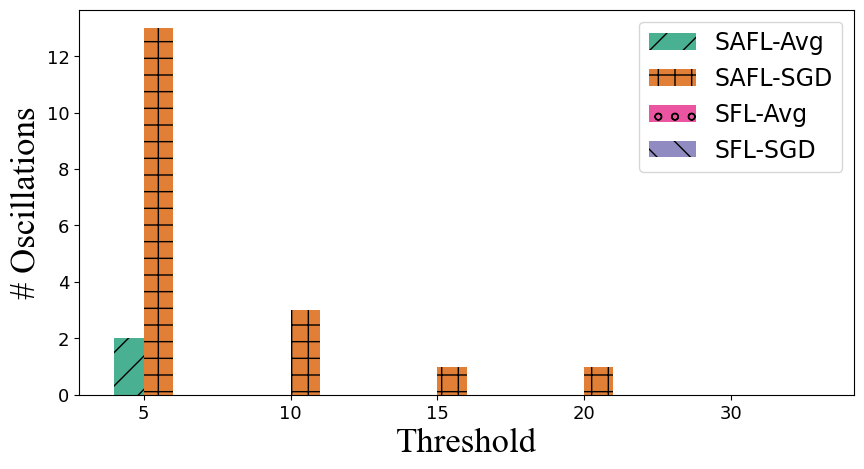}%
}
\hfil
\subfloat[$\alpha = 1$]{\includegraphics[width=1.5in]{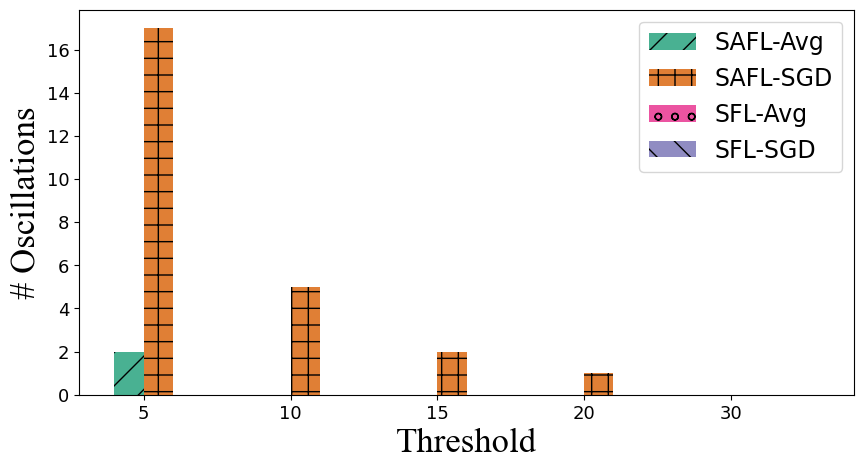}%
}    
% \hfil
% \subfloat[$N = 2$]{\includegraphics[width=1.5in]{pic/ret/resnet18_cifar100/shards/ots/2.png}%
% }   
\hfil
\subfloat[$N = 5$]{\includegraphics[width=1.5in]{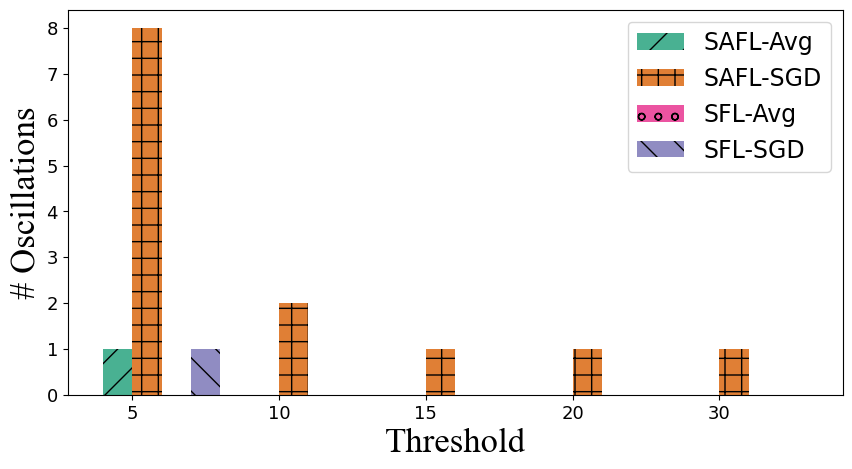}%
}   
\hfil
\subfloat[$N = 10$]{\includegraphics[width=1.5in]{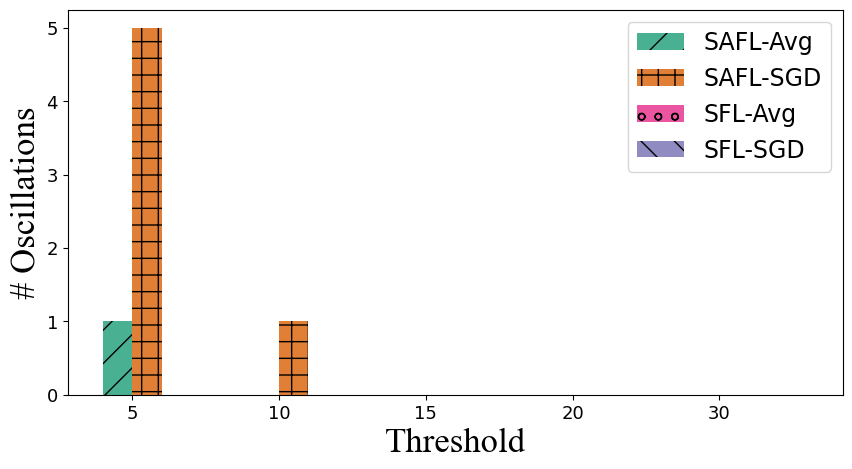}%
}   
\hfil
\subfloat[$\sigma = 0.1$]{\includegraphics[width=1.5in]{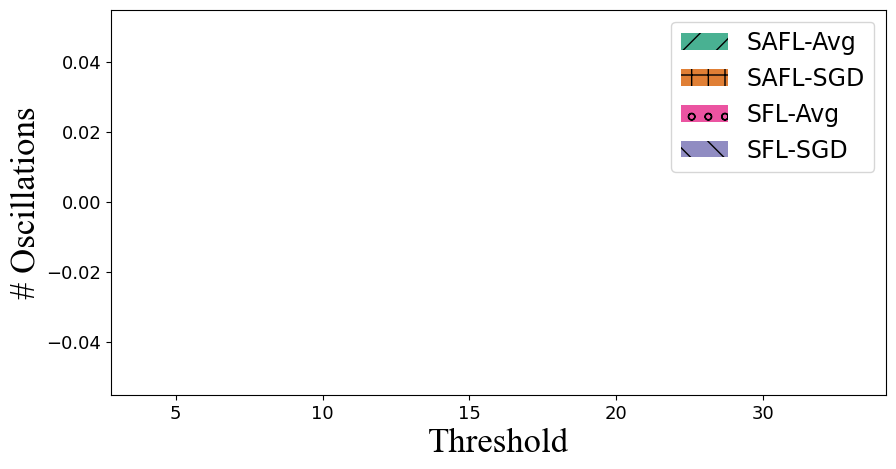}%
}   
\hfil
\subfloat[$\sigma = 0.5$]{\includegraphics[width=1.5in]{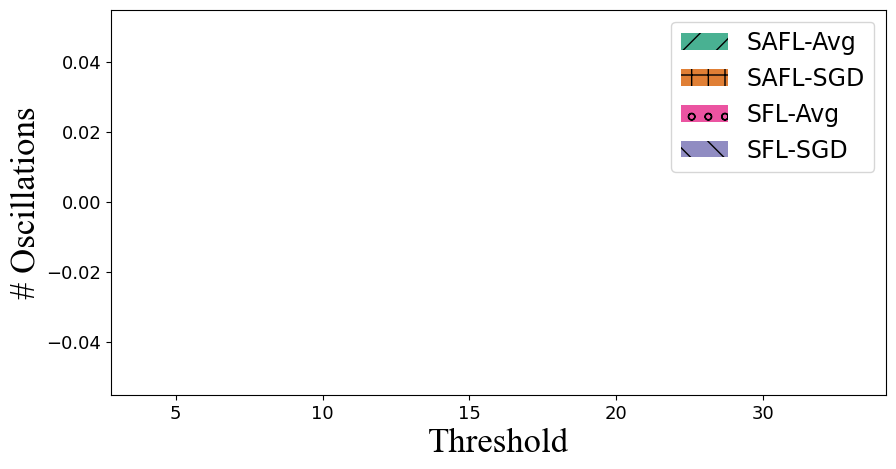}%
}   
\hfil
\subfloat[$\sigma = 1$]{\includegraphics[width=1.5in]{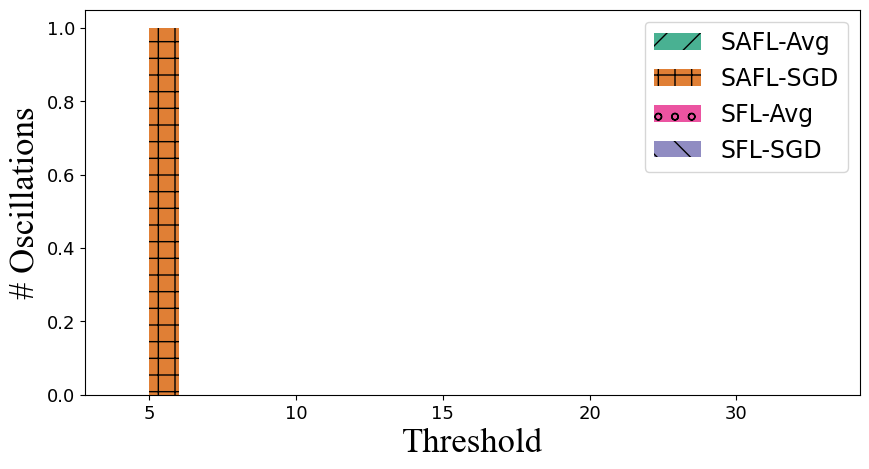}%
}   
\vspace{-1ex}
\caption{Statistics of severe oscillations of the ResNet-18 model under the CIFAR-100 dataset.}
\end{figure}

\newpage
\subsection{FEMNIST @ ResNet-18}
% The accuracy and loss under FEMNIST using ResNet-18 in SAFL:

\begin{figure}[!h]
\vspace{-2ex}
\centering
\subfloat[$\alpha = 0.1$]{\includegraphics[width=1.5in]{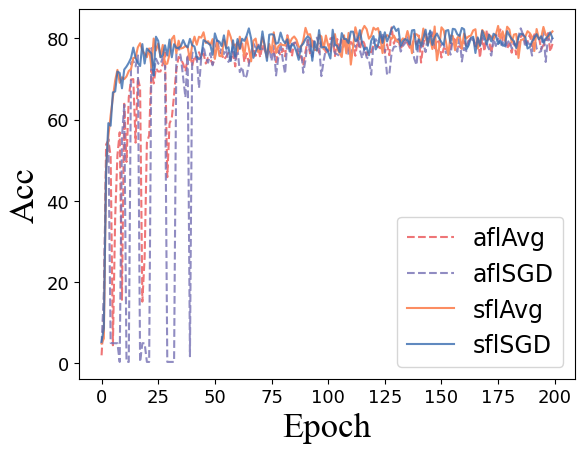}}
\hspace{1mm}
\subfloat[$\alpha = 0.1$]{\includegraphics[width=1.5in]{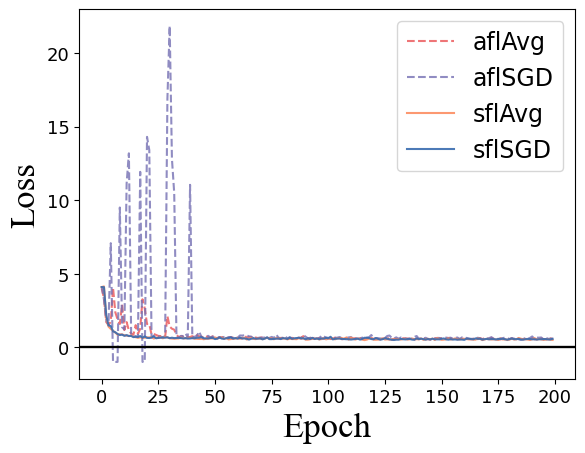}}
\hspace{1mm}
\subfloat[$\alpha = 0.5$]{\includegraphics[width=1.5in]{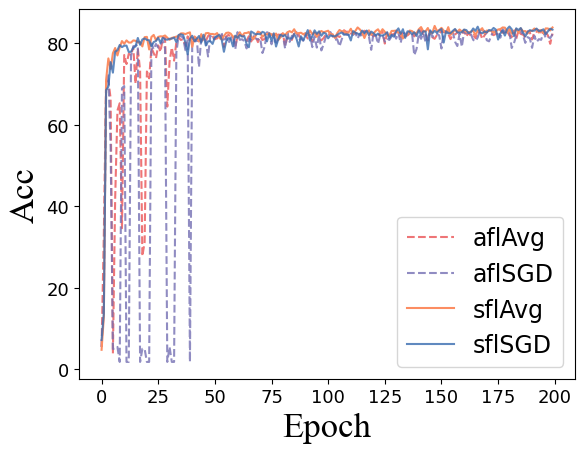}}
\hspace{1mm}
\subfloat[$\alpha = 0.5$]{\includegraphics[width=1.5in]{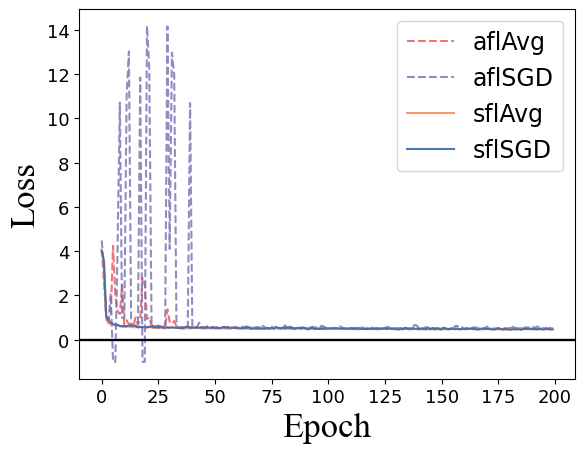}\label{fig_nan_concrete}}
\hspace{1mm}
\subfloat[$\alpha = 1$]{\includegraphics[width=1.5in]{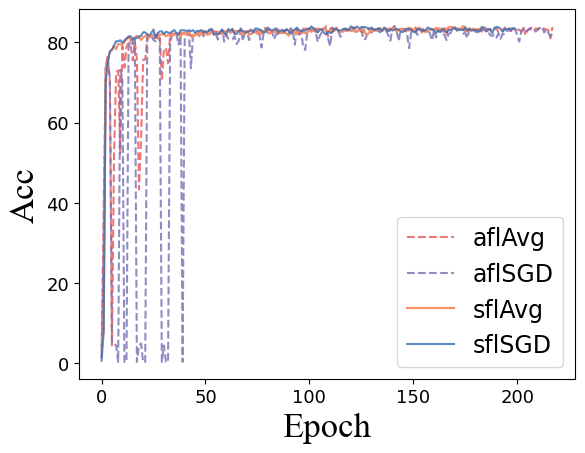}}
\hspace{1mm}
\subfloat[$\alpha = 1$]{\includegraphics[width=1.5in]{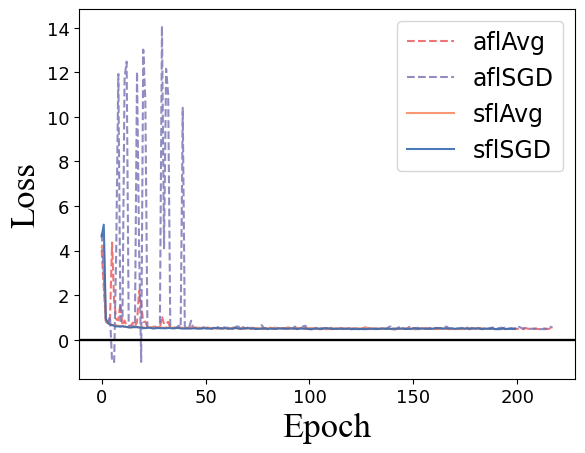}}
\hspace{1mm}
% \subfloat[$N = 2$]{\includegraphics[width=1.5in]{pic/ret/resnet18_femnist/shards/acc/2.png}}
% \hspace{1mm}
% \subfloat[$N = 2$]{\includegraphics[width=1.5in]{pic/ret/resnet18_femnist/shards/loss/2.png}}
% \hspace{1mm}
\subfloat[$N = 5$]{\includegraphics[width=1.5in]{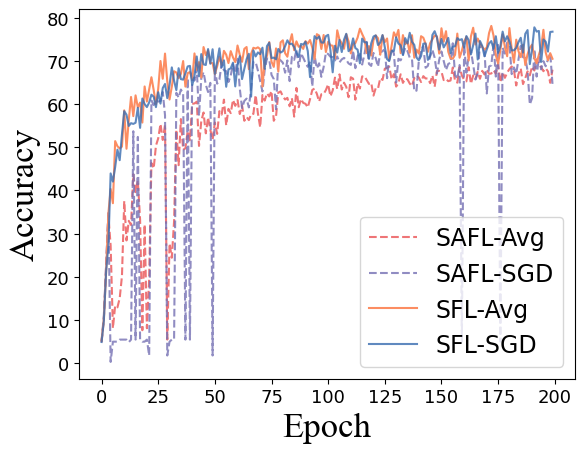}}
\hspace{1mm}
\subfloat[$N = 5$]{\includegraphics[width=1.5in]{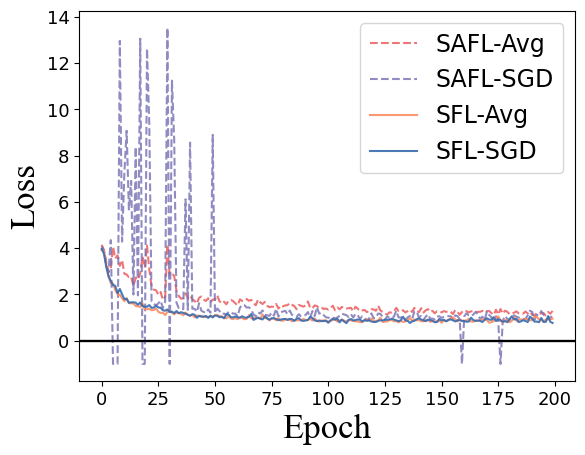}}
\hspace{1mm}
\subfloat[$N = 10$]{\includegraphics[width=1.5in]{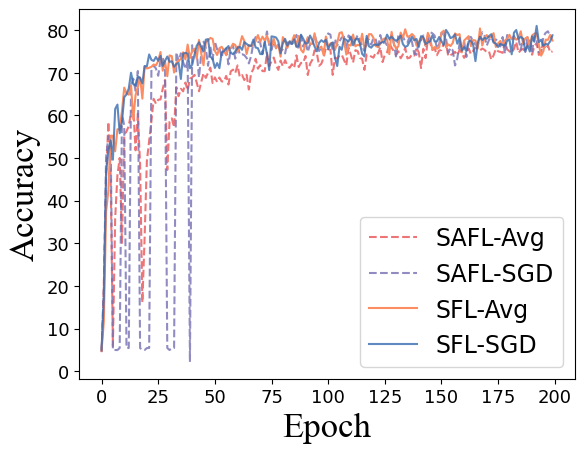}}
\hspace{1mm}
\subfloat[$N = 10$]{\includegraphics[width=1.5in]{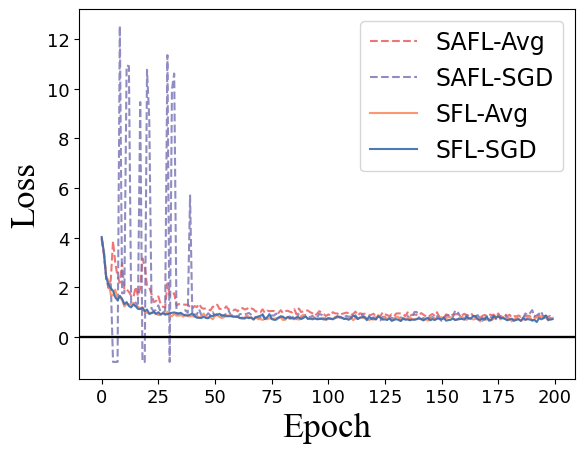}}
\hspace{1mm}
\subfloat[$\sigma = 0.1$]{\includegraphics[width=1.5in]{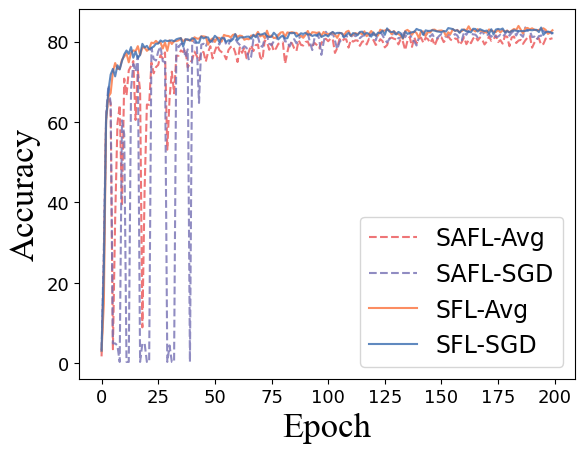}}
\hspace{1mm}
\subfloat[$\sigma = 0.1$]{\includegraphics[width=1.5in]{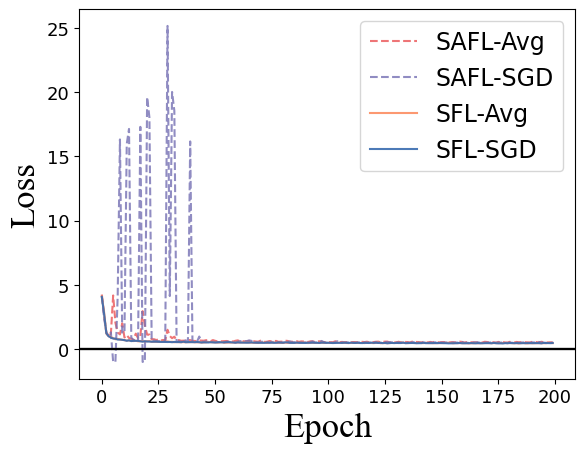}}
\hspace{1mm}
\subfloat[$\sigma = 0.5$]{\includegraphics[width=1.5in]{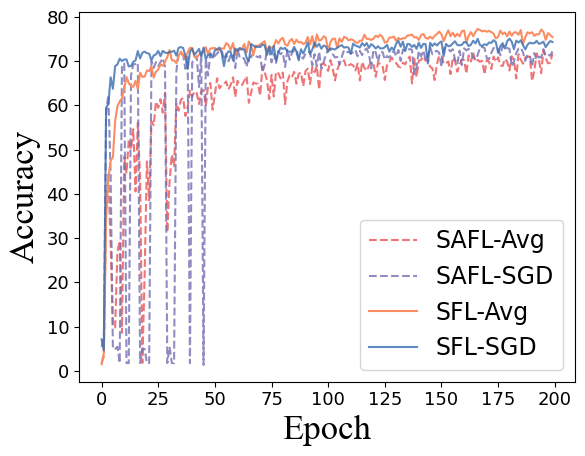}}
\hspace{1mm}
\subfloat[$\sigma = 0.5$]{\includegraphics[width=1.5in]{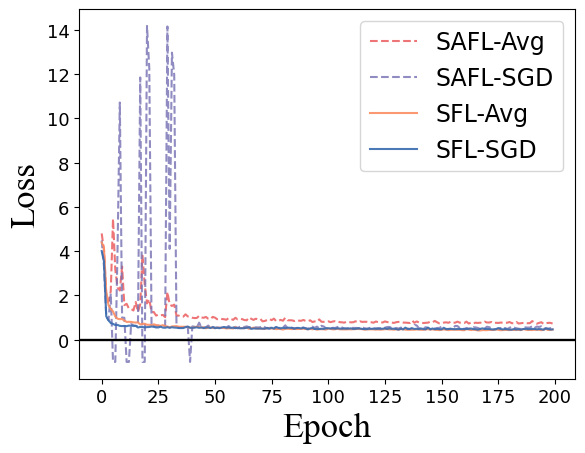}}
\hspace{1mm}
\subfloat[$\sigma = 1$]{\includegraphics[width=1.5in]{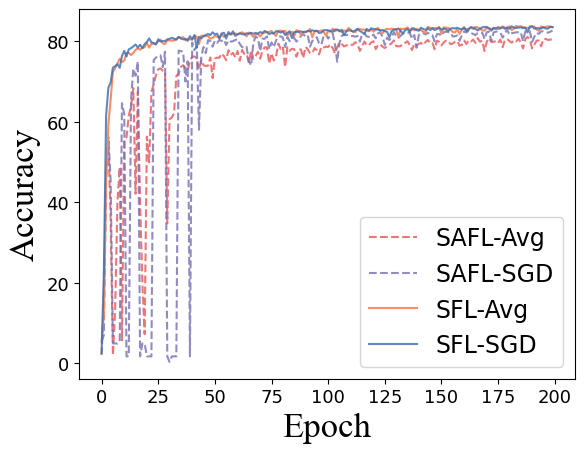}}
\hspace{1mm}
\subfloat[$\sigma = 1$]{\includegraphics[width=1.5in]{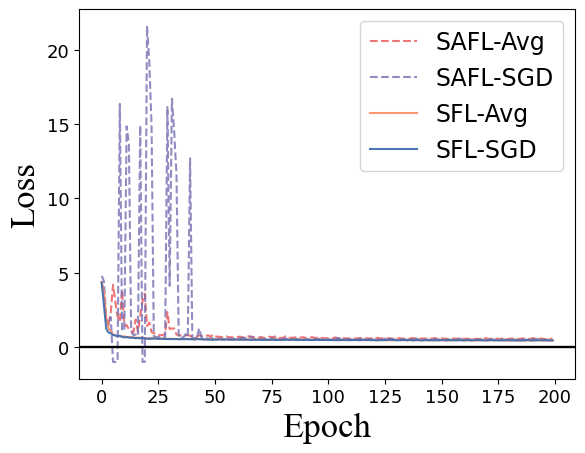}}
\caption{Global accuracy and loss of different models under FEMNIST dataset using ResNet-18 in SAFL. Note that -1 denotes the NAN value for loss. }
\vspace{-2ex}
\end{figure}
\begin{figure}[!h]
\vspace{-2ex}
\centering
\subfloat[$\alpha = 0.1$]{\includegraphics[width=1.5in]{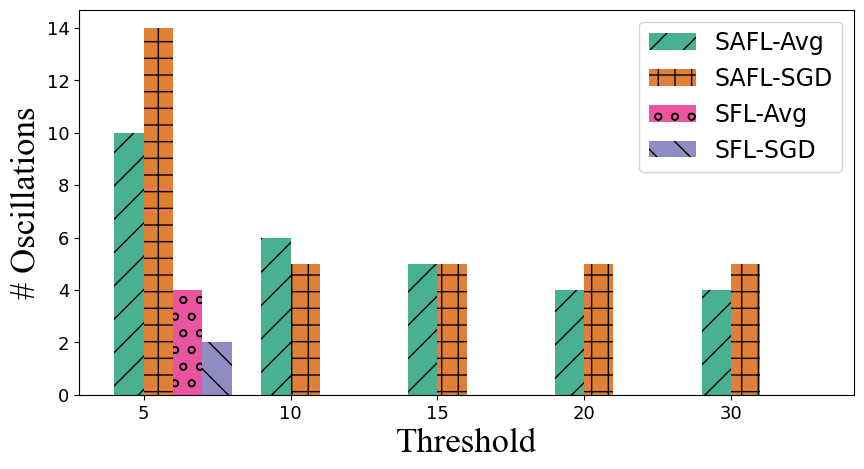}%
}
\hfil
\subfloat[$\alpha = 0.5$]{\includegraphics[width=1.5in]{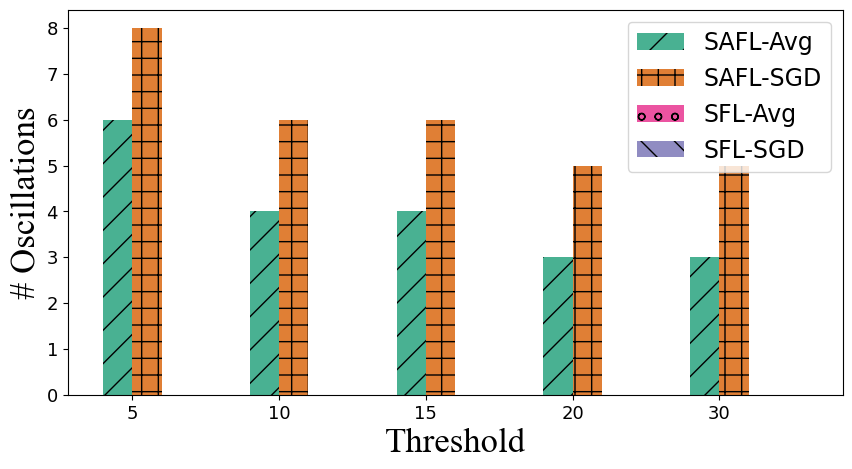}%
}
\hfil
\subfloat[$\alpha = 1$]{\includegraphics[width=1.5in]{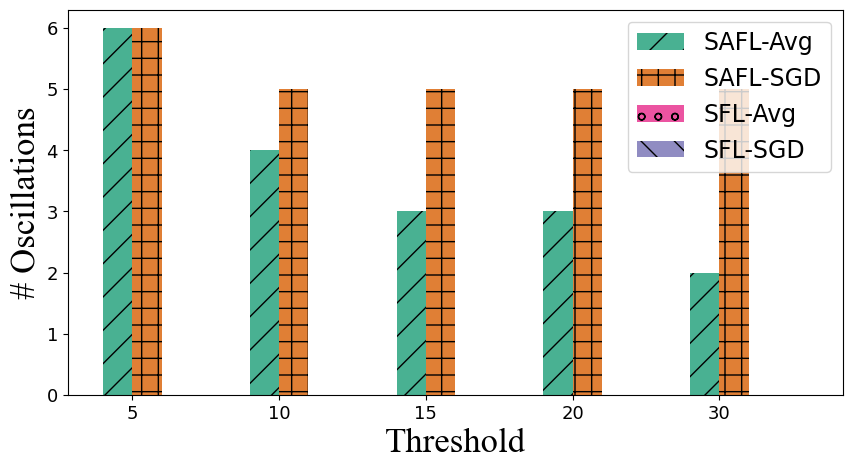}%
}    
% \hfil
% \subfloat[$N = 2$]{\includegraphics[width=1.5in]{pic/ret/resnet18_femnist/shards/ots/2.png}%
% }   
\hfil
\subfloat[$N = 5$]{\includegraphics[width=1.5in]{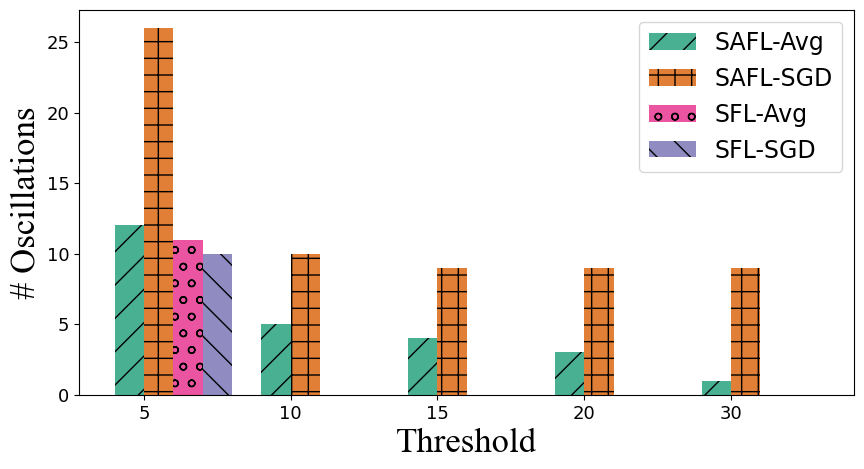}%
}   
\hfil
\subfloat[$N = 10$]{\includegraphics[width=1.5in]{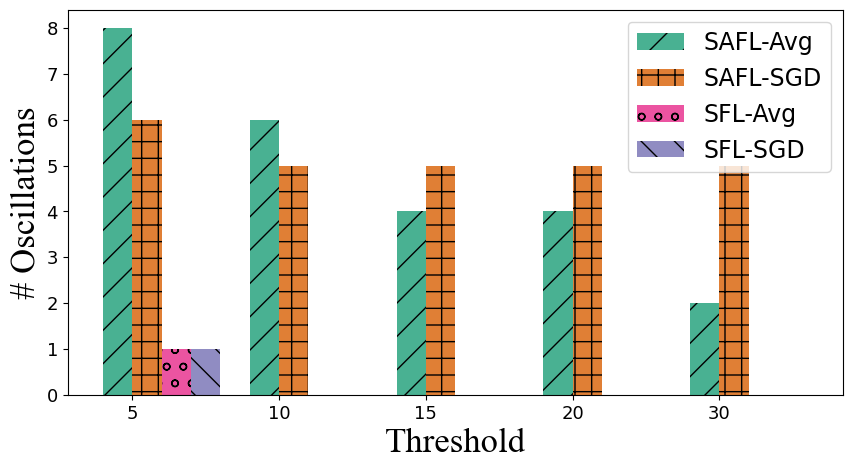}%
}   
\hfil
\subfloat[$\sigma = 0.1$]{\includegraphics[width=1.5in]{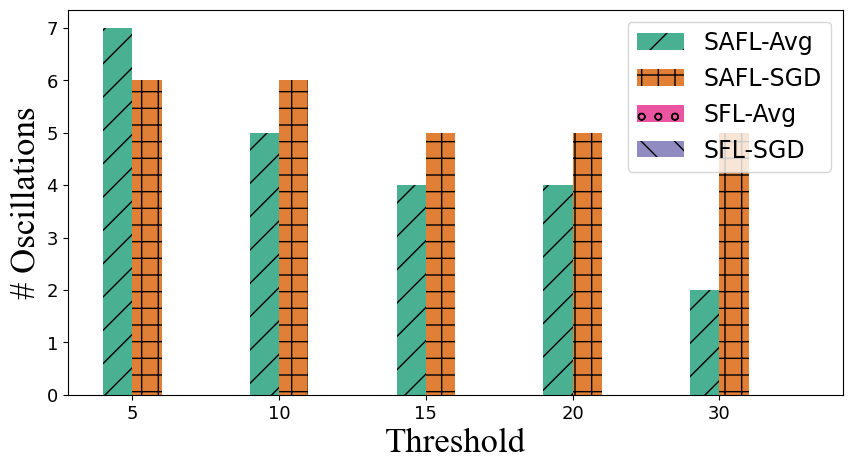}%
}   
\hfil
\subfloat[$\sigma = 0.5$]{\includegraphics[width=1.5in]{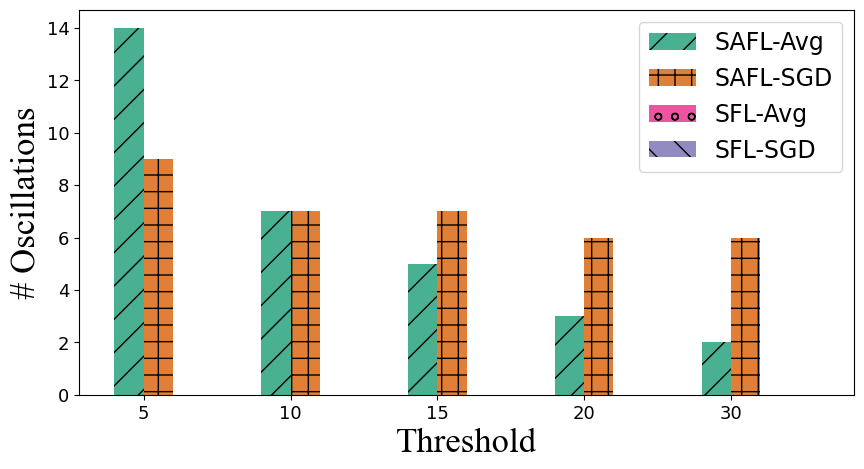}%
}   
\hfil
\subfloat[$\sigma = 1$]{\includegraphics[width=1.5in]{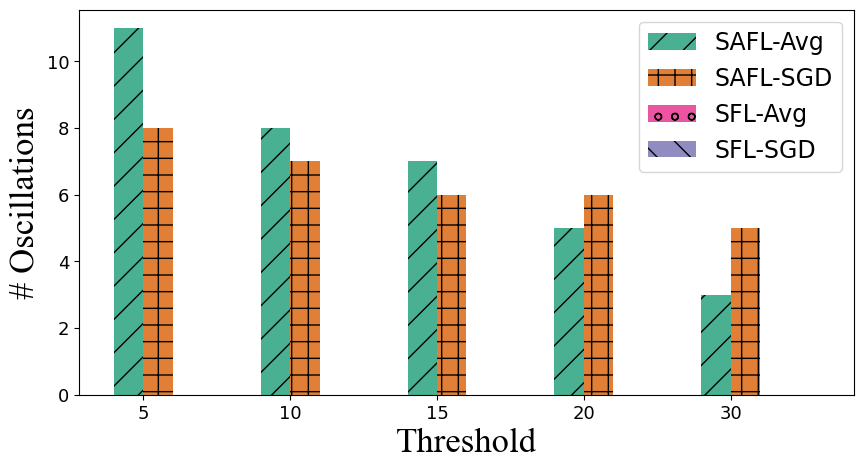}%
}   
\vspace{-1ex}
\caption{Statistics of severe oscillations of the ResNet-18 model under the FEMNIST dataset.}
\end{figure}

\newpage
\subsection{CIFAR-10 @ VGG-16}

\begin{figure}[!h]
\vspace{-2ex}
\centering
\subfloat[$\alpha = 0.1$]{\includegraphics[width=1.5in]{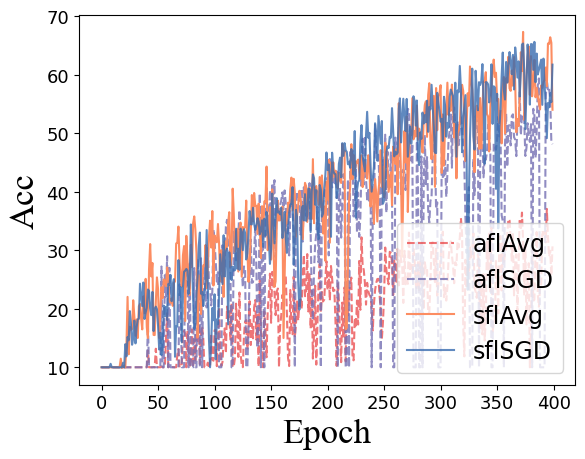}}
\hspace{1mm}
\subfloat[$\alpha = 0.1$]{\includegraphics[width=1.5in]{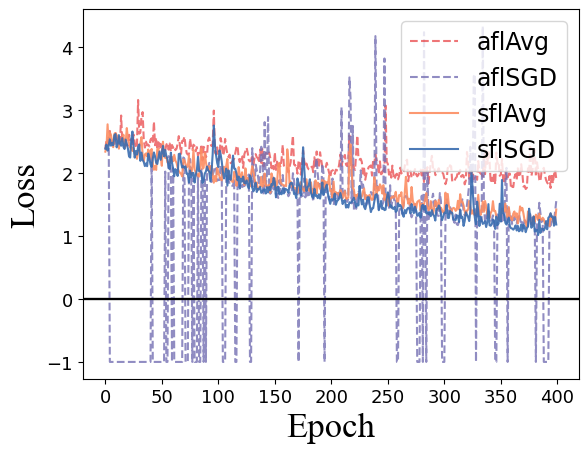}}
\hspace{1mm}
\subfloat[$\alpha = 0.5$]{\includegraphics[width=1.5in]{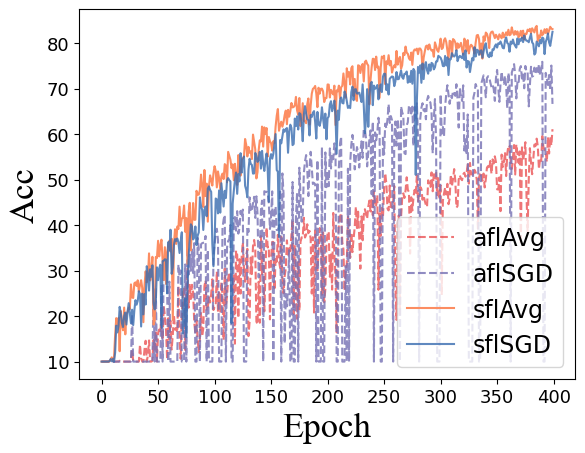}}
\hspace{1mm}
\subfloat[$\alpha = 0.5$]{\includegraphics[width=1.5in]{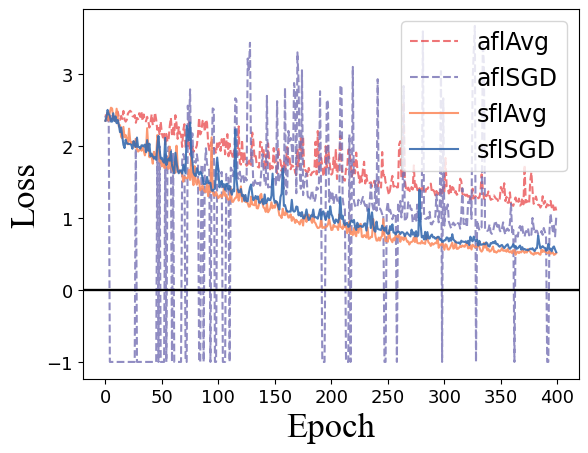}\label{fig_nan_concrete}}
\hspace{1mm}
\subfloat[$\alpha = 1$]{\includegraphics[width=1.5in]{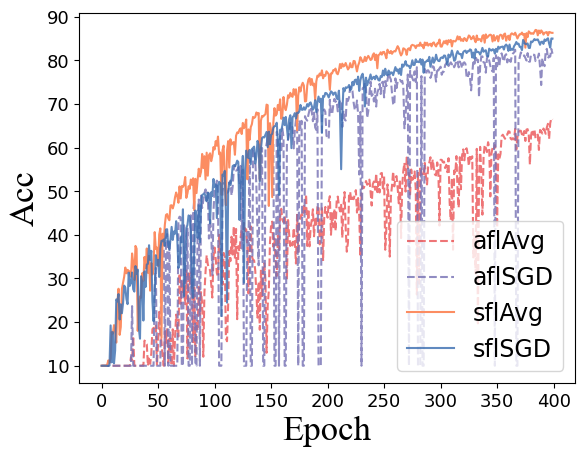}}
\hspace{1mm}
\subfloat[$\alpha = 1$]{\includegraphics[width=1.5in]{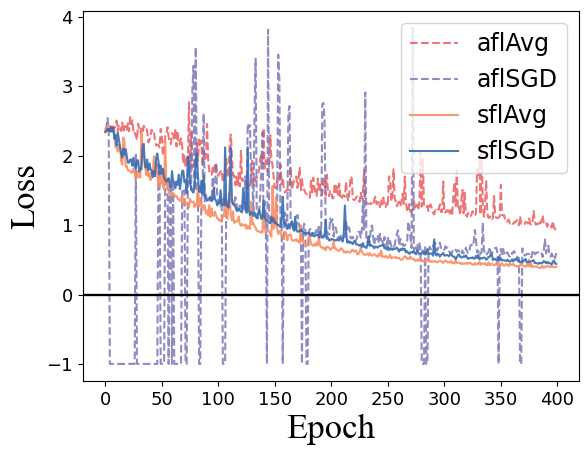}}
\hspace{1mm}
% \subfloat[$N = 2$]{\includegraphics[width=1.5in]{pic/ret/vgg16_cifar10/shards/acc/2.png}}
% \hspace{1mm}
% \subfloat[$N = 2$]{\includegraphics[width=1.5in]{pic/ret/vgg16_cifar10/shards/loss/2.png}}
% \hspace{1mm}
\subfloat[$N = 5$]{\includegraphics[width=1.5in]{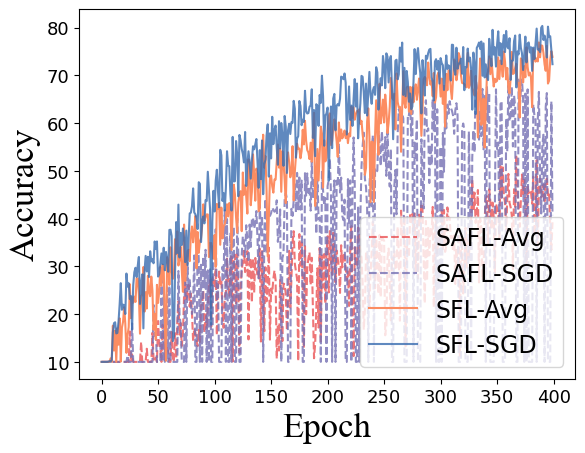}}
\hspace{1mm}
\subfloat[$N = 5$]{\includegraphics[width=1.5in]{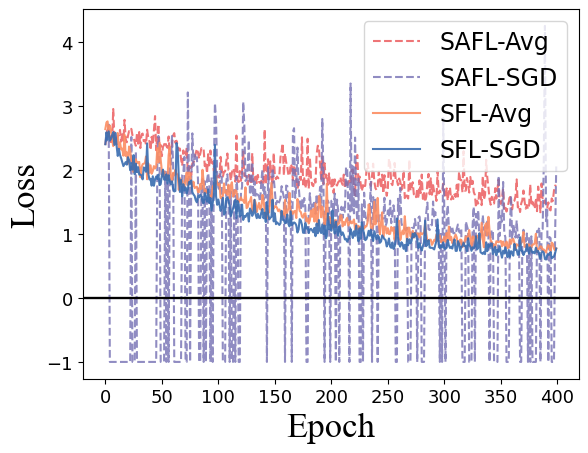}}
\hspace{1mm}
\subfloat[$N = 10$]{\includegraphics[width=1.5in]{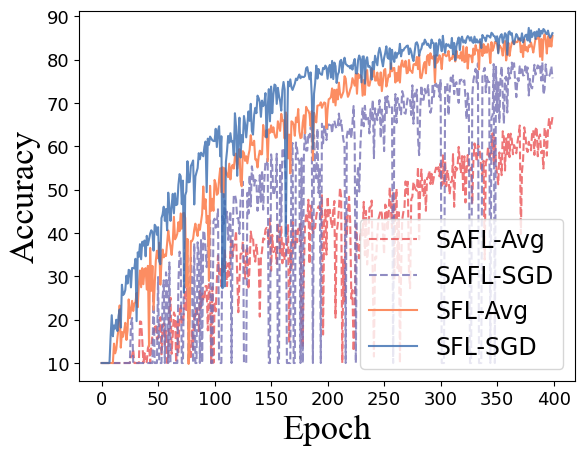}}
\hspace{1mm}
\subfloat[$N = 10$]{\includegraphics[width=1.5in]{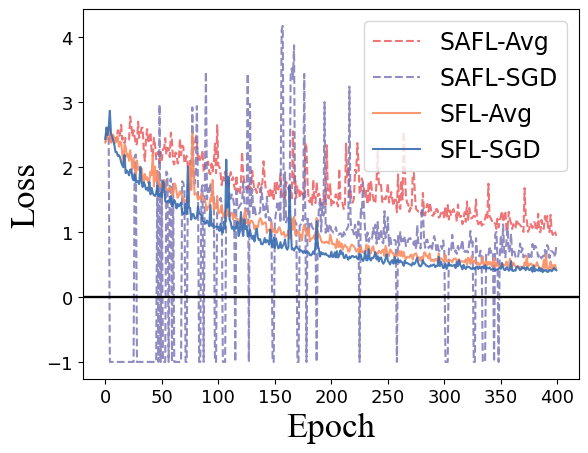}}
\hspace{1mm}
\subfloat[$\sigma = 0.1$]{\includegraphics[width=1.5in]{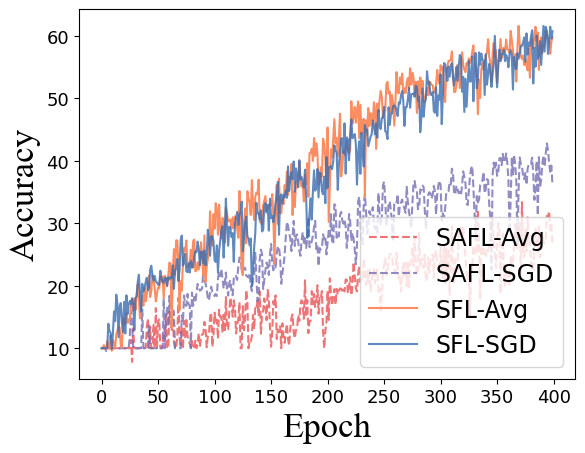}}
\hspace{1mm}
\subfloat[$\sigma = 0.1$]{\includegraphics[width=1.5in]{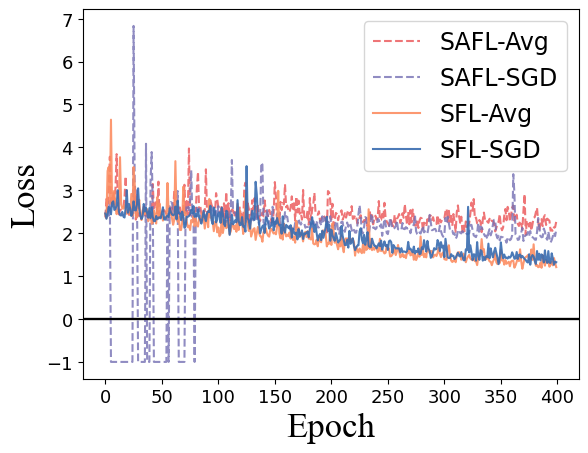}}
\hspace{1mm}
\subfloat[$\sigma = 0.5$]{\includegraphics[width=1.5in]{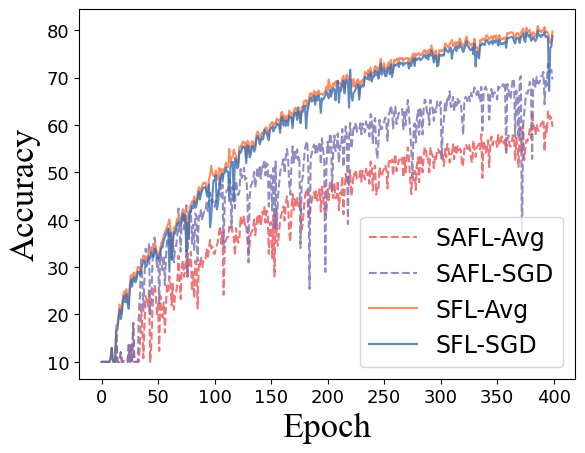}}
\hspace{1mm}
\subfloat[$\sigma = 0.5$]{\includegraphics[width=1.5in]{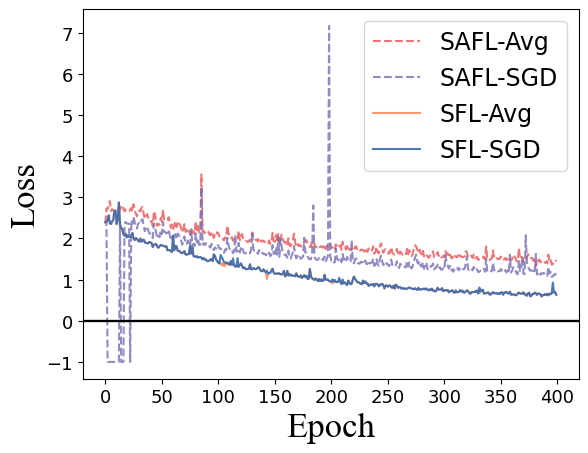}}
\hspace{1mm}
\subfloat[$\sigma = 1$]{\includegraphics[width=1.5in]{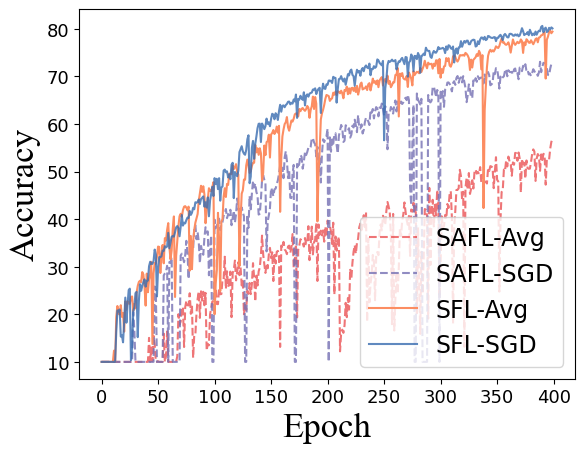}}
\hspace{1mm}
\subfloat[$\sigma = 1$]{\includegraphics[width=1.5in]{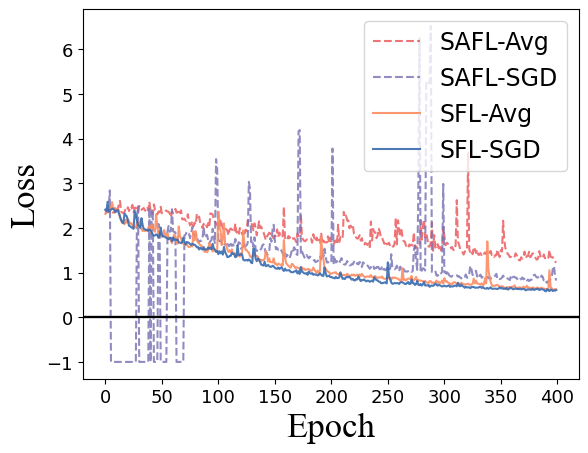}}
\caption{Global accuracy and loss of different models under CIFAR-10 dataset using VGG-16 in SAFL. Note that -1 denotes the NAN value for loss. }
\vspace{-2ex}
\end{figure}
\begin{figure}[!h]
\vspace{-2ex}
\centering
\subfloat[$\alpha = 0.1$]{\includegraphics[width=1.5in]{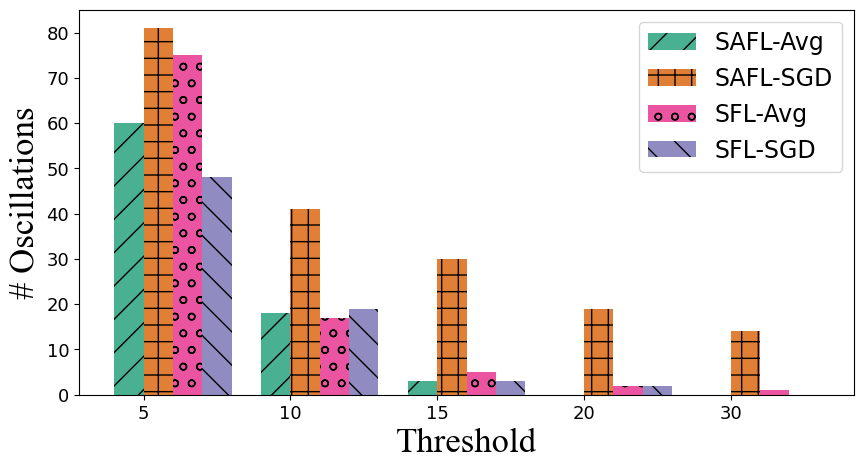}%
}
\hfil
\subfloat[$\alpha = 0.5$]{\includegraphics[width=1.5in]{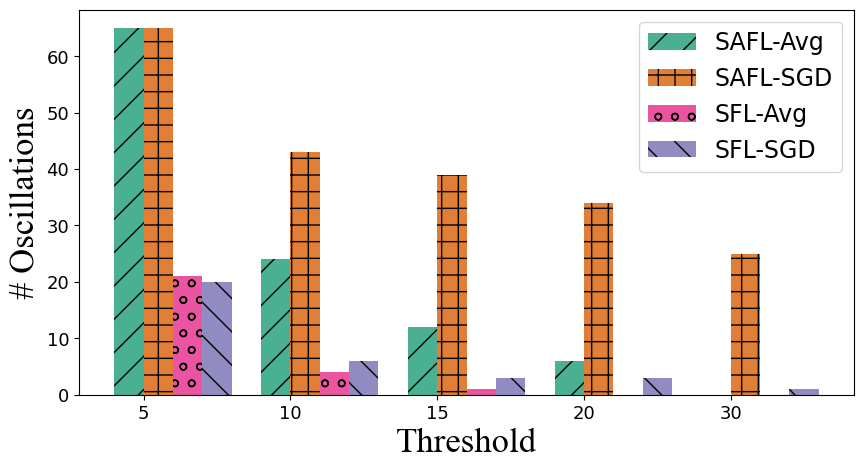}%
}
\hfil
\subfloat[$\alpha = 1$]{\includegraphics[width=1.5in]{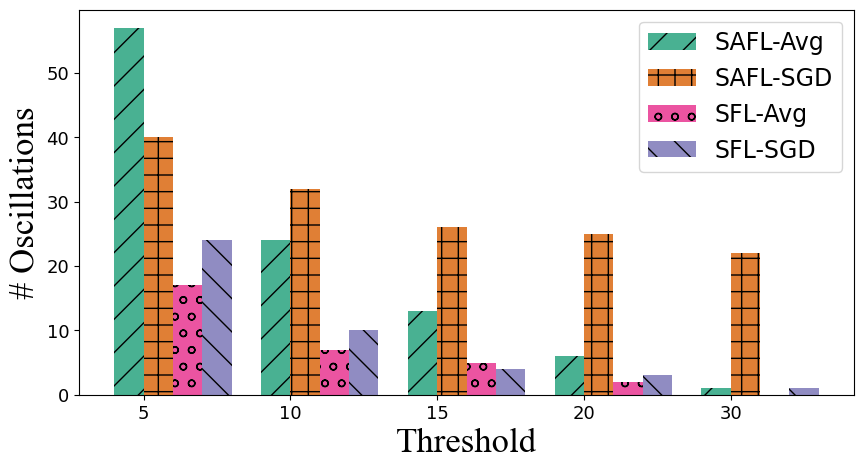}%
}    
% \hfil
% \subfloat[$N = 2$]{\includegraphics[width=1.5in]{pic/ret/vgg16_cifar10/shards/ots/2.png}%
% }   
\hfil
\subfloat[$N = 5$]{\includegraphics[width=1.5in]{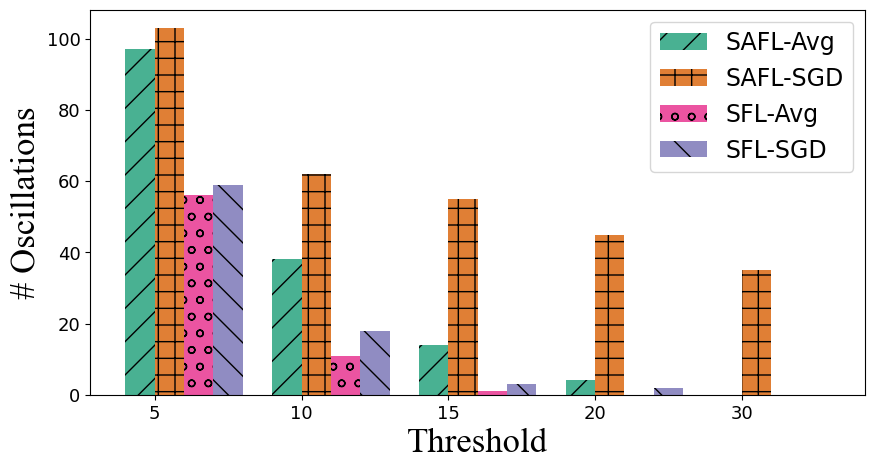}%
}   
\hfil
\subfloat[$N = 10$]{\includegraphics[width=1.5in]{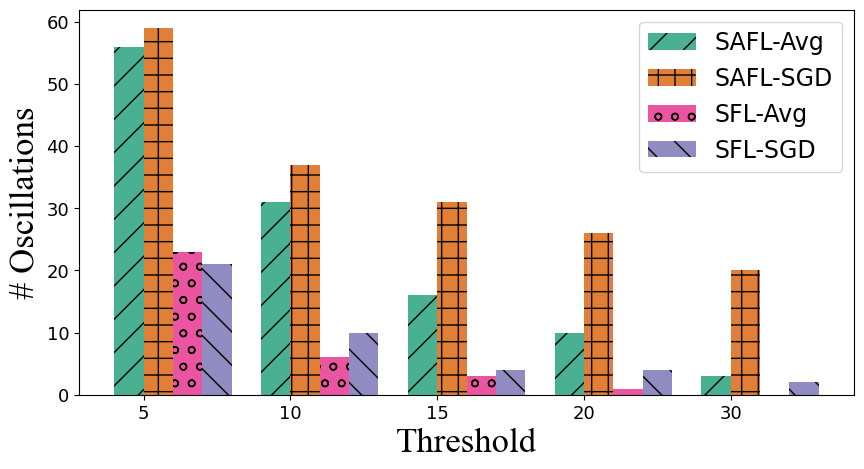}%
}   
\hfil
\subfloat[$\sigma = 0.1$]{\includegraphics[width=1.5in]{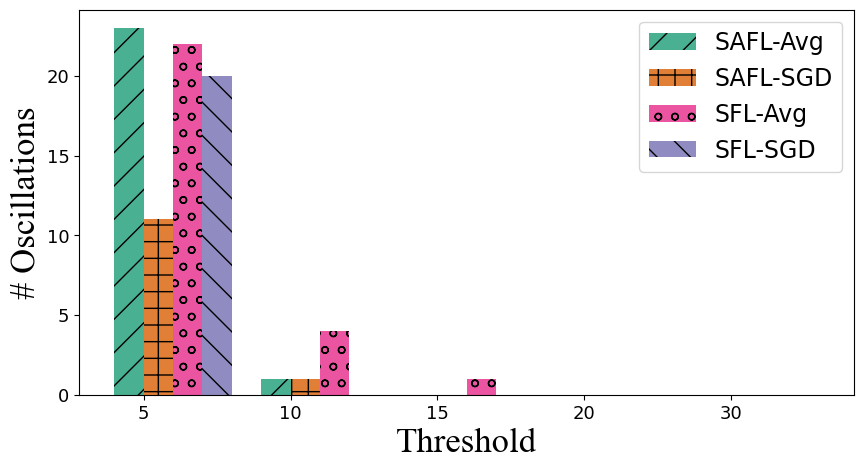}%
}   
\hfil
\subfloat[$\sigma = 0.5$]{\includegraphics[width=1.5in]{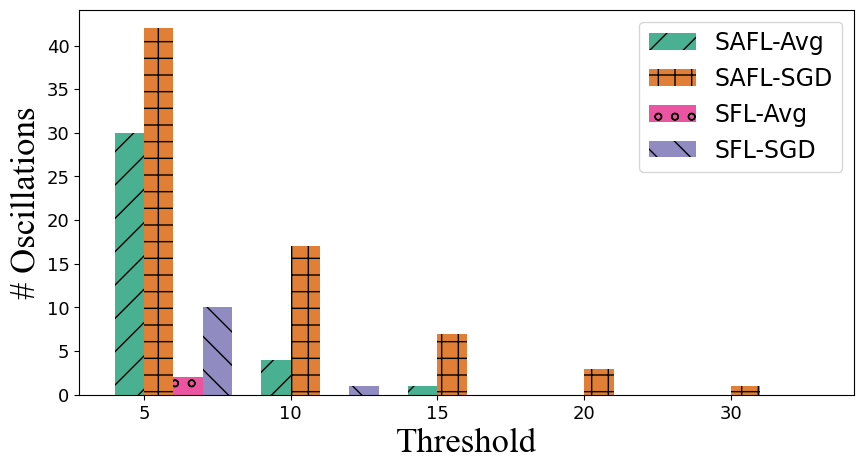}%
}   
\hfil
\subfloat[$\sigma = 1$]{\includegraphics[width=1.5in]{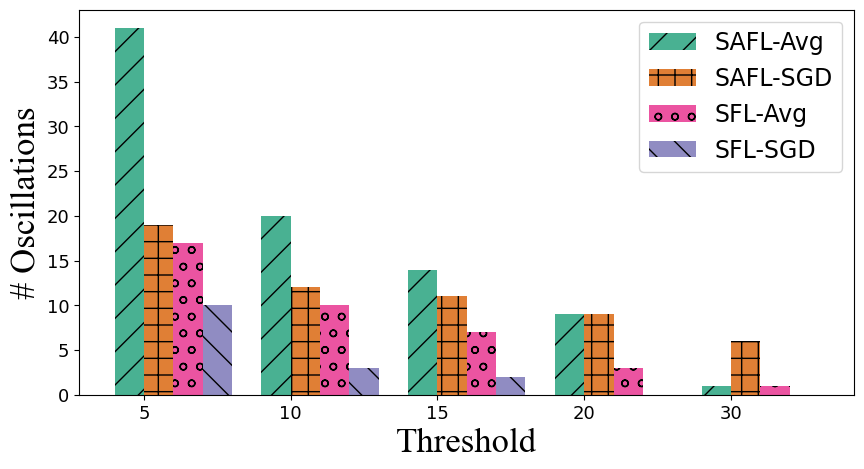}%
}   
\vspace{-1ex}
\caption{Statistics of severe oscillations of the VGG-16 model under the CIFAR-10 dataset.}
\end{figure}

\newpage
\subsection{CIFAR-100 @ VGG-16}

\begin{figure}[!h]
\vspace{-2ex}
\centering
\subfloat[$\alpha = 0.1$]{\includegraphics[width=1.5in]{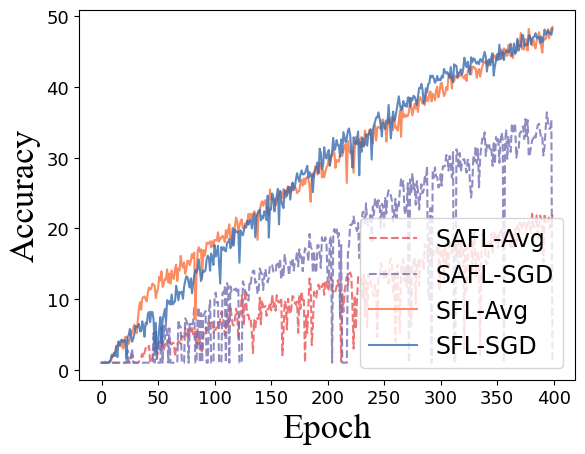}}
\hspace{1mm}
\subfloat[$\alpha = 0.1$]{\includegraphics[width=1.5in]{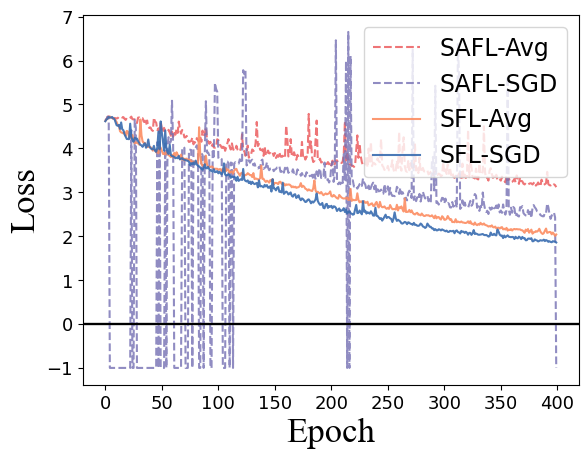}}
\hspace{1mm}
\subfloat[$\alpha = 0.5$]{\includegraphics[width=1.5in]{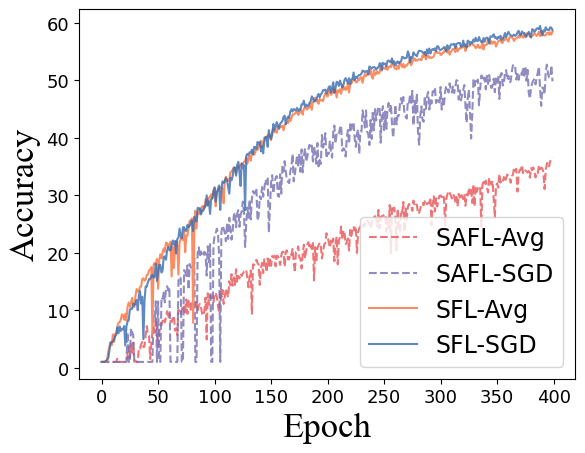}}
\hspace{1mm}
\subfloat[$\alpha = 0.5$]{\includegraphics[width=1.5in]{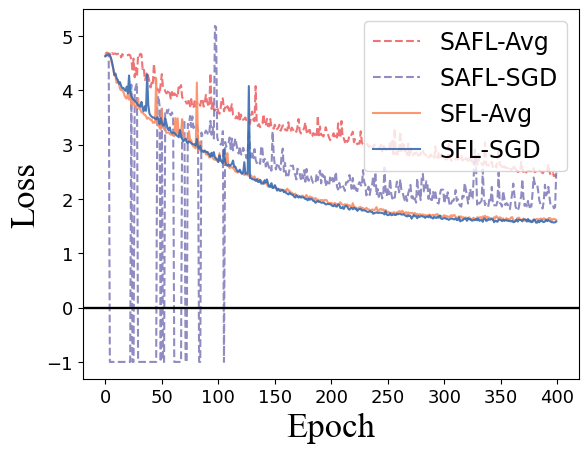}\label{fig_nan_concrete}}
\hspace{1mm}
\subfloat[$\alpha = 1$]{\includegraphics[width=1.5in]{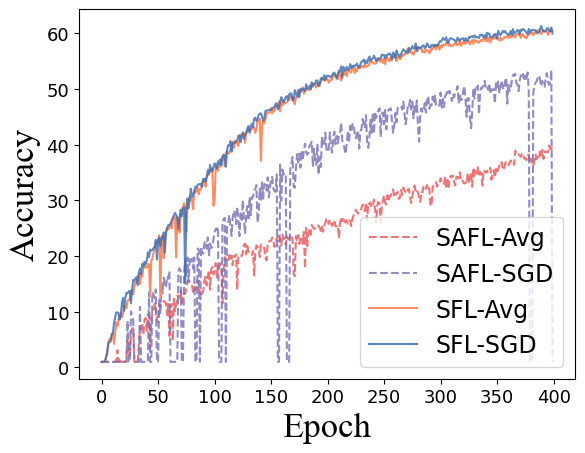}}
\hspace{1mm}
\subfloat[$\alpha = 1$]{\includegraphics[width=1.5in]{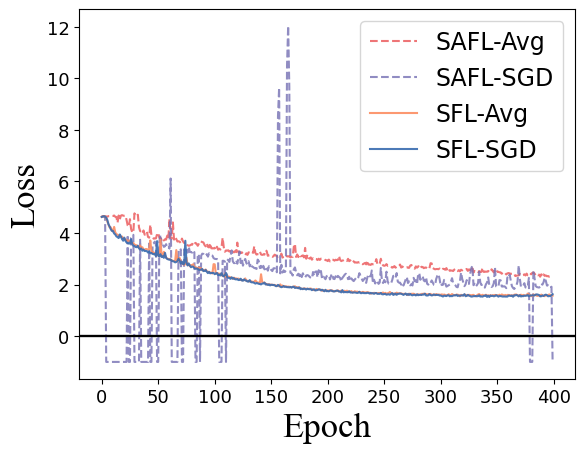}}
\hspace{1mm}
% \subfloat[$N = 2$]{\includegraphics[width=1.5in]{pic/ret/vgg16_cifar100/shards/acc/2.png}}
% \hspace{1mm}
% \subfloat[$N = 2$]{\includegraphics[width=1.5in]{pic/ret/vgg16_cifar100/shards/loss/2.png}}
% \hspace{1mm}
\subfloat[$N = 5$]{\includegraphics[width=1.5in]{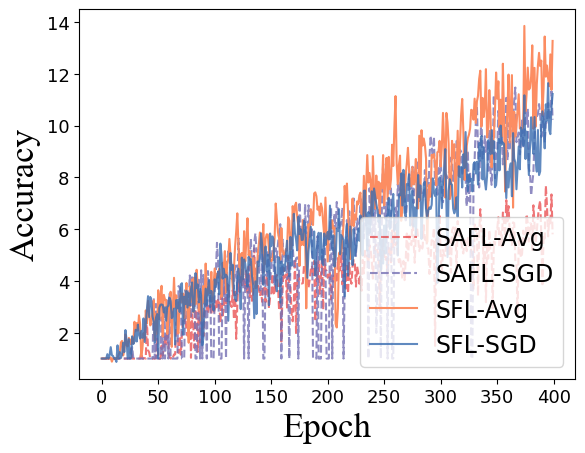}}
\hspace{1mm}
\subfloat[$N = 5$]{\includegraphics[width=1.5in]{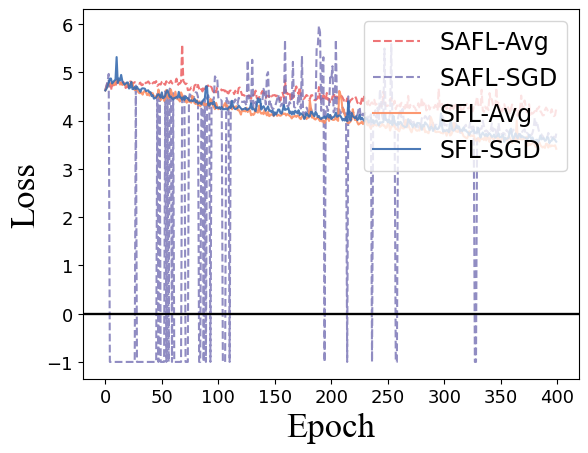}}
\hspace{1mm}
\subfloat[$N = 10$]{\includegraphics[width=1.5in]{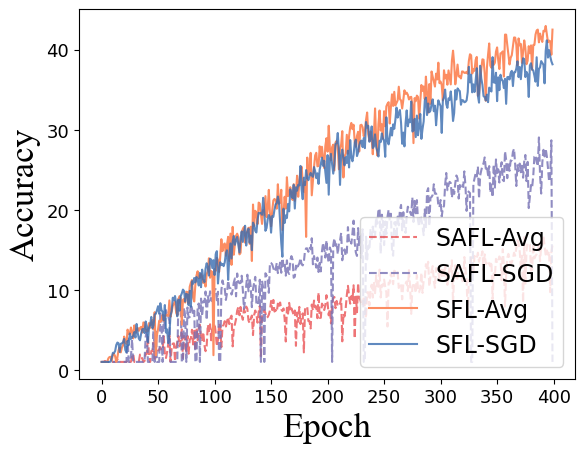}}
\hspace{1mm}
\subfloat[$N = 10$]{\includegraphics[width=1.5in]{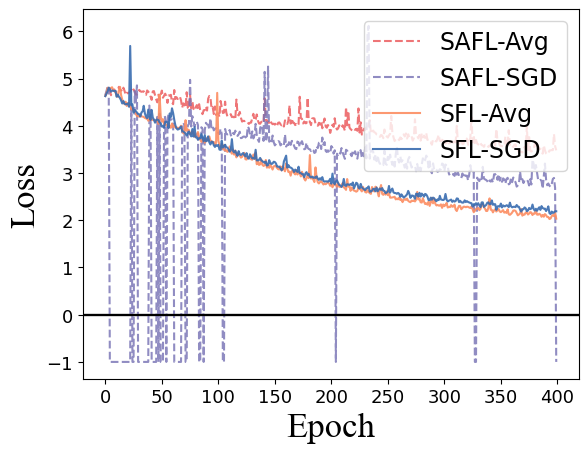}}
\hspace{1mm}
\subfloat[$\sigma = 0.1$]{\includegraphics[width=1.5in]{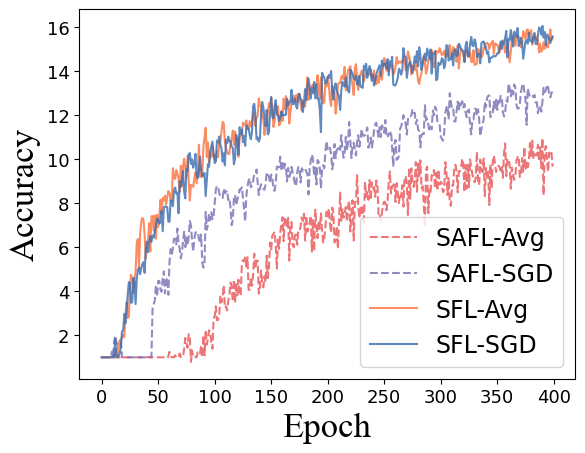}}
\hspace{1mm}
\subfloat[$\sigma = 0.1$]{\includegraphics[width=1.5in]{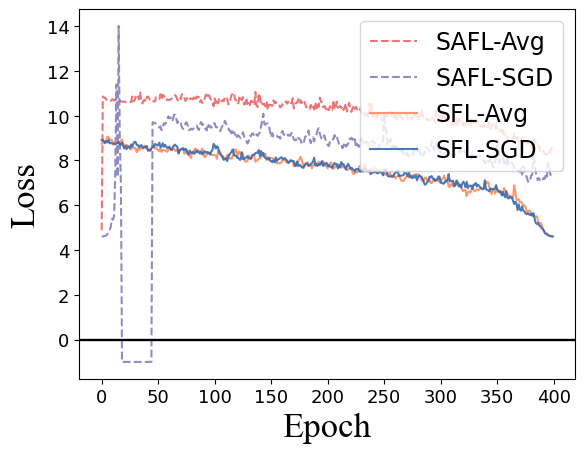}}
\hspace{1mm}
\subfloat[$\sigma = 0.5$]{\includegraphics[width=1.5in]{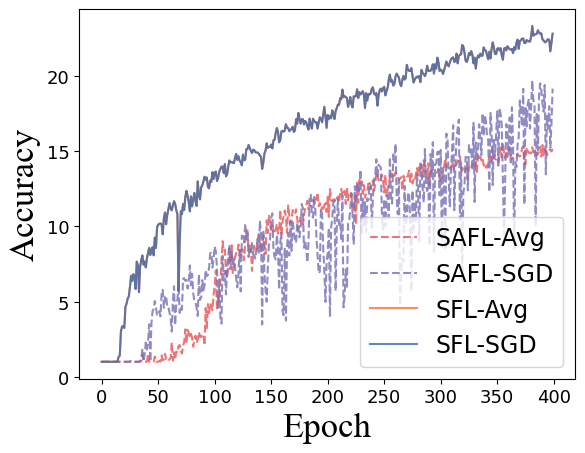}}
\hspace{1mm}
\subfloat[$\sigma = 0.5$]{\includegraphics[width=1.5in]{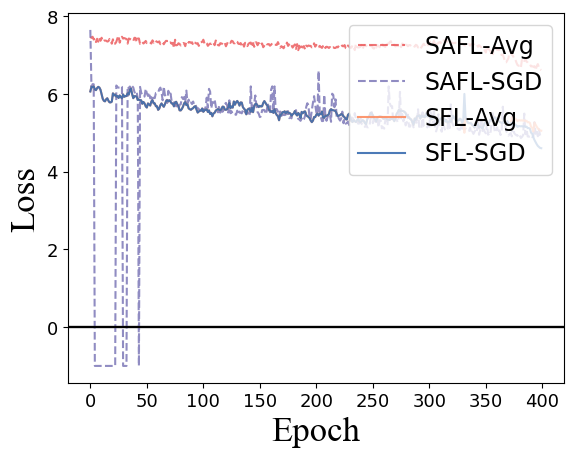}}
\hspace{1mm}
\subfloat[$\sigma = 1$]{\includegraphics[width=1.5in]{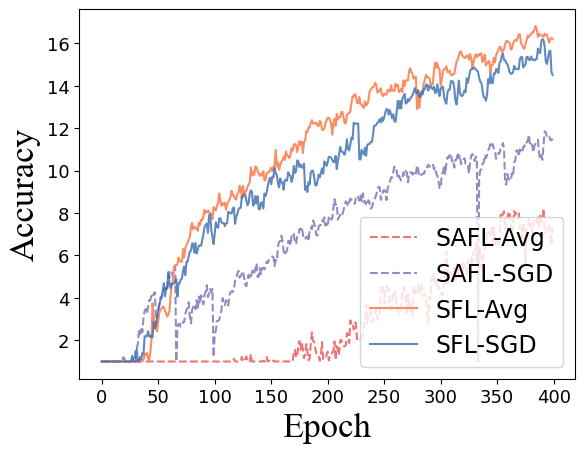}}
\hspace{1mm}
\subfloat[$\sigma = 1$]{\includegraphics[width=1.5in]{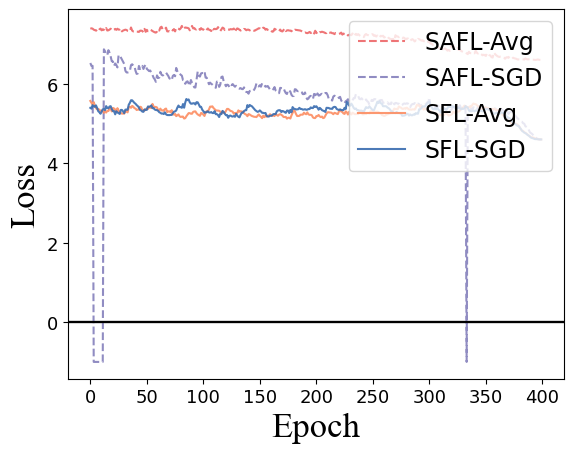}}
\caption{Global accuracy and loss of different models under CIFAR-100 dataset using VGG-16 in SAFL. Note that -1 denotes the NAN value for loss. }
\vspace{-2ex}
\end{figure}
\begin{figure}[!h]
\vspace{-2ex}
\centering
\subfloat[$\alpha = 0.1$]{\includegraphics[width=1.5in]{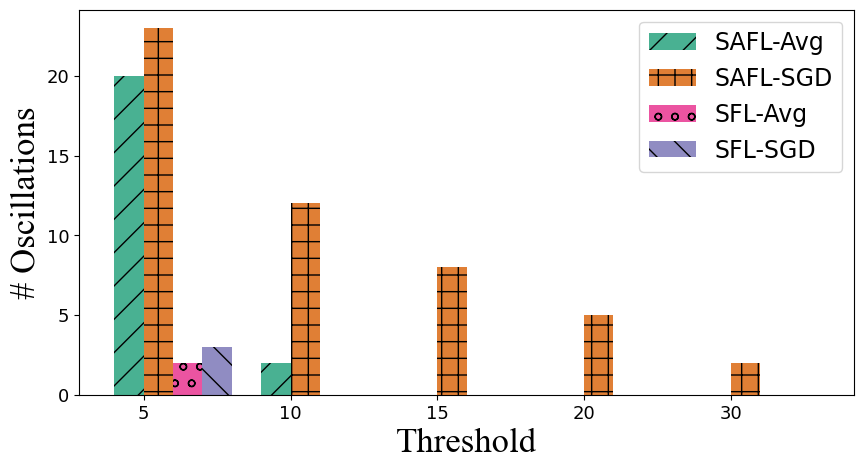}%
}
\hfil
\subfloat[$\alpha = 0.5$]{\includegraphics[width=1.5in]{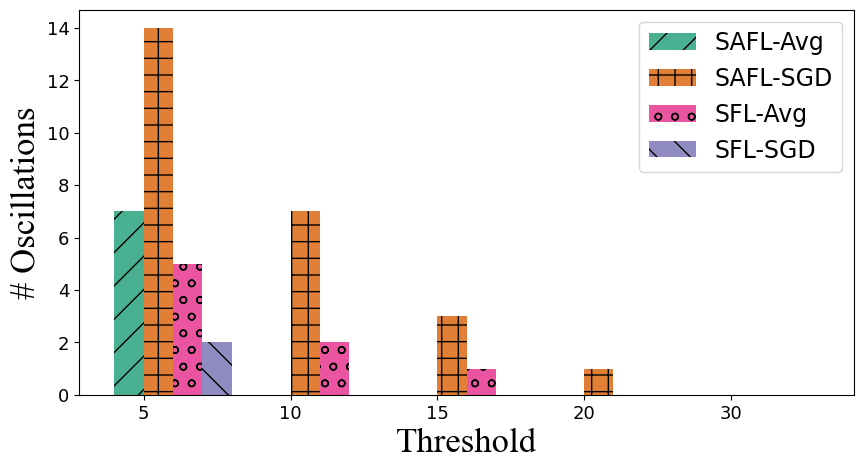}%
}
\hfil
\subfloat[$\alpha = 1$]{\includegraphics[width=1.5in]{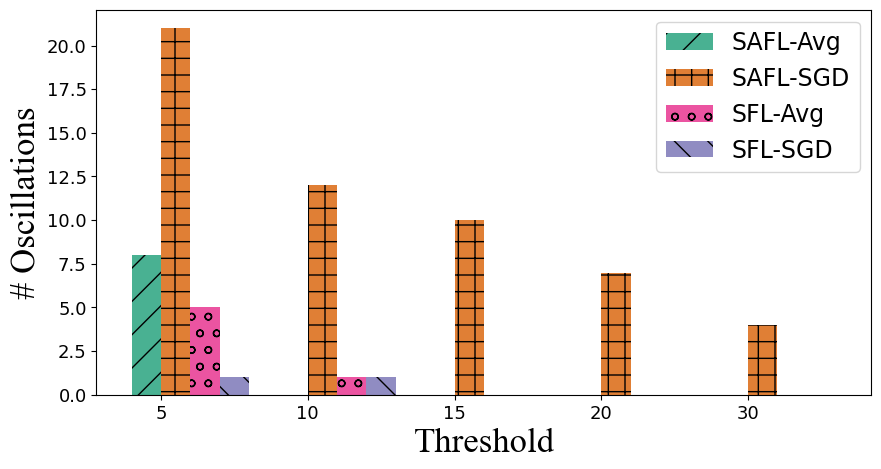}%
}    
% \hfil
% \subfloat[$N = 2$]{\includegraphics[width=1.5in]{pic/ret/vgg16_cifar100/shards/ots/2.png}%
% }   
\hfil
\subfloat[$N = 5$]{\includegraphics[width=1.5in]{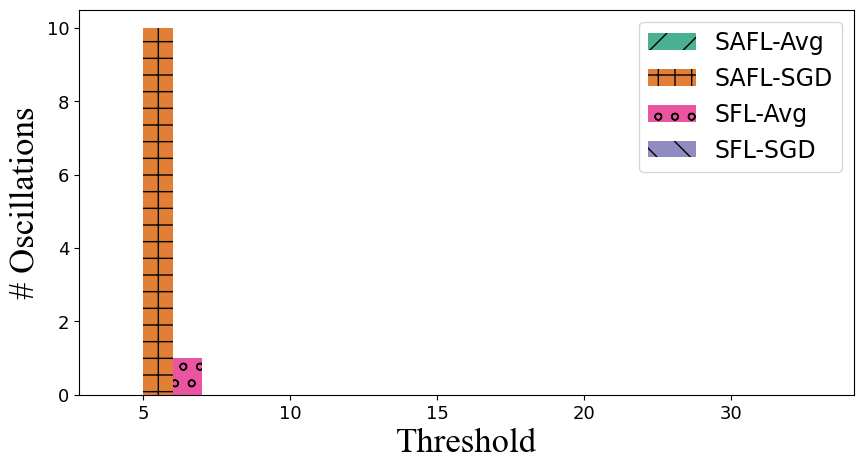}%
}   
\hfil
\subfloat[$N = 10$]{\includegraphics[width=1.5in]{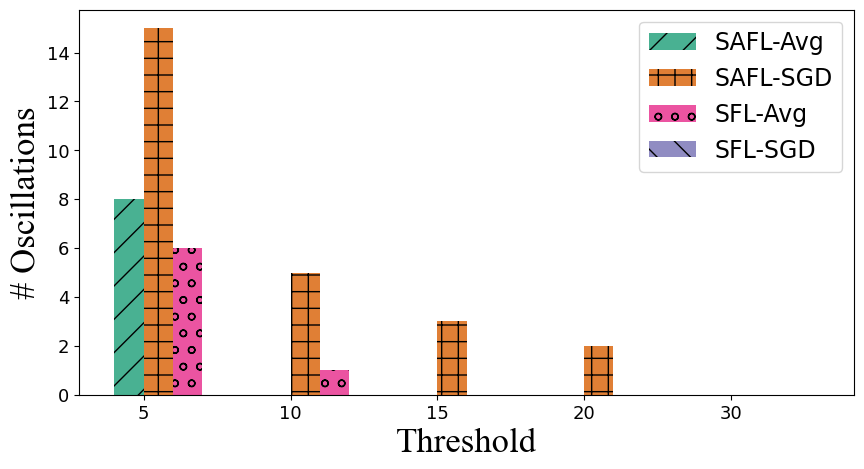}%
}   
\hfil
\subfloat[$\sigma = 0.1$]{\includegraphics[width=1.5in]{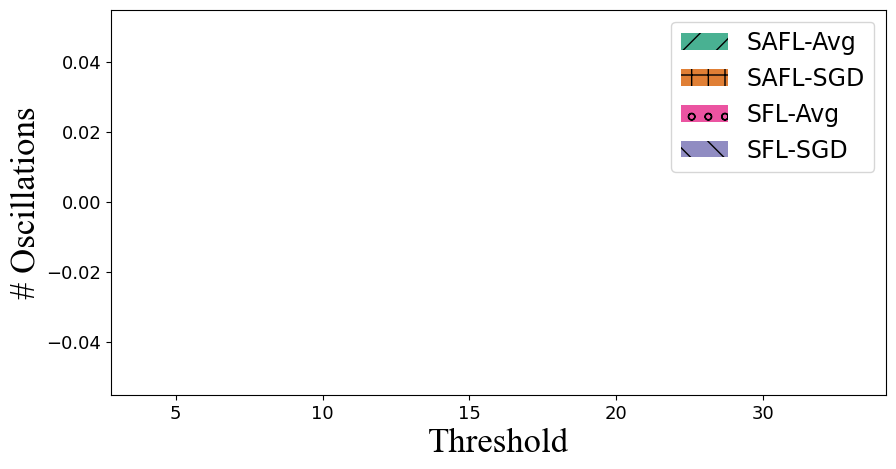}%
}   
\hfil
\subfloat[$\sigma = 0.5$]{\includegraphics[width=1.5in]{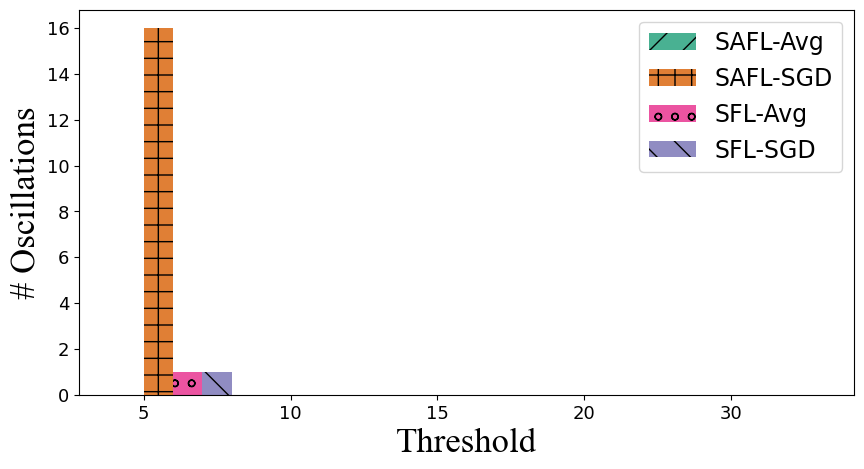}%
}   
\hfil
\subfloat[$\sigma = 1$]{\includegraphics[width=1.5in]{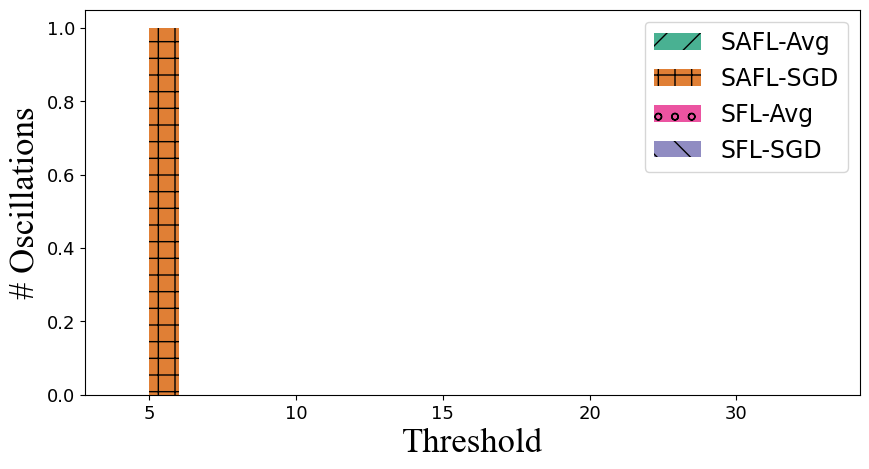}%
}   
\vspace{-1ex}
\caption{Statistics of severe oscillations of the VGG-16 model under the CIFAR-100 dataset.}
\end{figure}

\newpage
\subsection{FEMNIST @ VGG-16}

\begin{figure}[!h]
\vspace{-2ex}
\centering
\subfloat[$\alpha = 0.1$]{\includegraphics[width=1.5in]{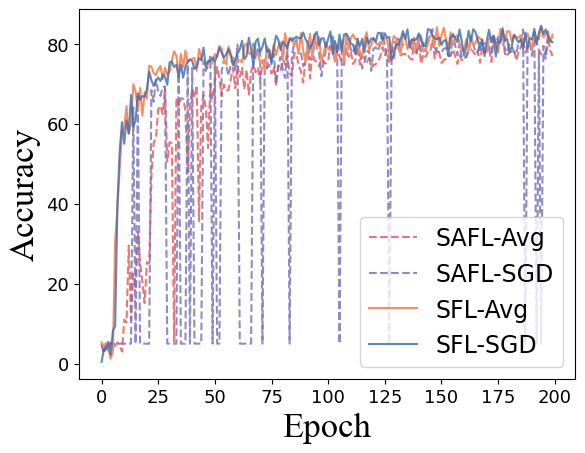}}
\hspace{1mm}
\subfloat[$\alpha = 0.1$]{\includegraphics[width=1.5in]{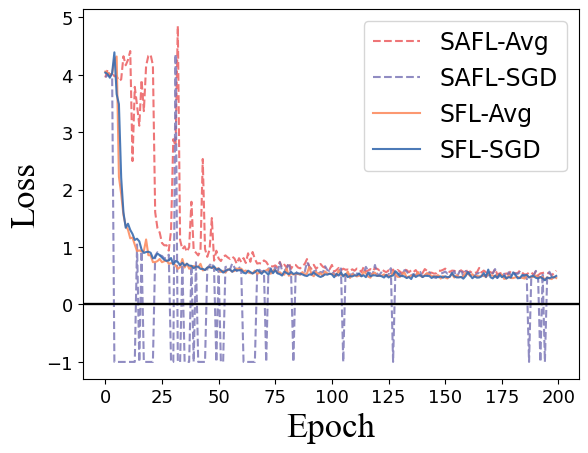}}
\hspace{1mm}
\subfloat[$\alpha = 0.5$]{\includegraphics[width=1.5in]{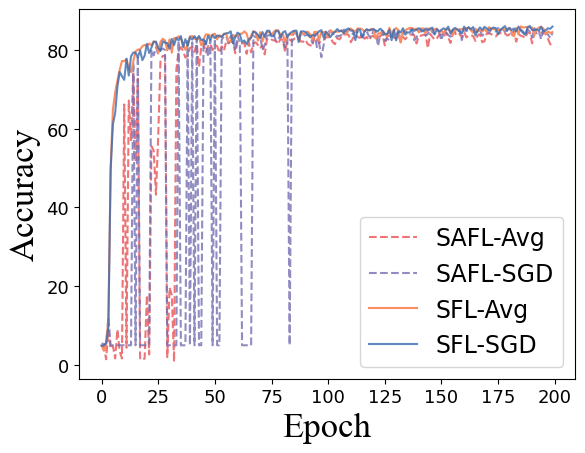}}
\hspace{1mm}
\subfloat[$\alpha = 0.5$]{\includegraphics[width=1.5in]{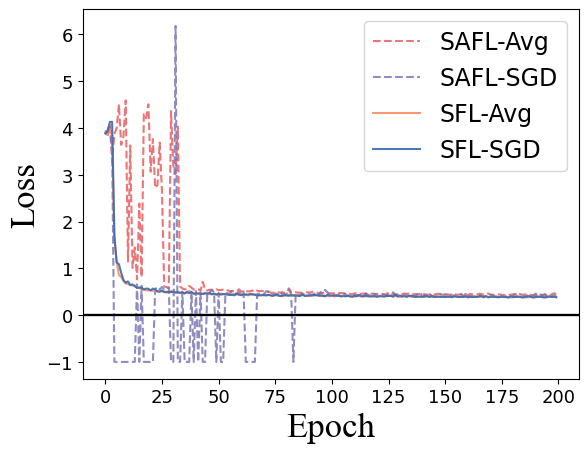}\label{fig_nan_concrete}}
\hspace{1mm}
\subfloat[$\alpha = 1$]{\includegraphics[width=1.5in]{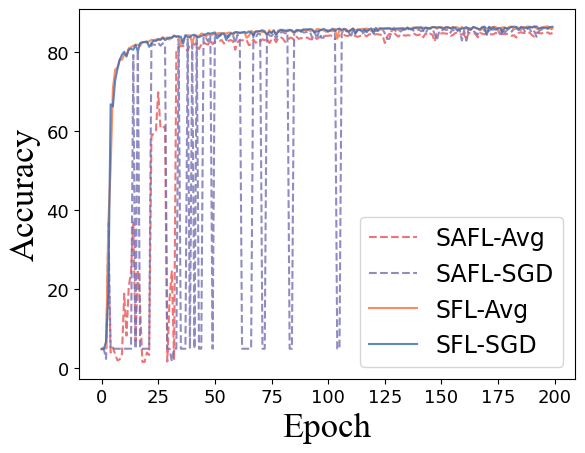}}
\hspace{1mm}
\subfloat[$\alpha = 1$]{\includegraphics[width=1.5in]{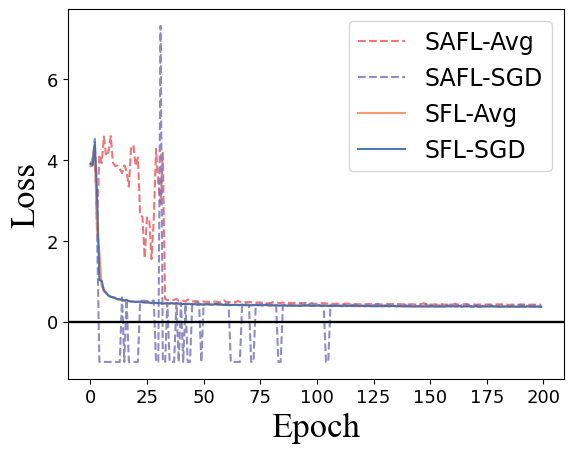}}
\hspace{1mm}
% \subfloat[$N = 2$]{\includegraphics[width=1.5in]{pic/ret/vgg16_femnist/shards/acc/2.png}}
% \hspace{1mm}
% \subfloat[$N = 2$]{\includegraphics[width=1.5in]{pic/ret/vgg16_femnist/shards/loss/2.png}}
% \hspace{1mm}
\subfloat[$N = 5$]{\includegraphics[width=1.5in]{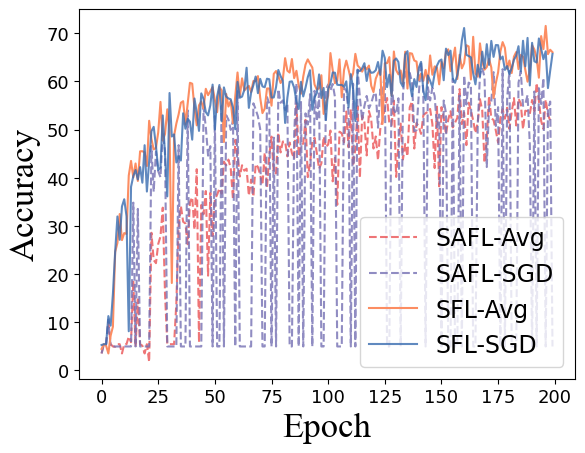}}
\hspace{1mm}
\subfloat[$N = 5$]{\includegraphics[width=1.5in]{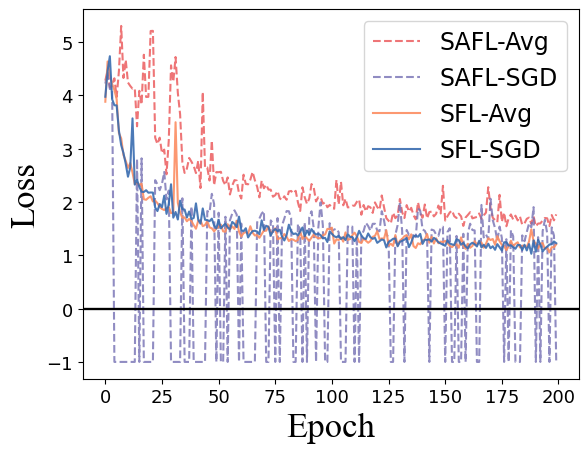}}
\hspace{1mm}
\subfloat[$N = 10$]{\includegraphics[width=1.5in]{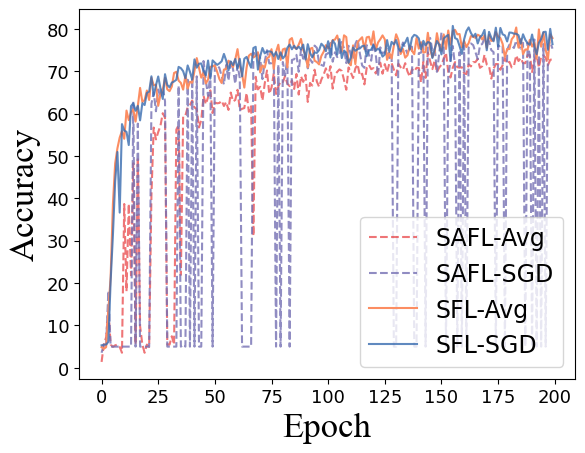}}
\hspace{1mm}
\subfloat[$N = 10$]{\includegraphics[width=1.5in]{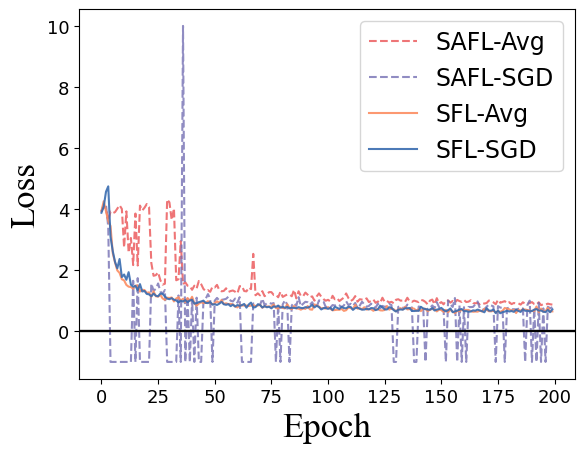}}
\hspace{1mm}
\subfloat[$\sigma = 0.1$]{\includegraphics[width=1.5in]{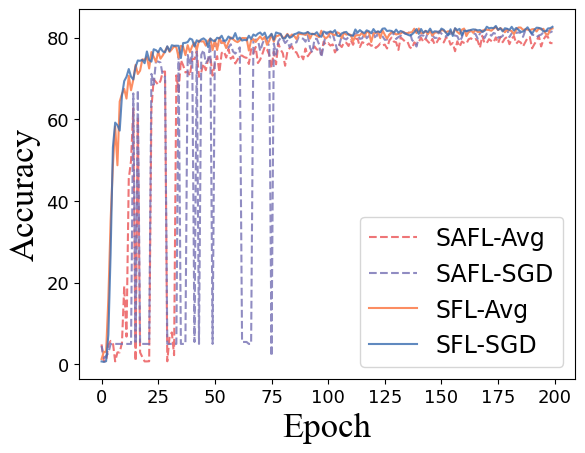}}
\hspace{1mm}
\subfloat[$\sigma = 0.1$]{\includegraphics[width=1.5in]{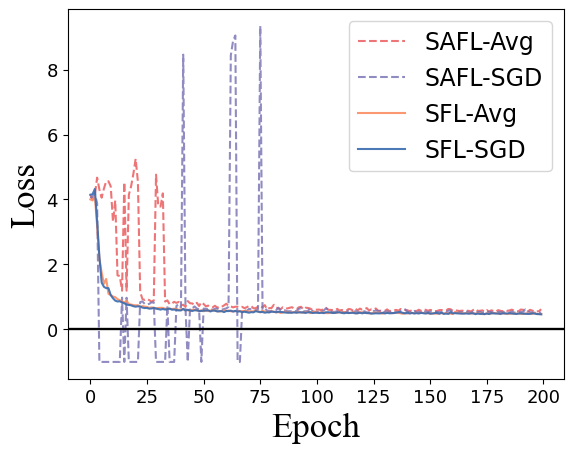}}
\hspace{1mm}
\subfloat[$\sigma = 0.5$]{\includegraphics[width=1.5in]{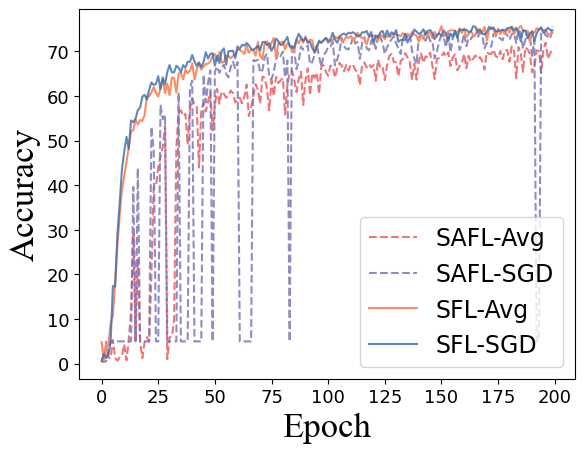}}
\hspace{1mm}
\subfloat[$\sigma = 0.5$]{\includegraphics[width=1.5in]{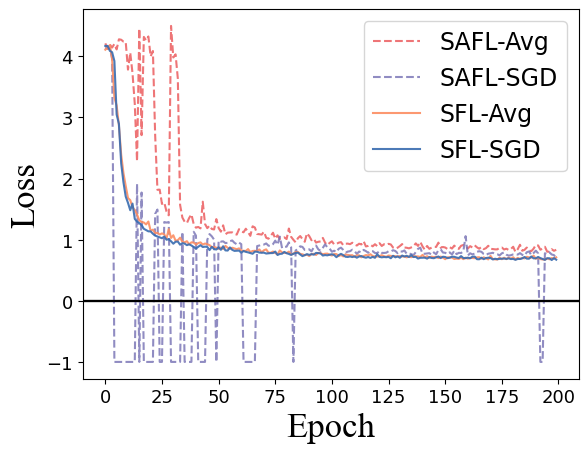}}
\hspace{1mm}
\subfloat[$\sigma = 1$]{\includegraphics[width=1.5in]{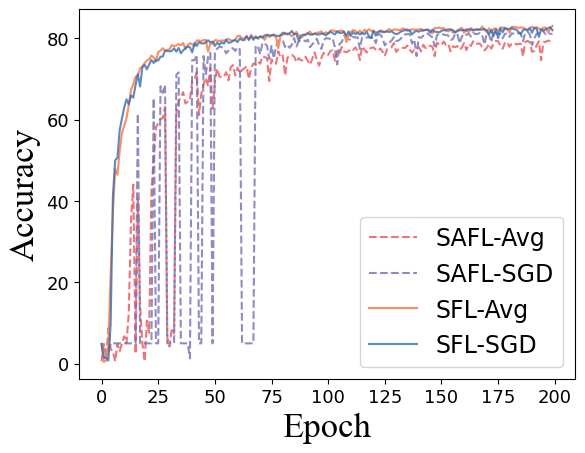}}
\hspace{1mm}
\subfloat[$\sigma = 1$]{\includegraphics[width=1.5in]{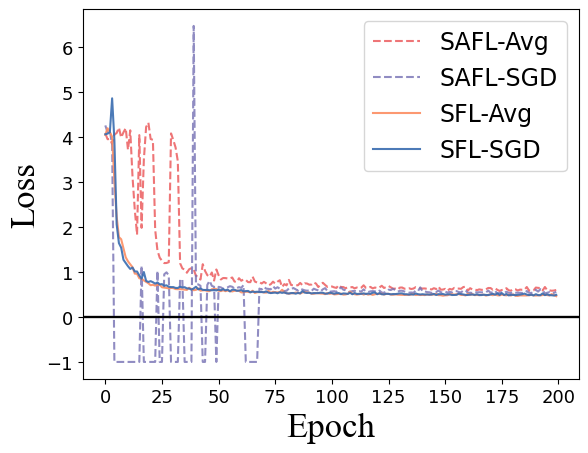}}
\caption{Global accuracy and loss of different models under FEMNIST dataset using VGG-16 in SAFL. Note that -1 denotes the NAN value for loss. }
\vspace{-2ex}
\end{figure}
\begin{figure}[!h]
\vspace{-2ex}
\centering
\subfloat[$\alpha = 0.1$]{\includegraphics[width=1.5in]{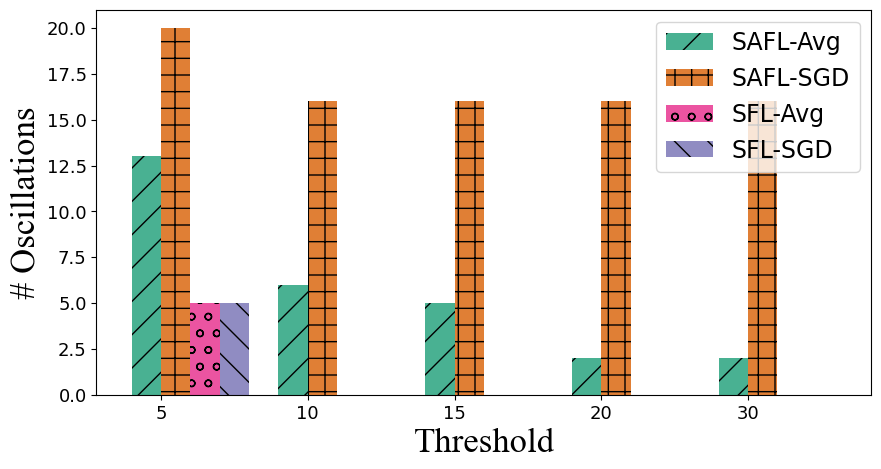}%
}
\hfil
\subfloat[$\alpha = 0.5$]{\includegraphics[width=1.5in]{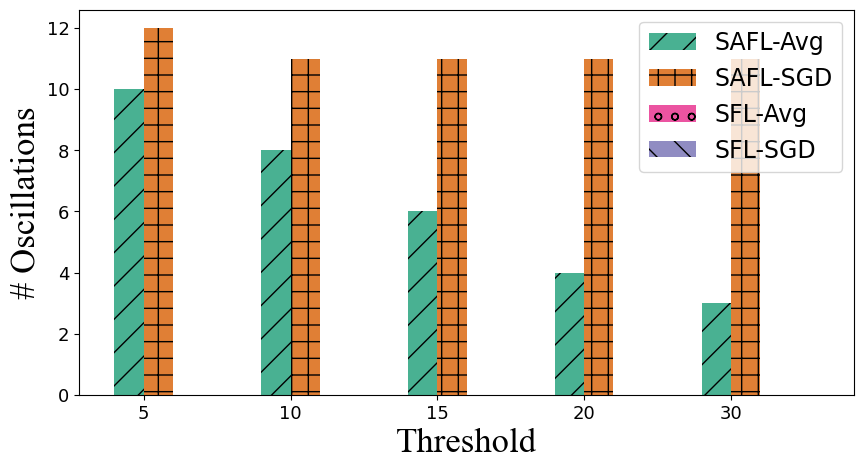}%
}
\hfil
\subfloat[$\alpha = 1$]{\includegraphics[width=1.5in]{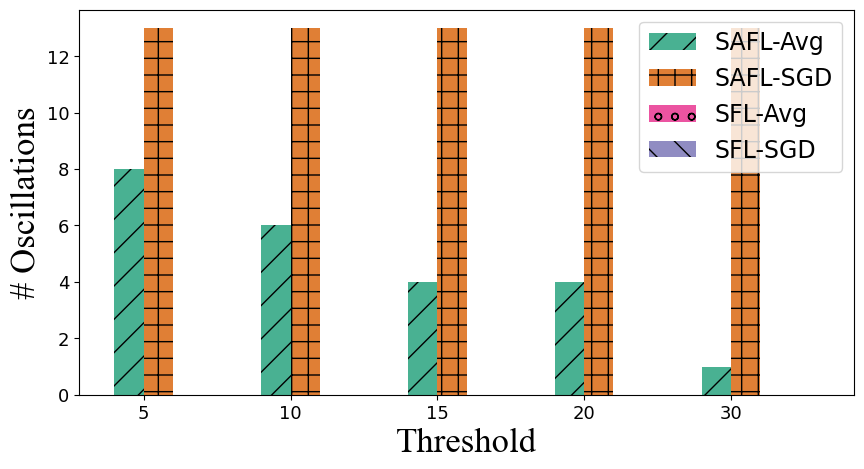}%
}    
% \hfil
% \subfloat[$N = 2$]{\includegraphics[width=1.5in]{pic/ret/vgg16_femnist/shards/ots/2.png}%
% }   
\hfil
\subfloat[$N = 5$]{\includegraphics[width=1.5in]{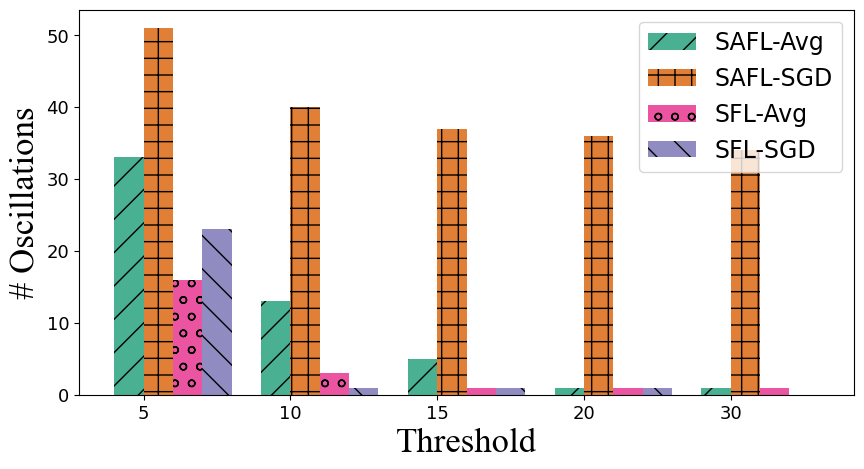}%
}   
\hfil
\subfloat[$N = 10$]{\includegraphics[width=1.5in]{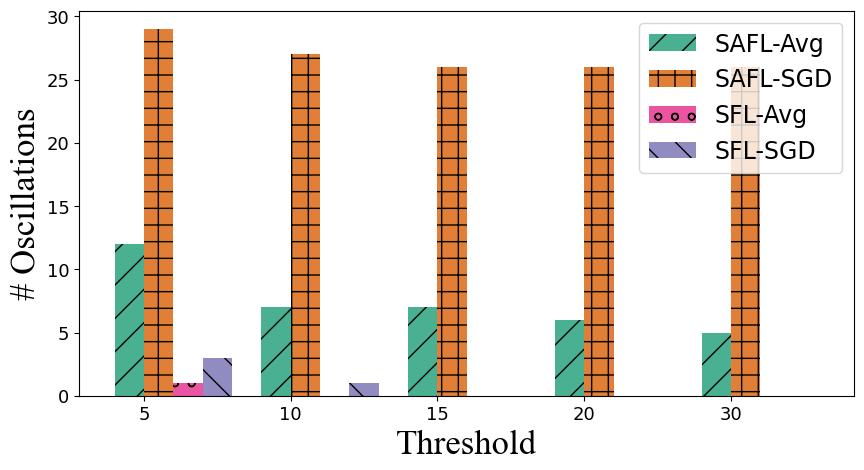}%
}   
\hfil
\subfloat[$\sigma = 0.1$]{\includegraphics[width=1.5in]{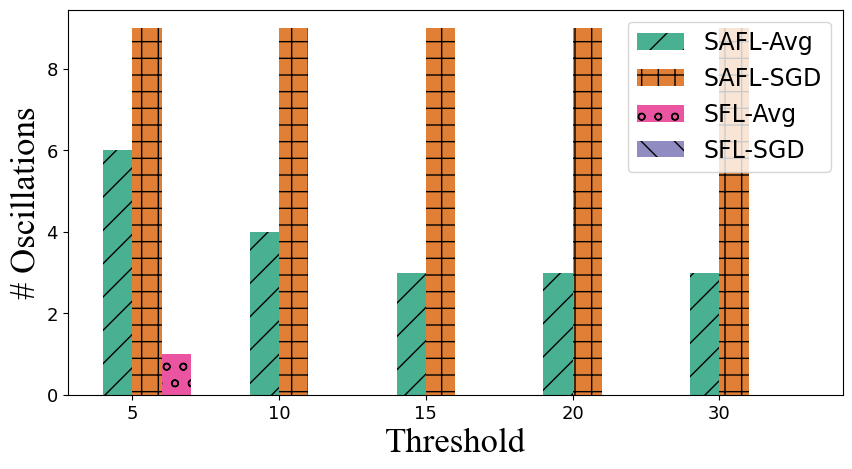}%
}   
\hfil
\subfloat[$\sigma = 0.5$]{\includegraphics[width=1.5in]{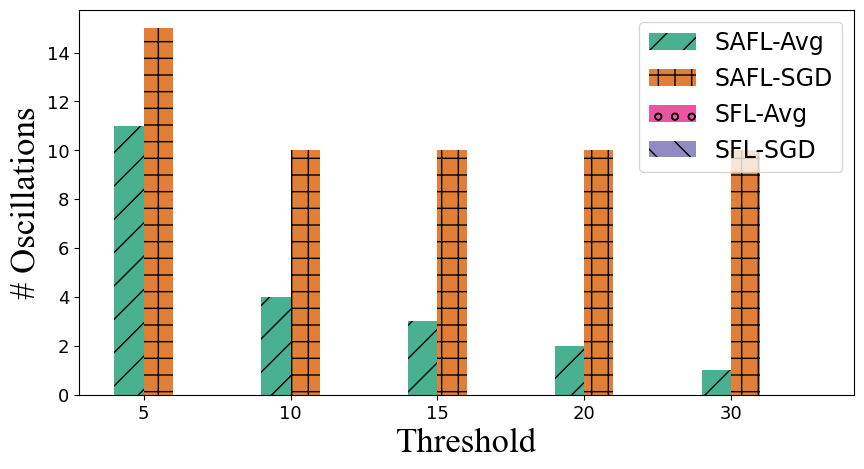}%
}   
\hfil
\subfloat[$\sigma = 1$]{\includegraphics[width=1.5in]{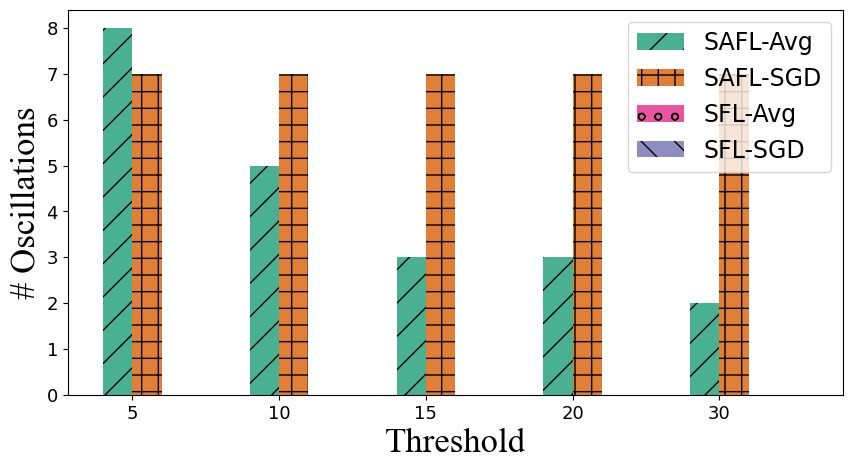}%
}   
\vspace{-1ex}
\caption{Statistics of severe oscillations of the VGG-16 model under the FEMNIST dataset.}
\end{figure}

\newpage
\subsection{CIFAR-10 @ CNN}

\begin{figure}[!h]
\vspace{-2ex}
\centering
\subfloat[$\alpha = 0.1$]{\includegraphics[width=1.5in]{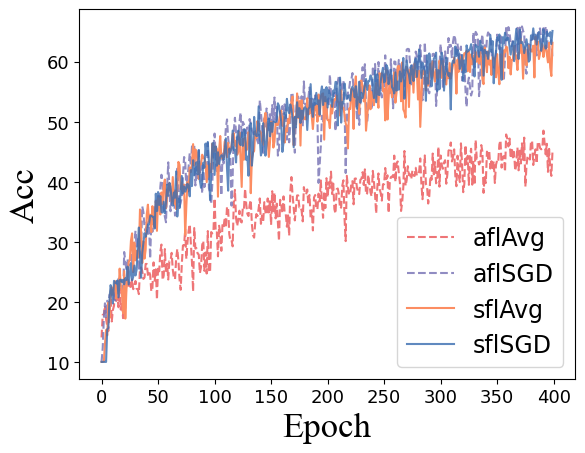}}
\hspace{1mm}
\subfloat[$\alpha = 0.1$]{\includegraphics[width=1.5in]{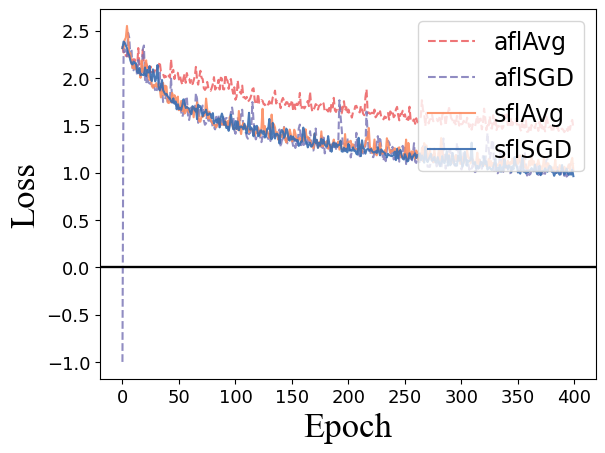}}
\hspace{1mm}
\subfloat[$\alpha = 0.5$]{\includegraphics[width=1.5in]{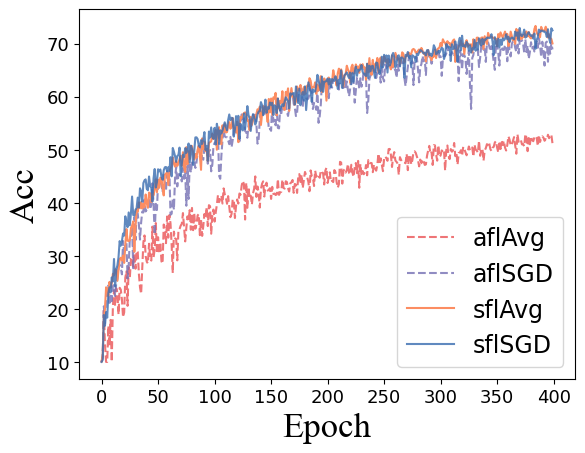}}
\hspace{1mm}
\subfloat[$\alpha = 0.5$]{\includegraphics[width=1.5in]{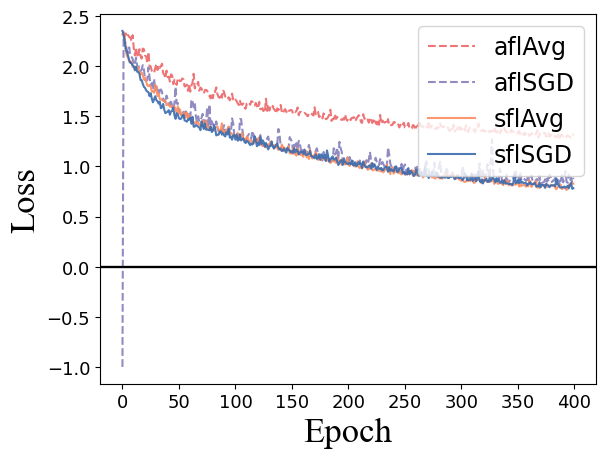}\label{fig_nan_concrete}}
\hspace{1mm}
\subfloat[$\alpha = 1$]{\includegraphics[width=1.5in]{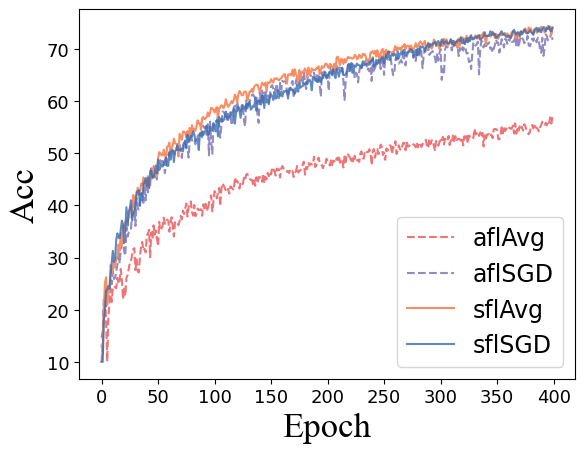}}
\hspace{1mm}
\subfloat[$\alpha = 1$]{\includegraphics[width=1.5in]{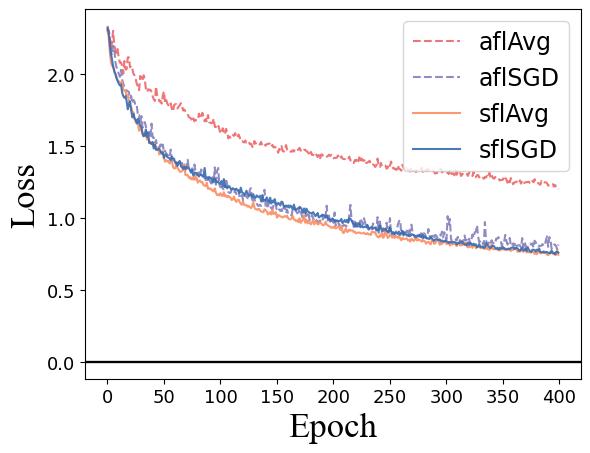}}
\hspace{1mm}
% \subfloat[$N = 2$]{\includegraphics[width=1.5in]{pic/ret/cnn_cifar10/shards/acc/2.png}}
% \hspace{1mm}
% \subfloat[$N = 2$]{\includegraphics[width=1.5in]{pic/ret/cnn_cifar10/shards/loss/2.png}}
% \hspace{1mm}
\subfloat[$N = 5$]{\includegraphics[width=1.5in]{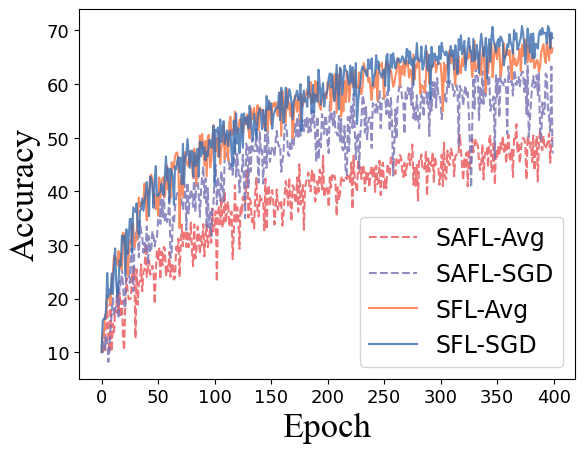}}
\hspace{1mm}
\subfloat[$N = 5$]{\includegraphics[width=1.5in]{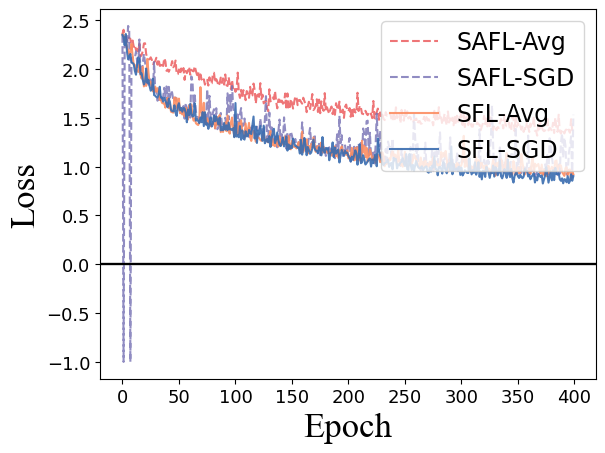}}
\hspace{1mm}
\subfloat[$N = 10$]{\includegraphics[width=1.5in]{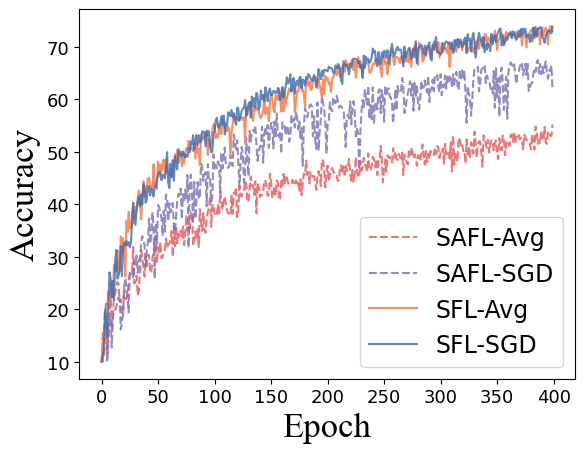}}
\hspace{1mm}
\subfloat[$N = 10$]{\includegraphics[width=1.5in]{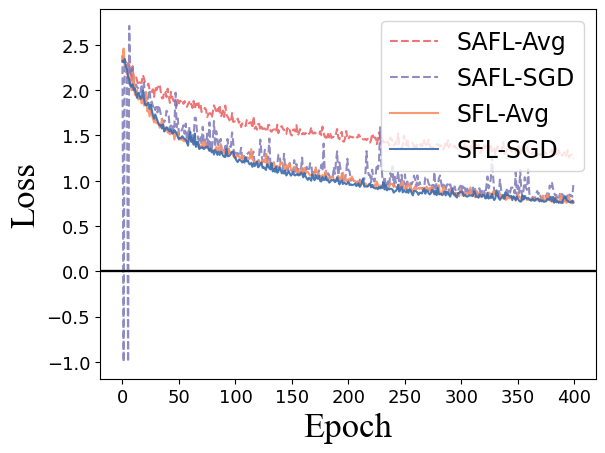}}
\hspace{1mm}
\subfloat[$\sigma = 0.1$]{\includegraphics[width=1.5in]{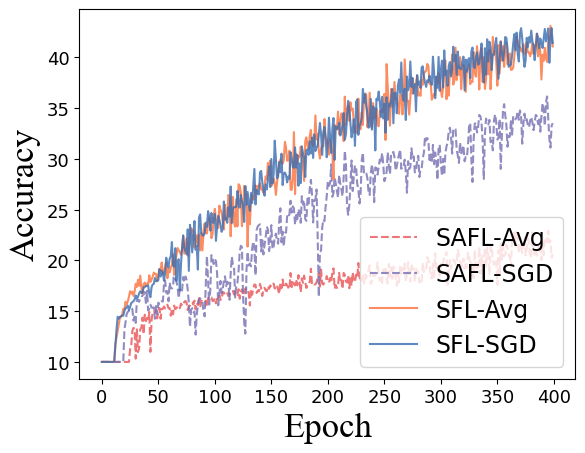}}
\hspace{1mm}
\subfloat[$\sigma = 0.1$]{\includegraphics[width=1.5in]{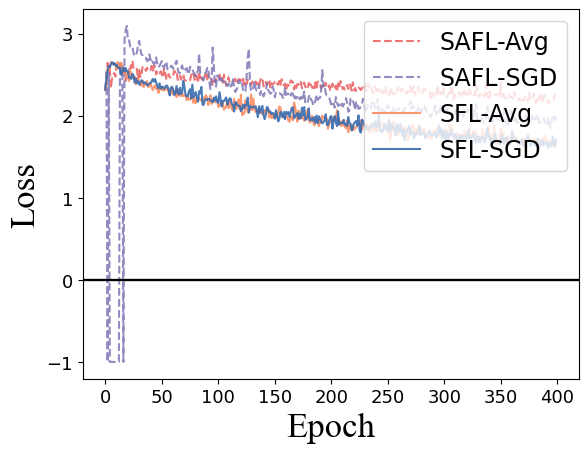}}
\hspace{1mm}
\subfloat[$\sigma = 0.5$]{\includegraphics[width=1.5in]{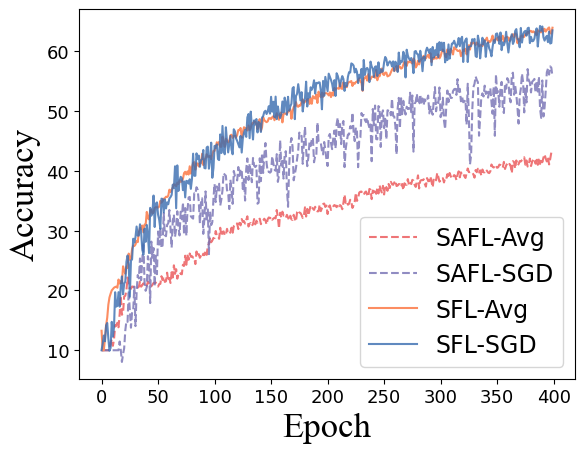}}
\hspace{1mm}
\subfloat[$\sigma = 0.5$]{\includegraphics[width=1.5in]{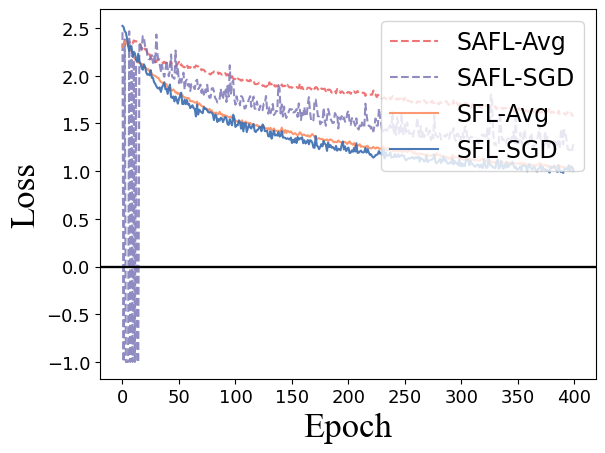}}
\hspace{1mm}
\subfloat[$\sigma = 1$]{\includegraphics[width=1.5in]{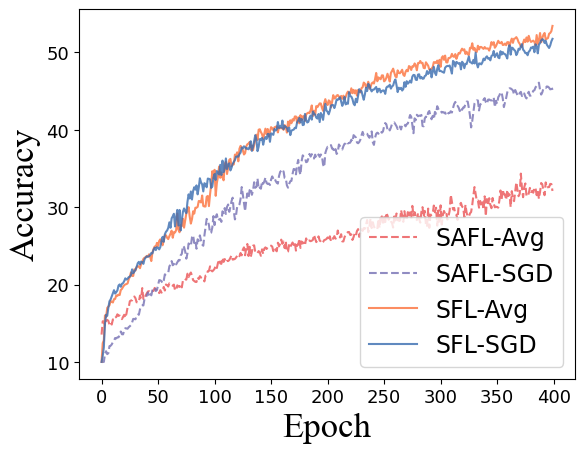}}
\hspace{1mm}
\subfloat[$\sigma = 1$]{\includegraphics[width=1.5in]{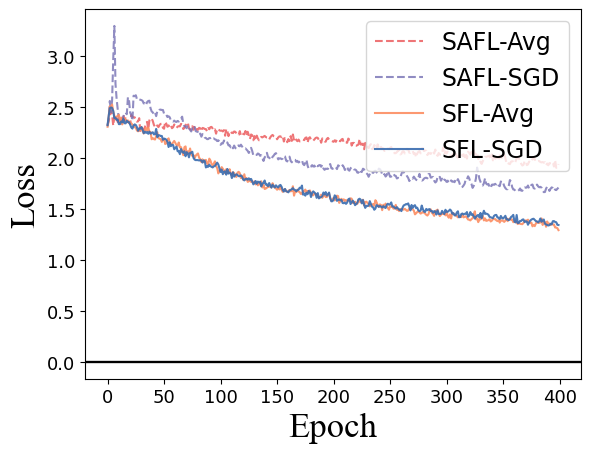}}
\caption{Global accuracy and loss of different models under CIFAR-10 dataset using CNN in SAFL. Note that -1 denotes the NAN value for loss. }
\vspace{-2ex}
\end{figure}
\begin{figure}[!h]
\vspace{-2ex}
\centering
\subfloat[$\alpha = 0.1$]{\includegraphics[width=1.5in]{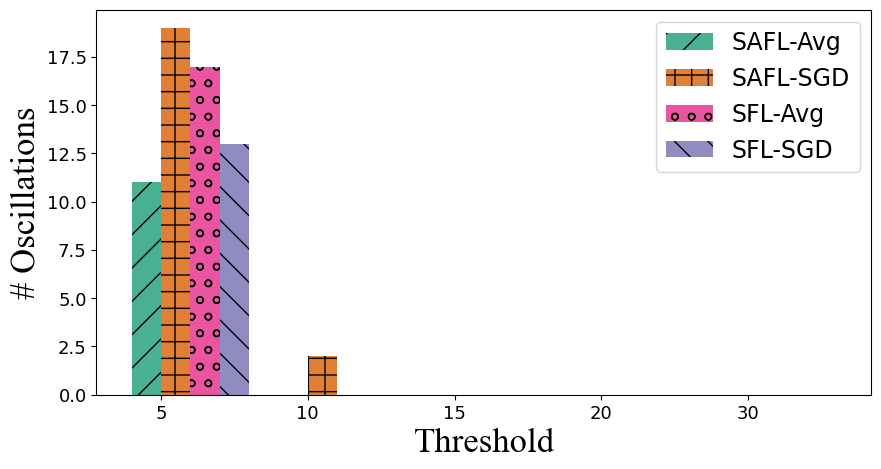}%
}
\hfil
\subfloat[$\alpha = 0.5$]{\includegraphics[width=1.5in]{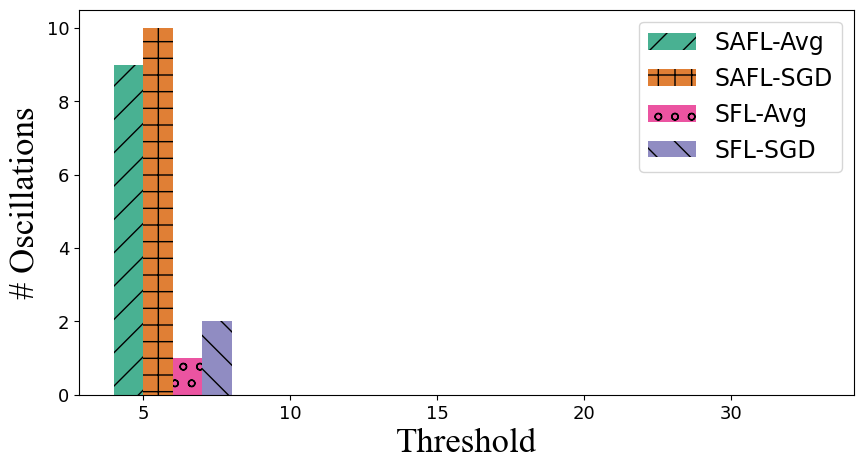}%
}
\hfil
\subfloat[$\alpha = 1$]{\includegraphics[width=1.5in]{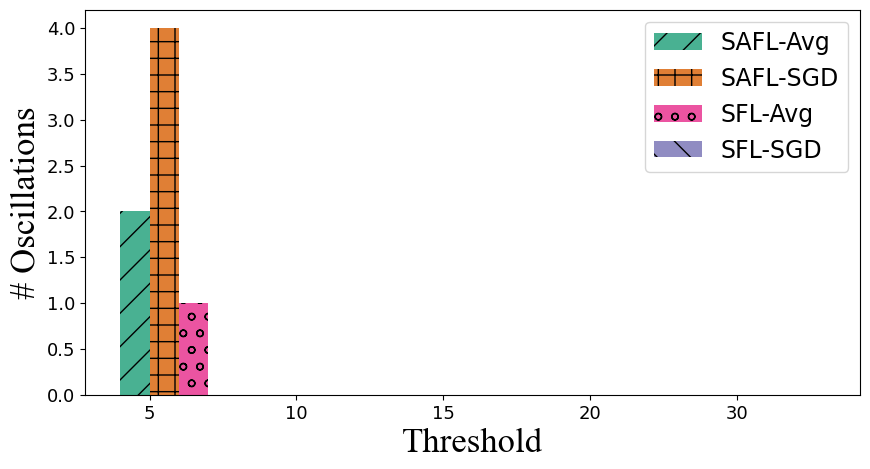}%
}    
% \hfil
% \subfloat[$N = 2$]{\includegraphics[width=1.5in]{pic/ret/cnn_cifar10/shards/ots/2.png}%
% }   
\hfil
\subfloat[$N = 5$]{\includegraphics[width=1.5in]{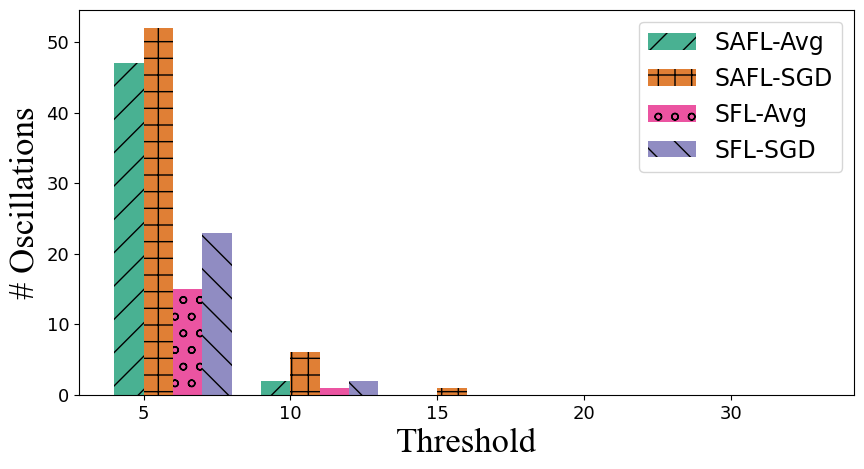}%
}   
\hfil
\subfloat[$N = 10$]{\includegraphics[width=1.5in]{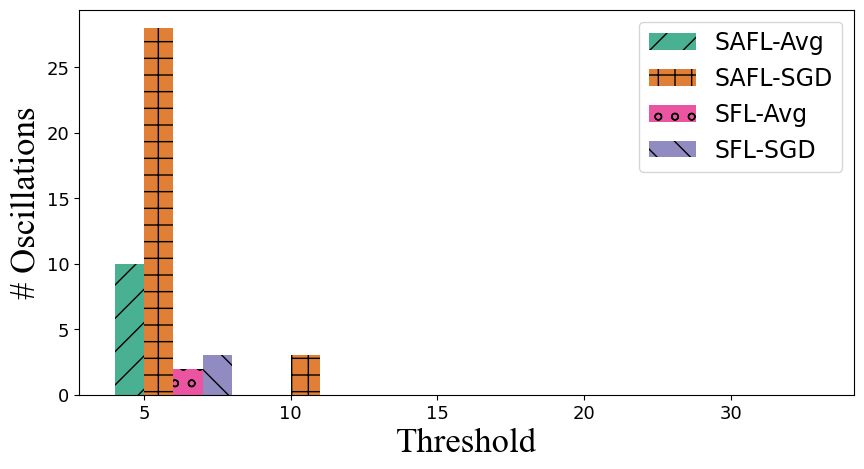}%
}   
\hfil
\subfloat[$\sigma = 0.1$]{\includegraphics[width=1.5in]{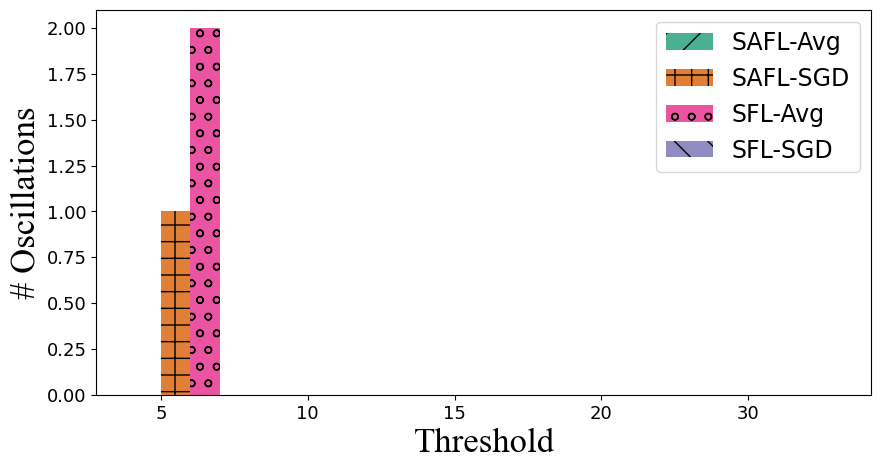}%
}   
\hfil
\subfloat[$\sigma = 0.5$]{\includegraphics[width=1.5in]{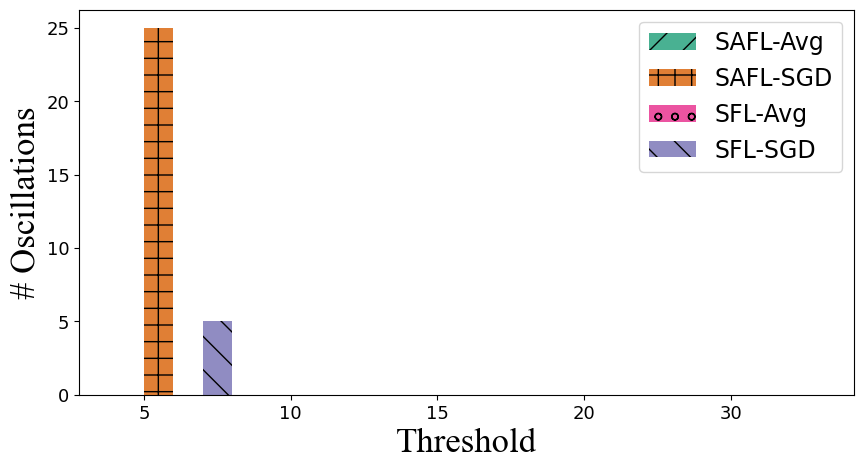}%
}   
\hfil
\subfloat[$\sigma = 1$]{\includegraphics[width=1.5in]{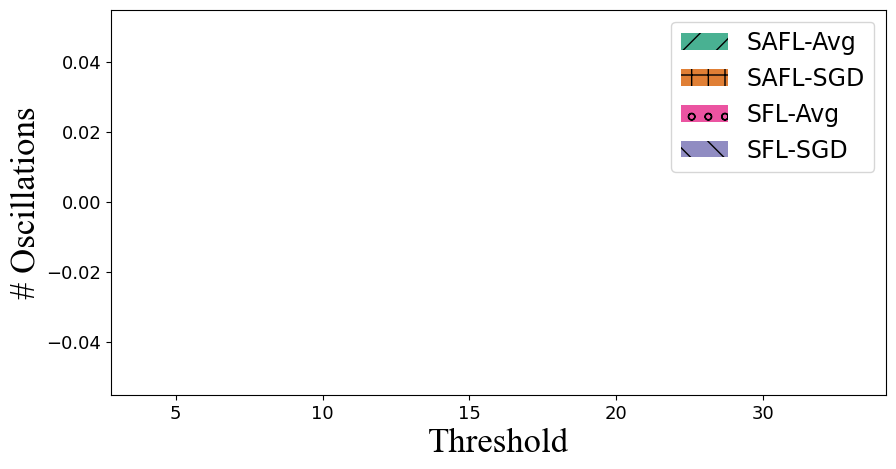}%
}   
\vspace{-1ex}
\caption{Statistics of severe oscillations of the CNN model under the CIFAR-10 dataset.}
\end{figure}

\newpage
\subsection{CIFAR-100 @ CNN}

\begin{figure}[!h]
\vspace{-2ex}
\centering
\subfloat[$\alpha = 0.1$]{\includegraphics[width=1.5in]{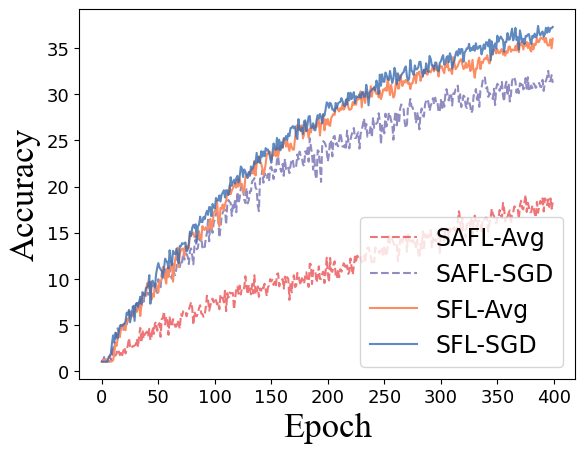}}
\hspace{1mm}
\subfloat[$\alpha = 0.1$]{\includegraphics[width=1.5in]{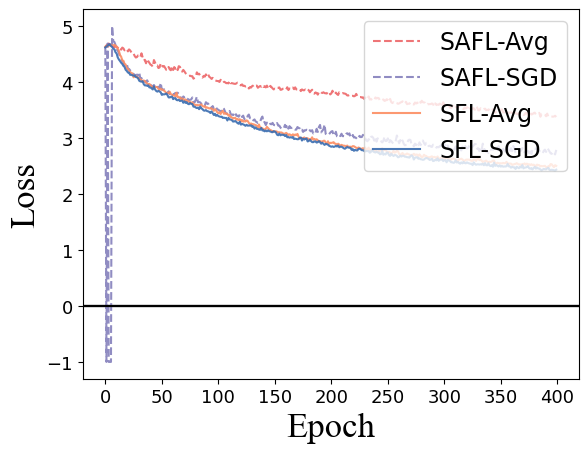}}
\hspace{1mm}
\subfloat[$\alpha = 0.5$]{\includegraphics[width=1.5in]{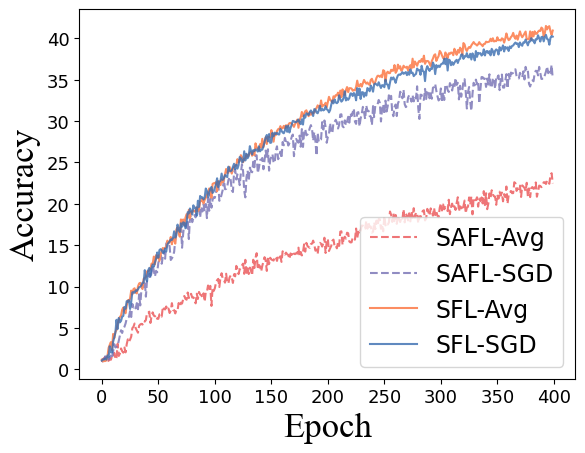}}
\hspace{1mm}
\subfloat[$\alpha = 0.5$]{\includegraphics[width=1.5in]{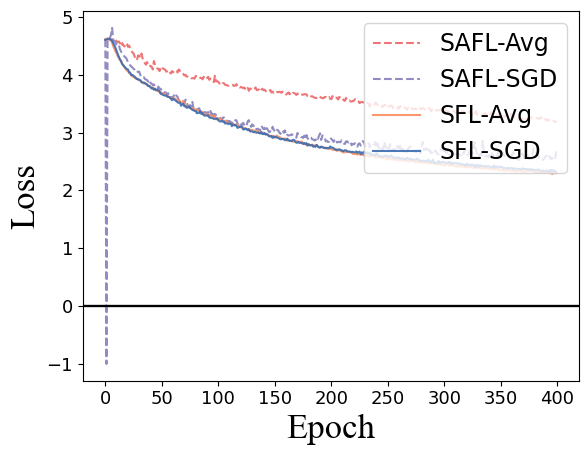}\label{fig_nan_concrete}}
\hspace{1mm}
\subfloat[$\alpha = 1$]{\includegraphics[width=1.5in]{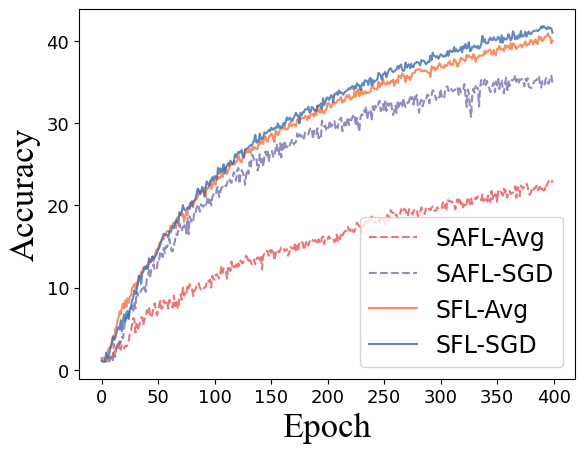}}
\hspace{1mm}
\subfloat[$\alpha = 1$]{\includegraphics[width=1.5in]{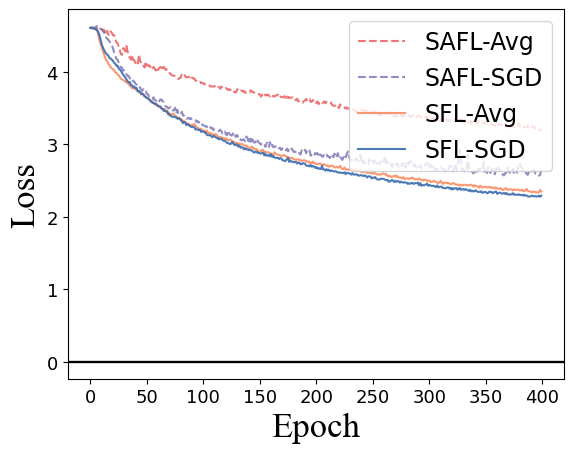}}
\hspace{1mm}
% \subfloat[$N = 2$]{\includegraphics[width=1.5in]{pic/ret/cnn_cifar100/shards/acc/2.png}}
% \hspace{1mm}
% \subfloat[$N = 2$]{\includegraphics[width=1.5in]{pic/ret/cnn_cifar100/shards/loss/2.png}}
% \hspace{1mm}
\subfloat[$N = 5$]{\includegraphics[width=1.5in]{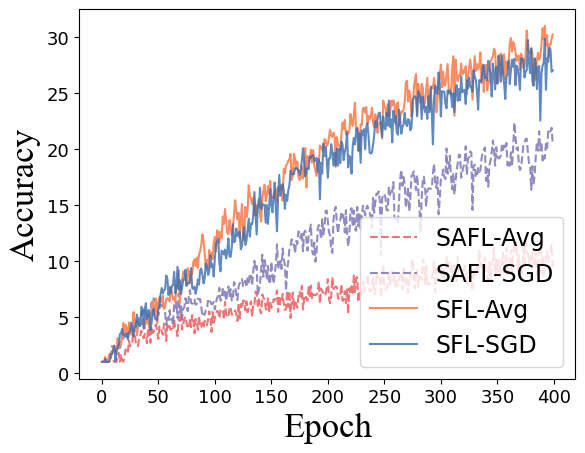}}
\hspace{1mm}
\subfloat[$N = 5$]{\includegraphics[width=1.5in]{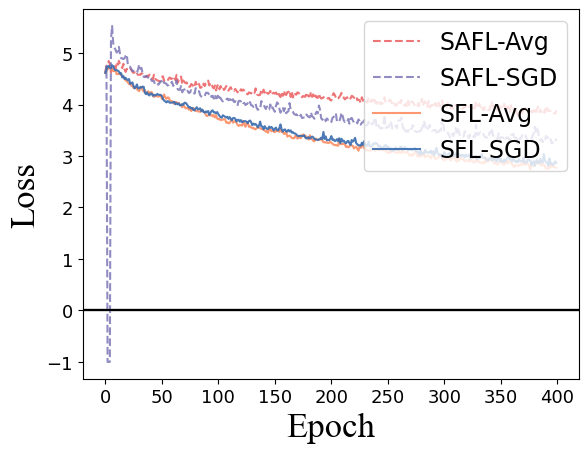}}
\hspace{1mm}
\subfloat[$N = 10$]{\includegraphics[width=1.5in]{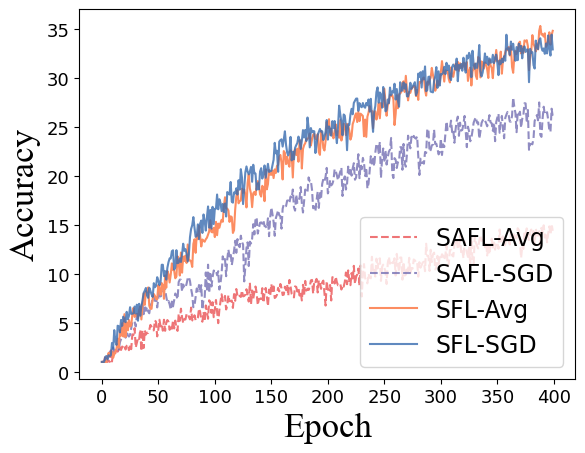}}
\hspace{1mm}
\subfloat[$N = 10$]{\includegraphics[width=1.5in]{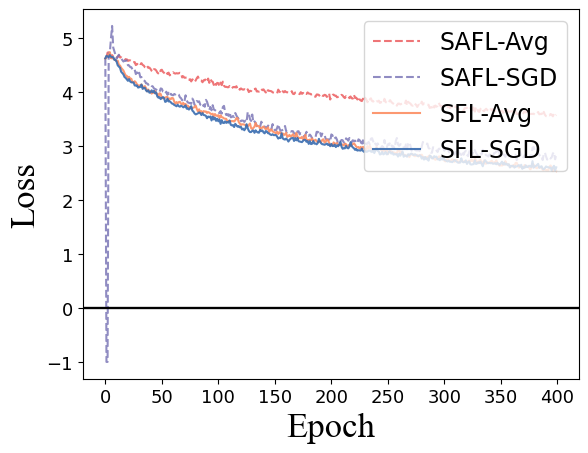}}
\hspace{1mm}
\subfloat[$\sigma = 0.1$]{\includegraphics[width=1.5in]{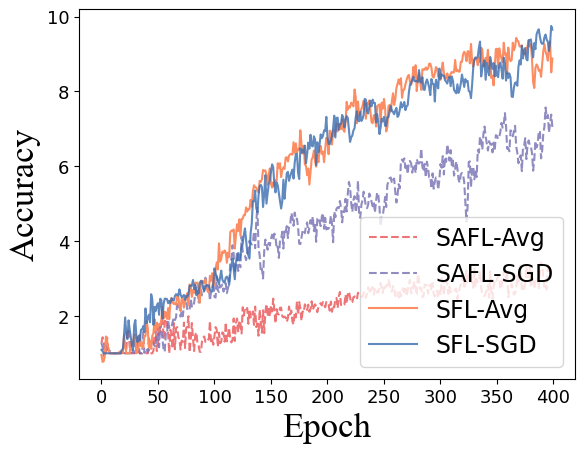}}
\hspace{1mm}
\subfloat[$\sigma = 0.1$]{\includegraphics[width=1.5in]{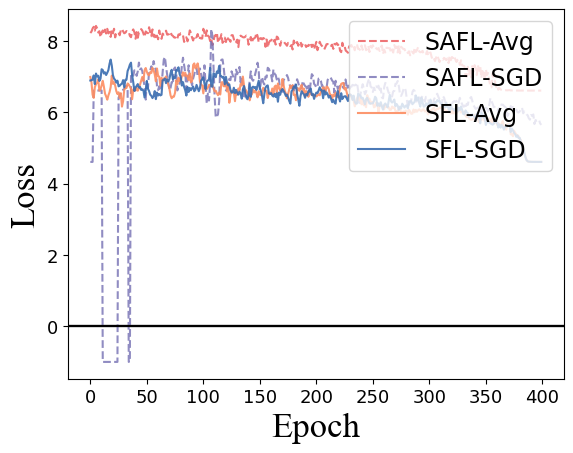}}
\hspace{1mm}
\subfloat[$\sigma = 0.5$]{\includegraphics[width=1.5in]{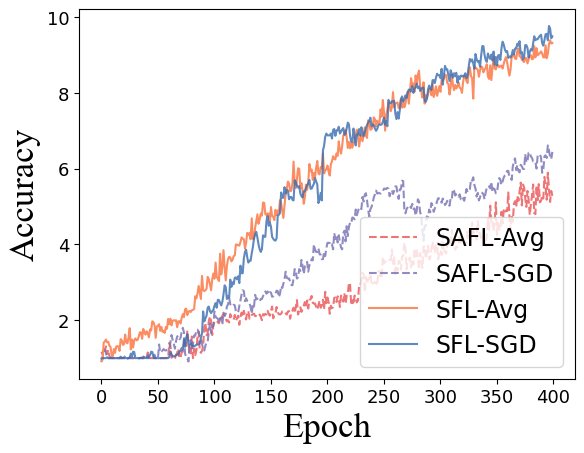}}
\hspace{1mm}
\subfloat[$\sigma = 0.5$]{\includegraphics[width=1.5in]{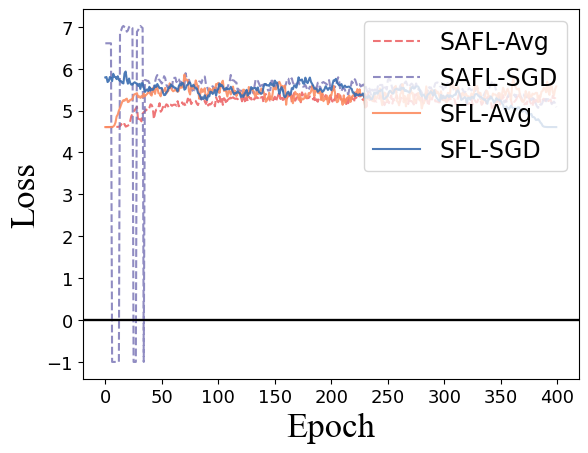}}
\hspace{1mm}
\subfloat[$\sigma = 1$]{\includegraphics[width=1.5in]{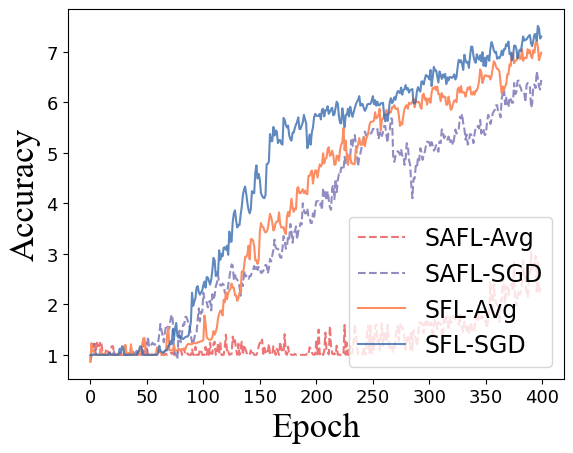}}
\hspace{1mm}
\subfloat[$\sigma = 1$]{\includegraphics[width=1.5in]{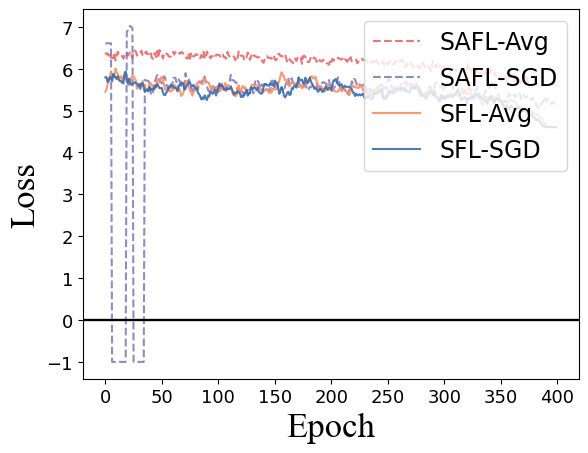}}
\caption{Global accuracy and loss of different models under CIFAR-100 dataset using CNN in SAFL. Note that -1 denotes the NAN value for loss. }
\vspace{-2ex}
\end{figure}
\begin{figure}[!h]
\vspace{-2ex}
\centering
\subfloat[$\alpha = 0.1$]{\includegraphics[width=1.5in]{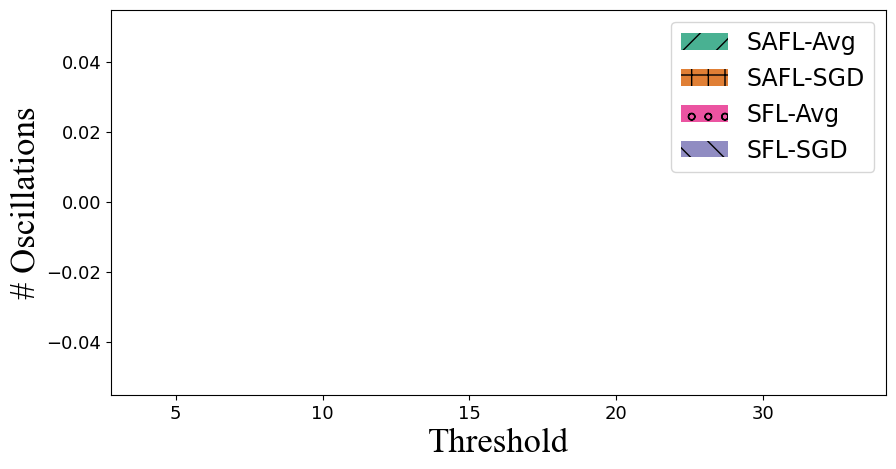}%
}
\hfil
\subfloat[$\alpha = 0.5$]{\includegraphics[width=1.5in]{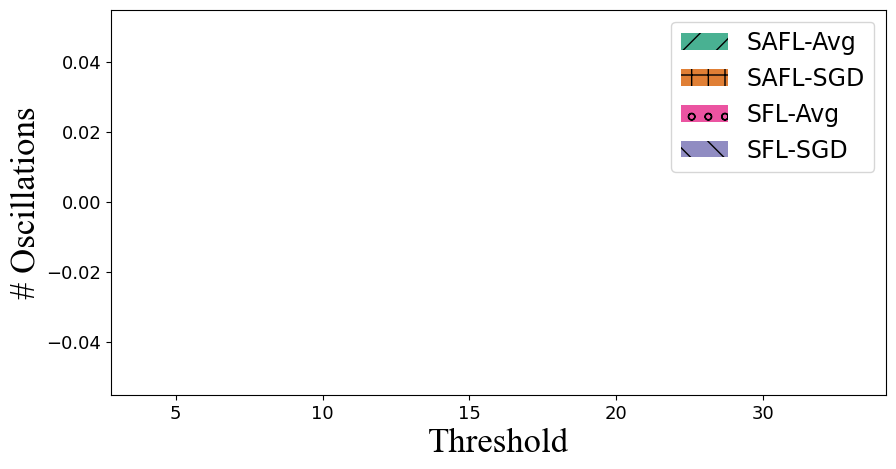}%
}
\hfil
\subfloat[$\alpha = 1$]{\includegraphics[width=1.5in]{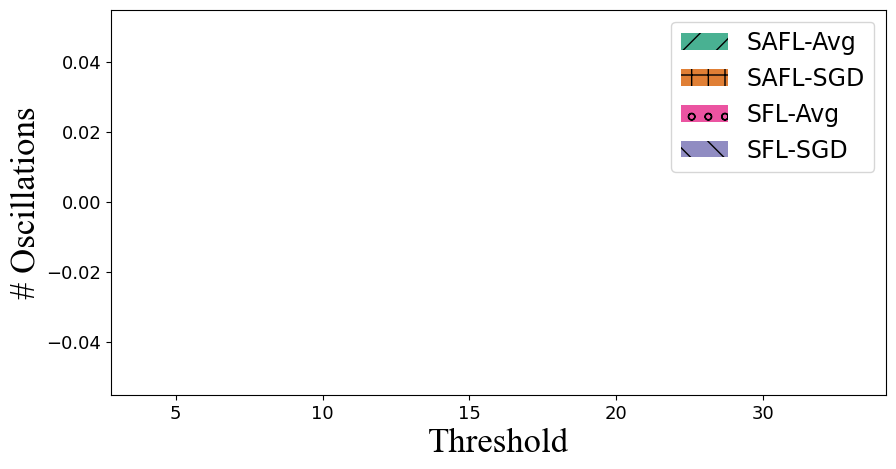}%
}    
% \hfil
% \subfloat[$N = 2$]{\includegraphics[width=1.5in]{pic/ret/cnn_cifar100/shards/ots/2.png}%
% }   
\hfil
\subfloat[$N = 5$]{\includegraphics[width=1.5in]{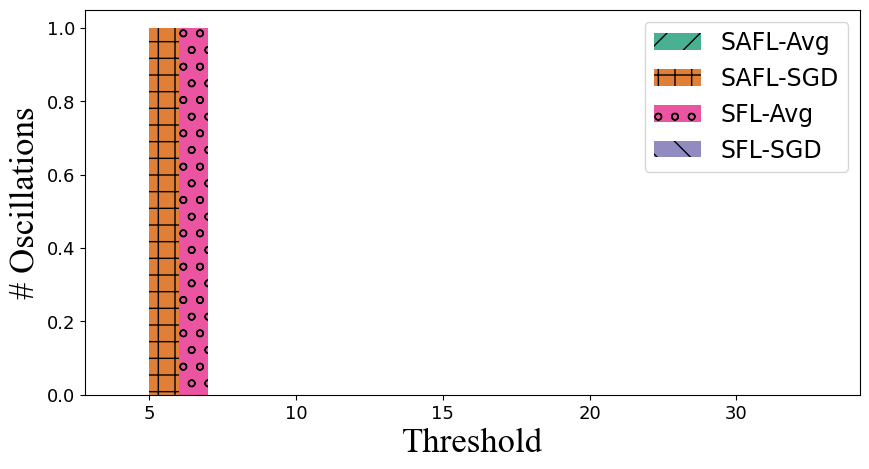}%
}   
\hfil
\subfloat[$N = 10$]{\includegraphics[width=1.5in]{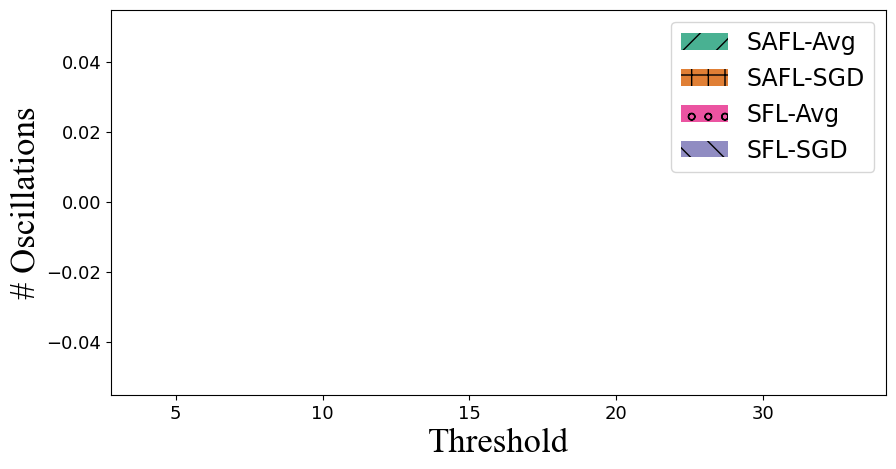}%
}   
\hfil
\subfloat[$\sigma = 0.1$]{\includegraphics[width=1.5in]{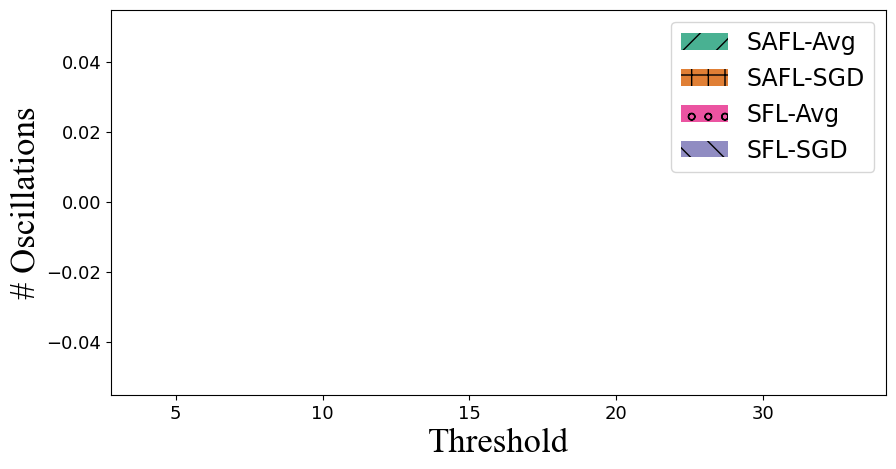}%
}   
\hfil
\subfloat[$\sigma = 0.5$]{\includegraphics[width=1.5in]{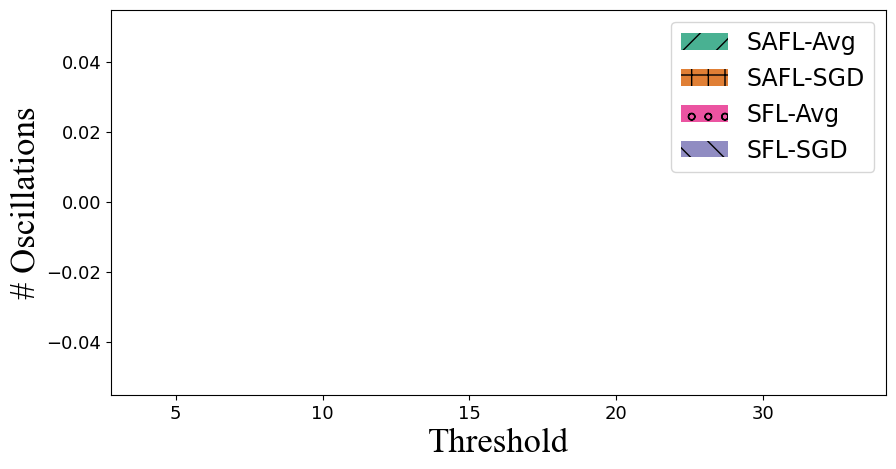}%
}   
\hfil
\subfloat[$\sigma = 1$]{\includegraphics[width=1.5in]{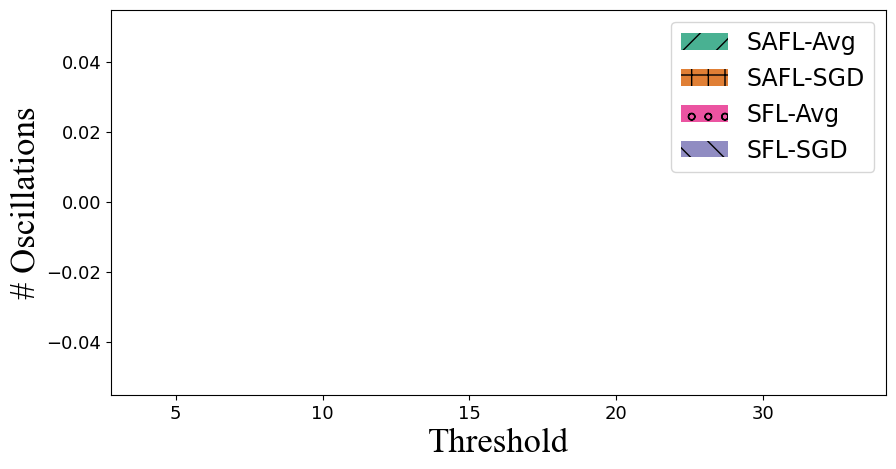}%
}   
\vspace{-1ex}
\caption{Statistics of severe oscillations of the CNN model under the CIFAR-100 dataset.}
\end{figure}

\newpage
\subsection{FEMNIST @ CNN}

\begin{figure}[!h]
\vspace{-3ex}
\centering
\subfloat[$\alpha = 0.1$]{\includegraphics[width=1.5in]{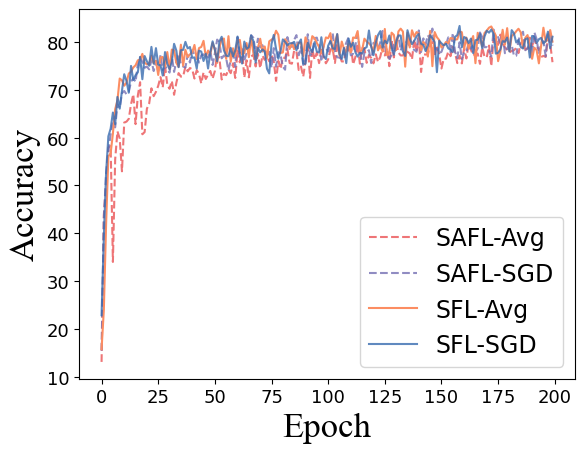}}
\hspace{1mm}
\subfloat[$\alpha = 0.1$]{\includegraphics[width=1.5in]{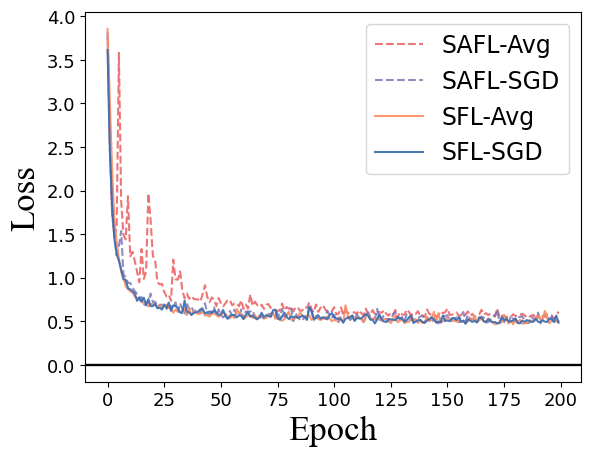}}
\hspace{1mm}
\subfloat[$\alpha = 0.5$]{\includegraphics[width=1.5in]{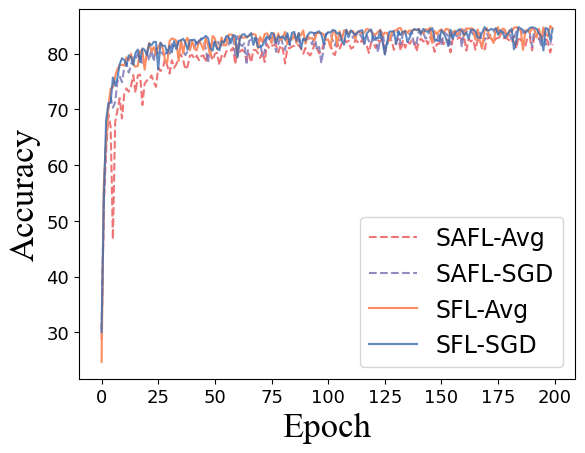}}
\hspace{1mm}
\subfloat[$\alpha = 0.5$]{\includegraphics[width=1.5in]{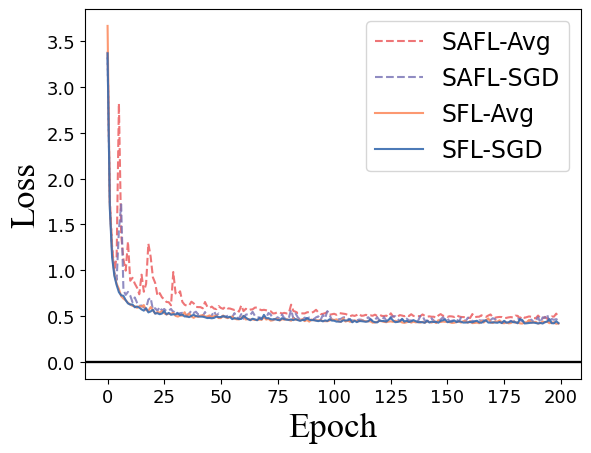}\label{fig_nan_concrete}}
\hspace{1mm}
\subfloat[$\alpha = 1$]{\includegraphics[width=1.5in]{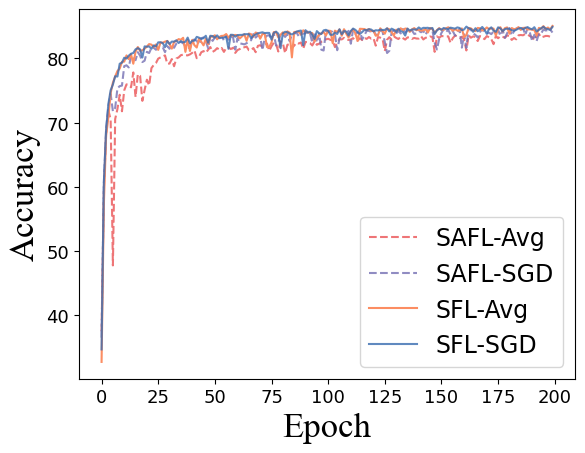}}
\hspace{1mm}
\subfloat[$\alpha = 1$]{\includegraphics[width=1.5in]{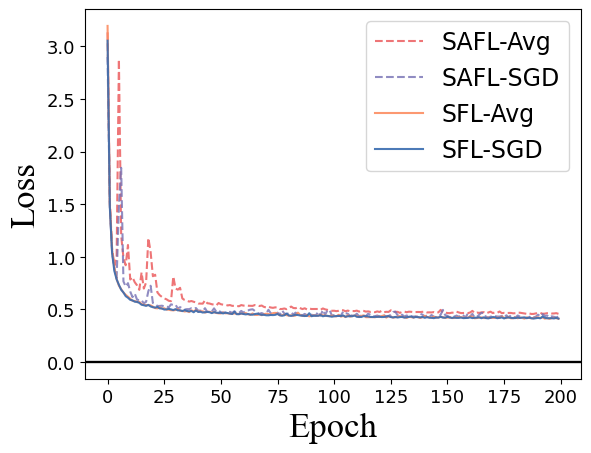}}
\hspace{1mm}
% \subfloat[$N = 2$]{\includegraphics[width=1.5in]{pic/ret/cnn_femnist/shards/acc/2.png}}
% \hspace{1mm}
% \subfloat[$N = 2$]{\includegraphics[width=1.5in]{pic/ret/cnn_femnist/shards/loss/2.png}}
% \hspace{1mm}
\subfloat[$N = 5$]{\includegraphics[width=1.5in]{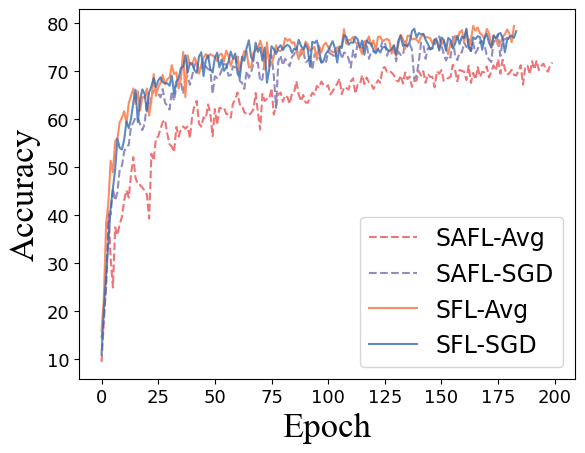}}
\hspace{1mm}
\subfloat[$N = 5$]{\includegraphics[width=1.5in]{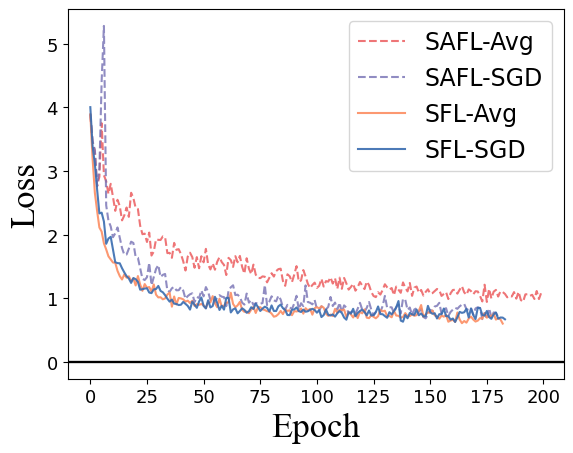}}
\hspace{1mm}
\subfloat[$N = 10$]{\includegraphics[width=1.5in]{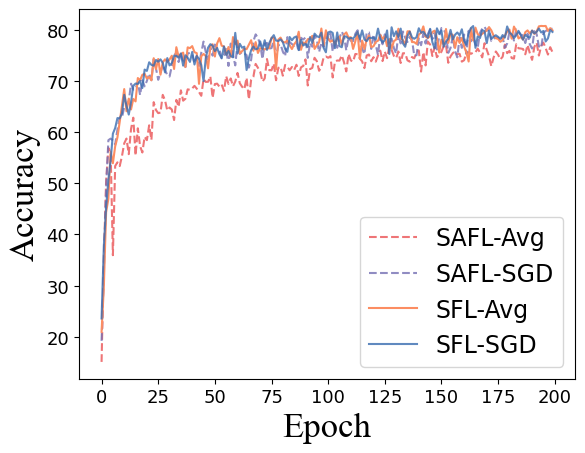}}
\hspace{1mm}
\subfloat[$N = 10$]{\includegraphics[width=1.5in]{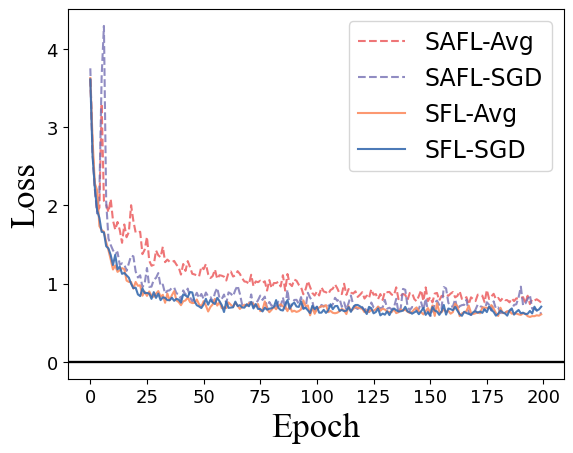}}
\hspace{1mm}
\subfloat[$\sigma = 0.1$]{\includegraphics[width=1.5in]{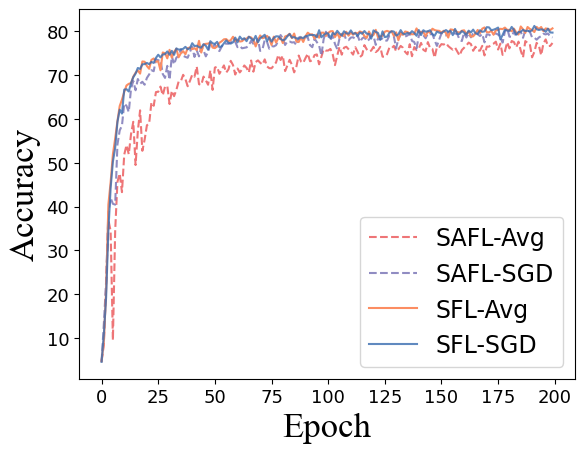}}
\hspace{1mm}
\subfloat[$\sigma = 0.1$]{\includegraphics[width=1.5in]{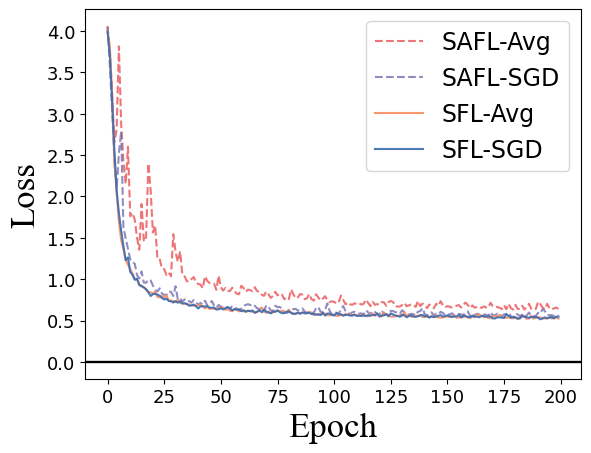}}
\hspace{1mm}
\subfloat[$\sigma = 0.5$]{\includegraphics[width=1.5in]{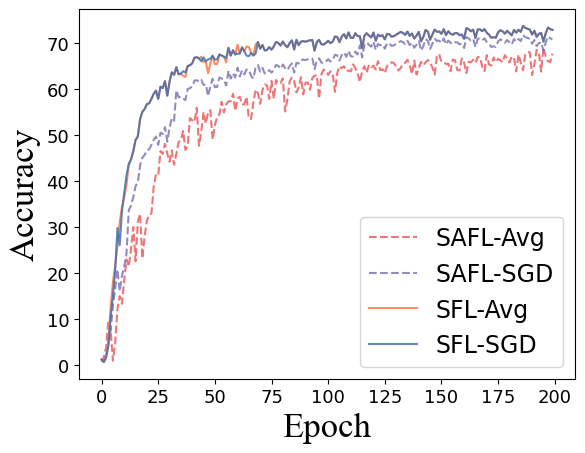}}
\hspace{1mm}
\subfloat[$\sigma = 0.5$]{\includegraphics[width=1.5in]{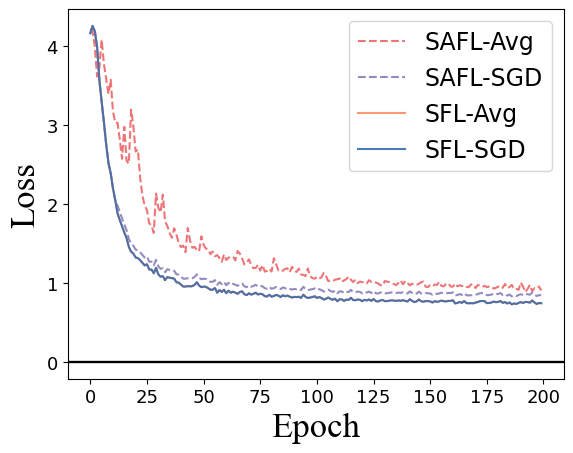}}
\hspace{1mm}
\subfloat[$\sigma = 1$]{\includegraphics[width=1.5in]{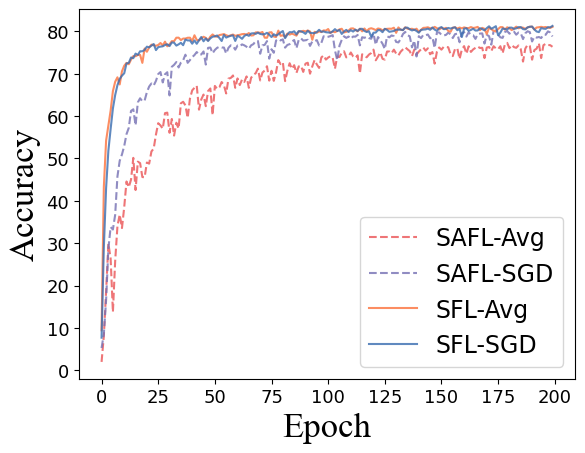}}
\hspace{1mm}
\subfloat[$\sigma = 1$]{\includegraphics[width=1.5in]{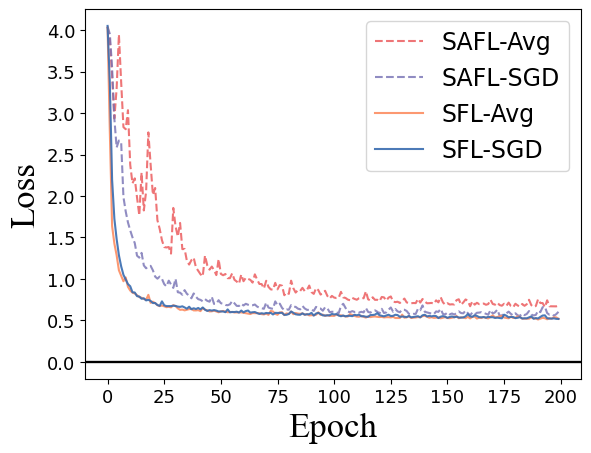}}
\caption{Global accuracy and loss of different models under FEMNIST dataset using CNN in SAFL. Note that -1 denotes the NAN value for loss. }
\vspace{-2ex}
\end{figure}
\begin{figure}[!h]
\vspace{-2ex}
\centering
\subfloat[$\alpha = 0.1$]{\includegraphics[width=1.5in]{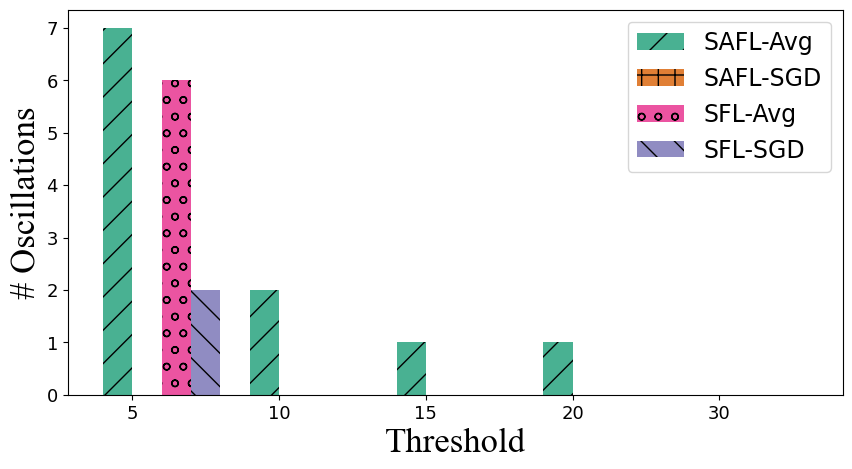}%
}
\hfil
\subfloat[$\alpha = 0.5$]{\includegraphics[width=1.5in]{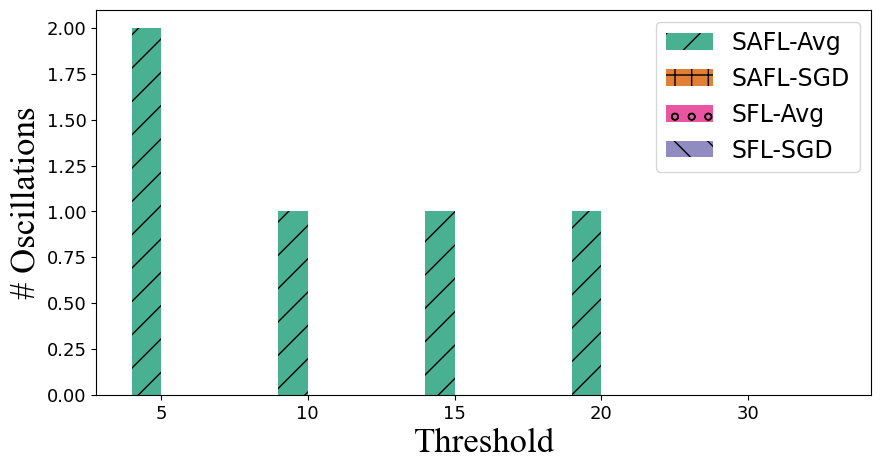}%
}
\hfil
\subfloat[$\alpha = 1$]{\includegraphics[width=1.5in]{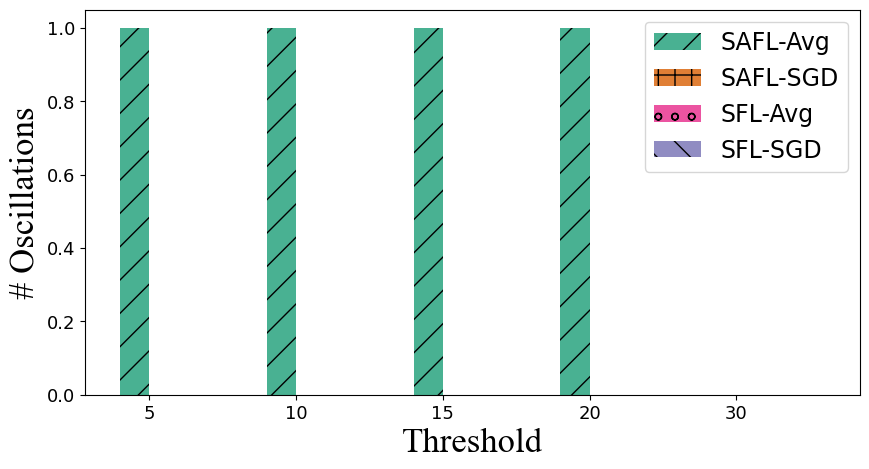}%
}    
% \hfil
% \subfloat[$N = 2$]{\includegraphics[width=1.5in]{pic/ret/cnn_femnist/shards/ots/2.png}%
% }   
\hfil
\subfloat[$N = 5$]{\includegraphics[width=1.5in]{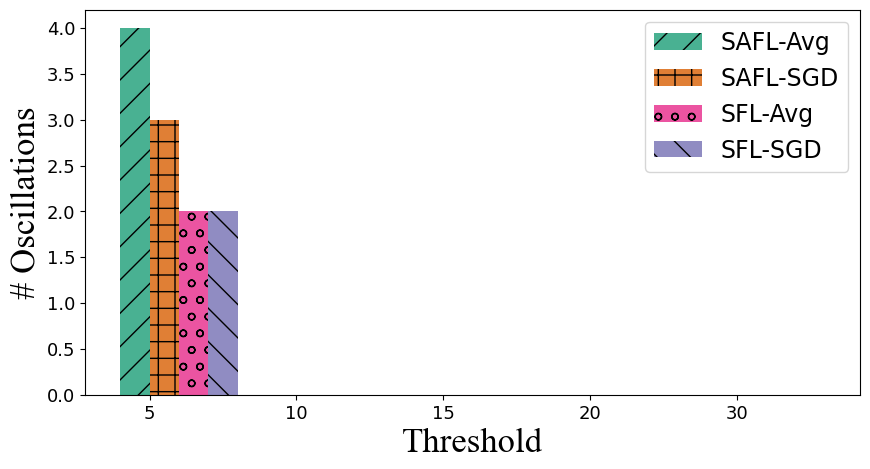}%
}   
\hfil
\subfloat[$N = 10$]{\includegraphics[width=1.5in]{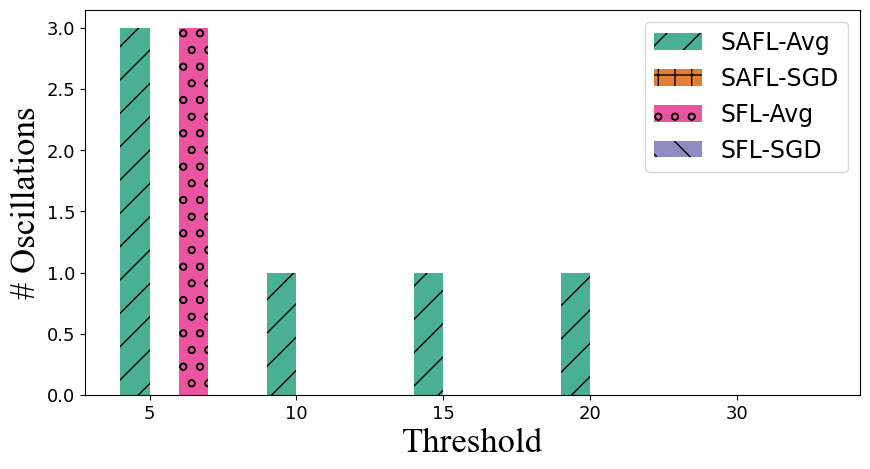}%
}   
\hfil
\subfloat[$\sigma = 0.1$]{\includegraphics[width=1.5in]{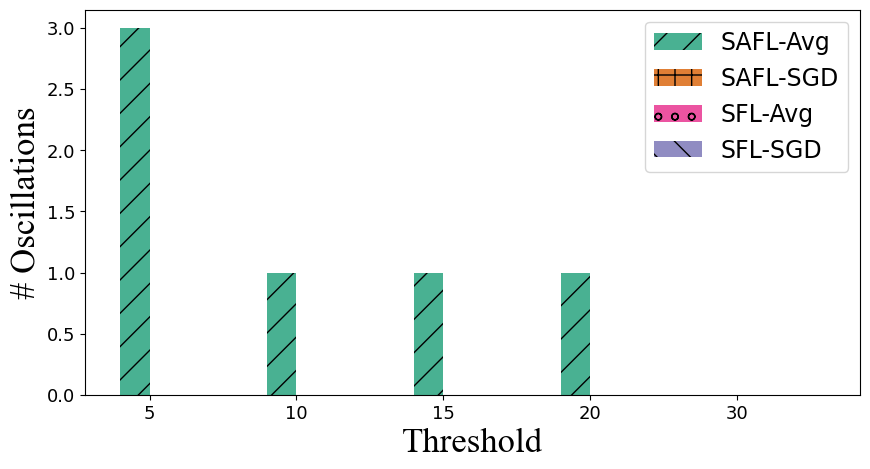}%
}   
\hfil
\subfloat[$\sigma = 0.5$]{\includegraphics[width=1.5in]{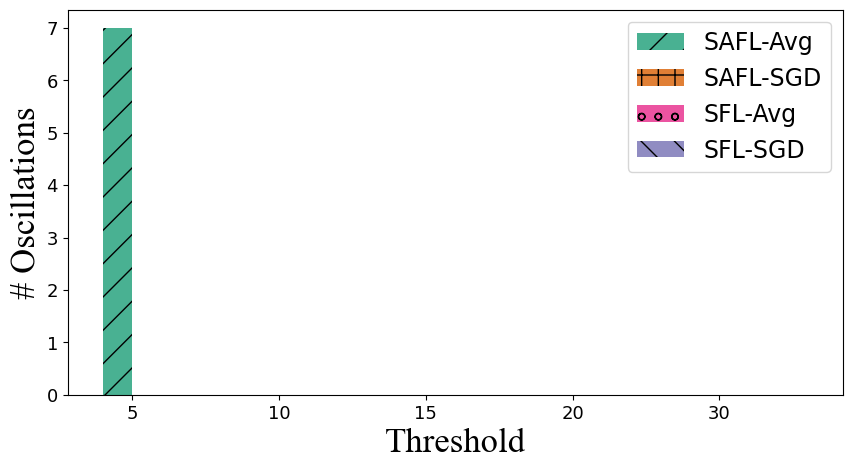}%
}   
\hfil
\subfloat[$\sigma = 1$]{\includegraphics[width=1.5in]{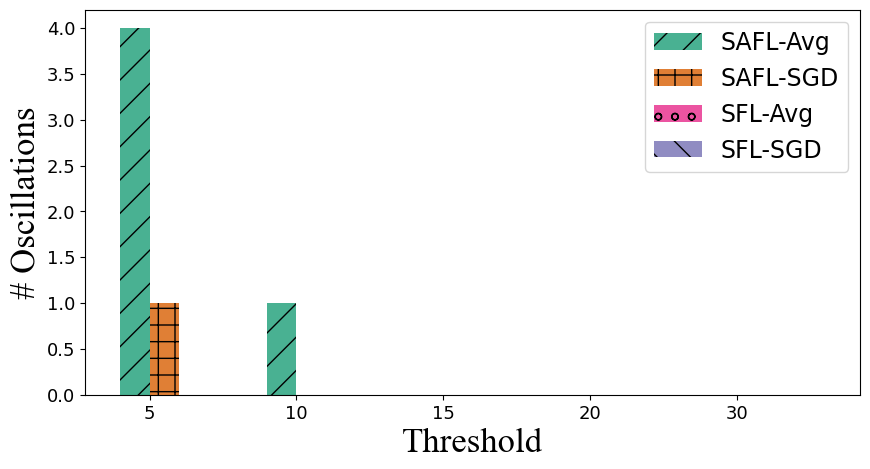}%
}   
\vspace{-1ex}
\caption{Statistics of severe oscillations of the CNN model under the FEMNIST dataset.}
\end{figure}

\newpage
\subsection{Shakespeare @ LSTM}

\begin{figure}[!h]
\vspace{-1ex}
\centering
\subfloat[$N = 100$]{\includegraphics[width=1.5in]{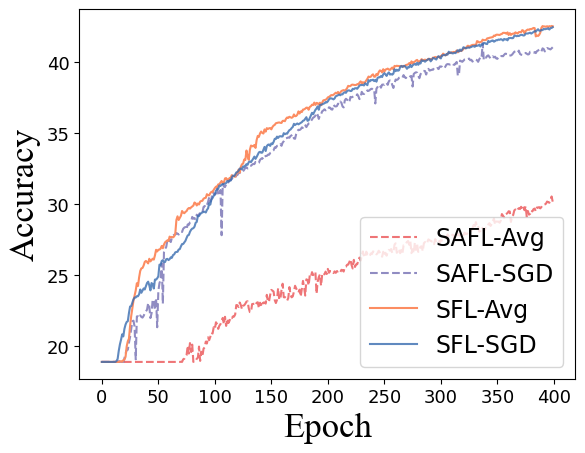}}
\hspace{1mm}
\subfloat[$N = 100$]{\includegraphics[width=1.5in]{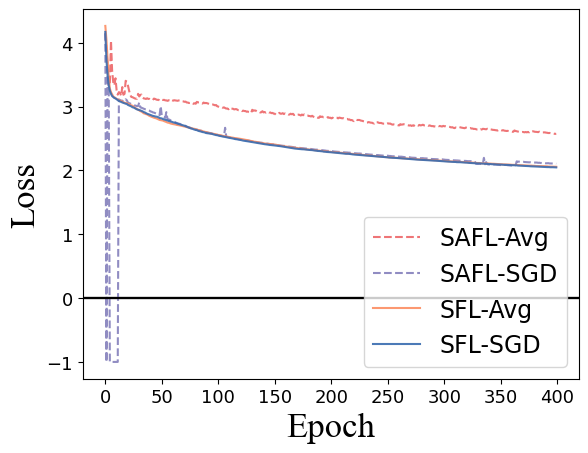}}
\hspace{1mm}
\subfloat[$N = 300$]{\includegraphics[width=1.5in]{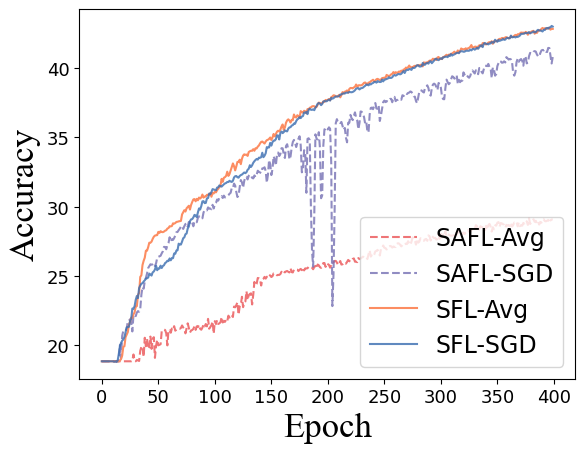}}
\hspace{1mm}
\subfloat[$N = 300$]{\includegraphics[width=1.5in]{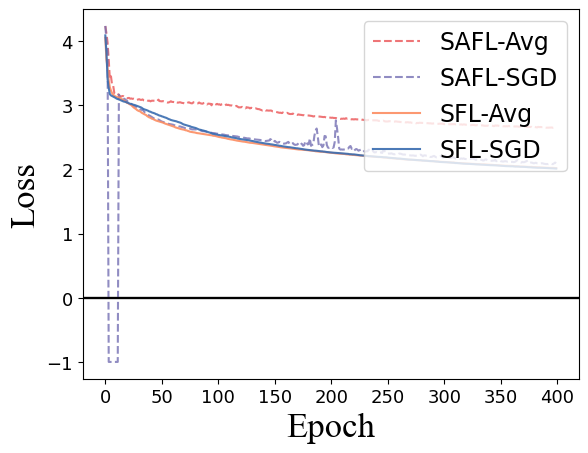}\label{fig_nan_concrete}}
\caption{Global accuracy and loss of different models under Shakespeare dataset using LSTM in SAFL. Note that -1 denotes the NAN value for loss. }
\vspace{-1ex}
\end{figure}
\begin{figure}[!h]
\vspace{-2ex}
\centering
\subfloat[$N = 100$]{\includegraphics[width=1.5in]{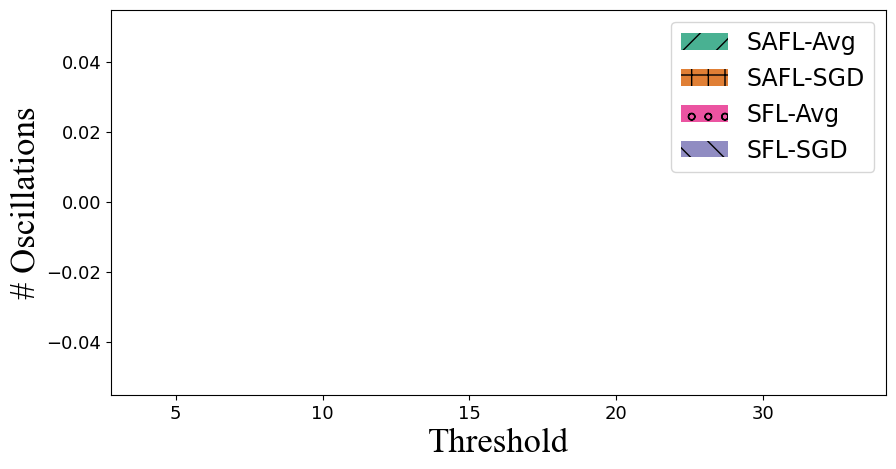}%
}
\hfil
\subfloat[$N = 300$]{\includegraphics[width=1.5in]{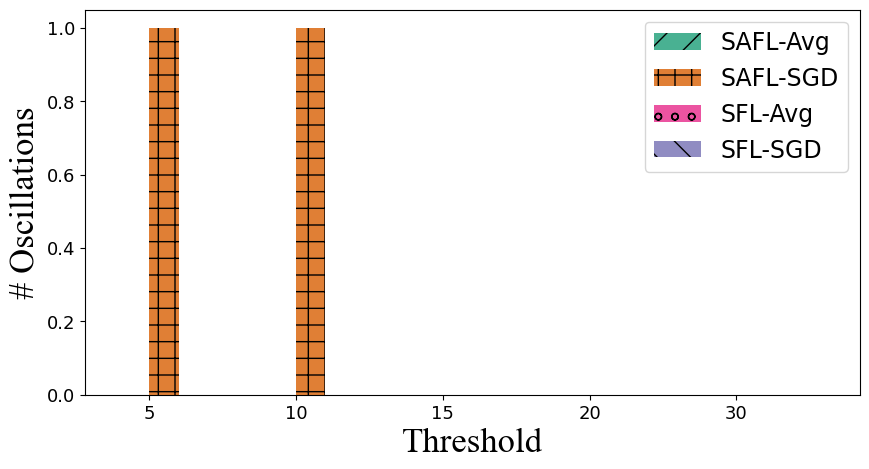}%
}
\caption{Statistics of severe oscillations of the LSTM model under the Shakespeare dataset.}
\end{figure}

\subsection{Sentiment140 @ LSTM}

\begin{figure}[!h]
\vspace{-1ex}
\centering
\subfloat[$\sigma = 0.1$]{\includegraphics[width=1.5in]{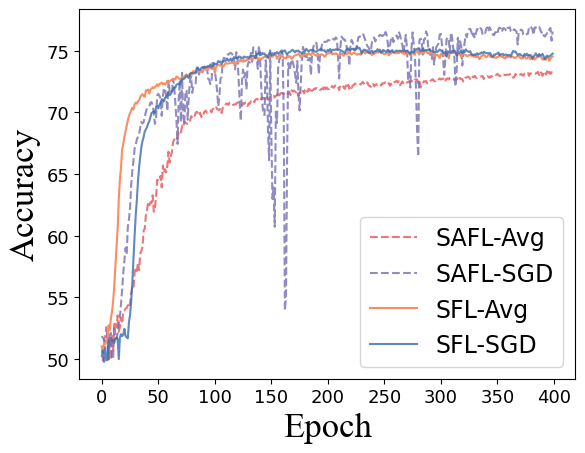}}
\hspace{1mm}
\subfloat[$\sigma = 0.1$]{\includegraphics[width=1.5in]{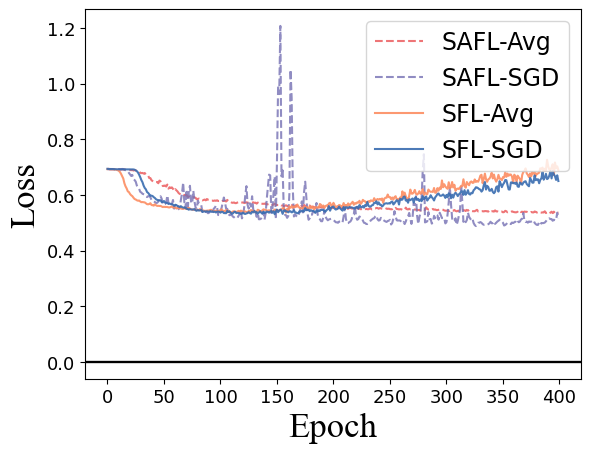}}
\hspace{1mm}
\subfloat[$\sigma = 0.5$]{\includegraphics[width=1.5in]{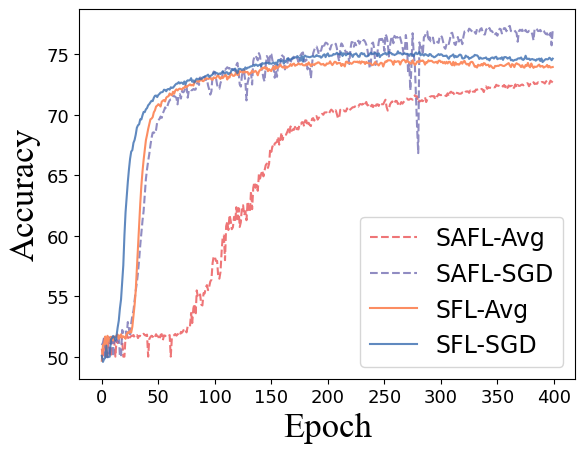}}
\hspace{1mm}
\subfloat[$\sigma = 0.5$]{\includegraphics[width=1.5in]{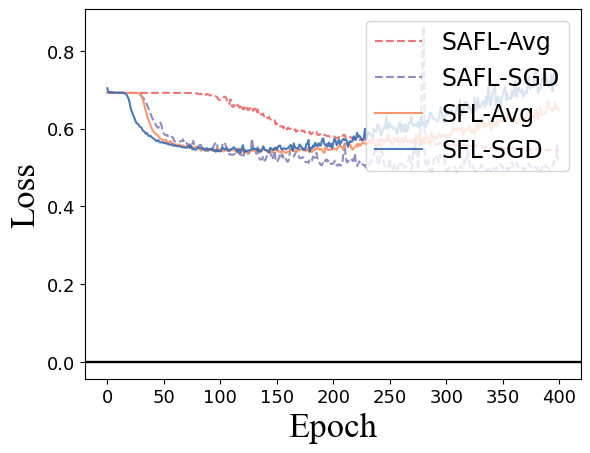}\label{fig_nan_concrete}}
\hspace{1mm}
\subfloat[$\sigma = 1$]{\includegraphics[width=1.5in]{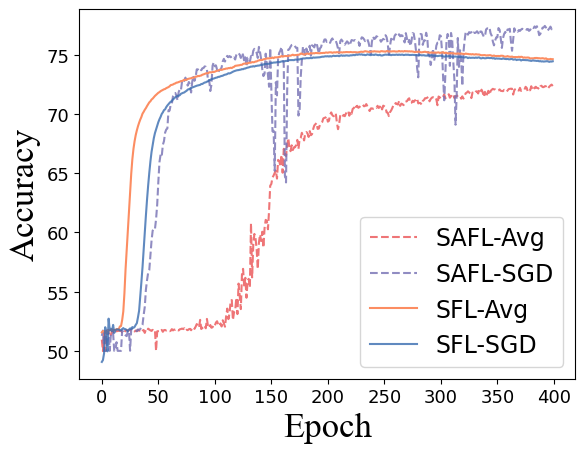}}
\hspace{1mm}
\subfloat[$\sigma = 1$]{\includegraphics[width=1.5in]{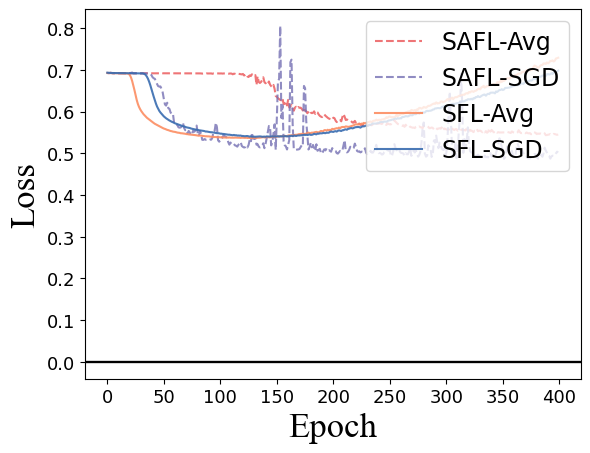}}
\caption{Global accuracy and loss of different models under Sentiment140 dataset using LSTM in SAFL. Note that -1 denotes the NAN value for loss. }
\vspace{-1ex}
\end{figure}
\begin{figure}[!h]
\vspace{-2ex}
\centering
\subfloat[$\sigma = 0.1$]{\includegraphics[width=1.5in]{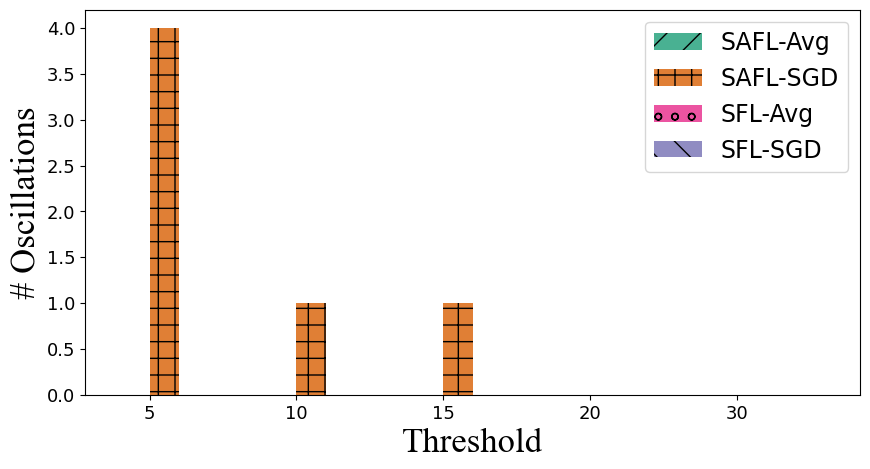}%
}
\hfil
\subfloat[$\sigma = 0.5$]{\includegraphics[width=1.5in]{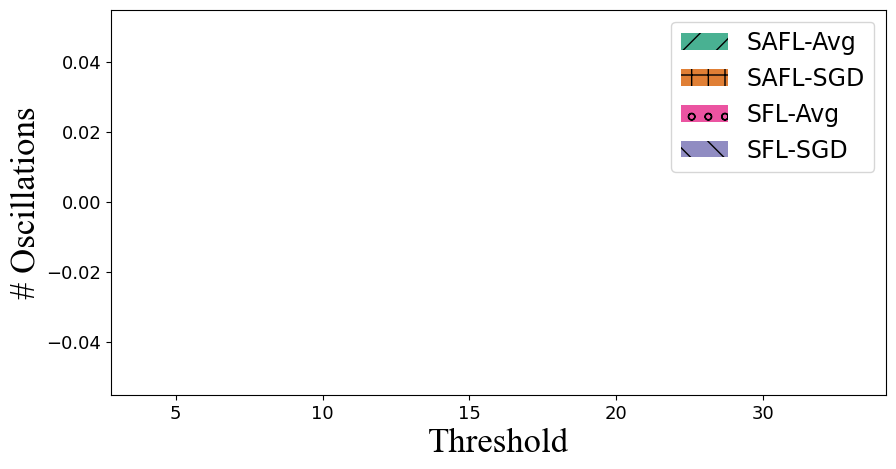}%
}
\hfil
\subfloat[$\sigma = 1$]{\includegraphics[width=1.5in]{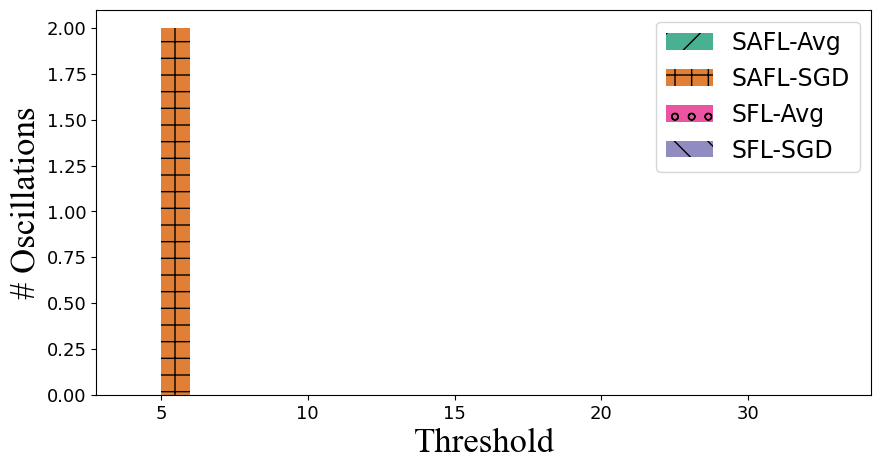}%
}
\caption{Statistics of severe oscillations of the LSTM model under the Sentiment140 dataset.}
\end{figure}
\end{document}